\documentclass[unnumsec,webpdf,contemporary,large,numbered]{oup-authoring-template}

\usepackage{graphicx}
\usepackage{amsmath}
\usepackage{booktabs} 
\usepackage{nameref}
\usepackage{float} 
\usepackage{placeins}
\usepackage{flafter} 
\usepackage{array}
\usepackage{multirow}
\usepackage{xspace}
\AtBeginDocument{%
  \let\STATE\State
  \let\RETURN\Return
  
  \let\ENDIF\EndIf
  \let\ENDFOR\EndFor
  \newcommand{\IF}[1]{\If{#1}}%
  \newcommand{\FOR}[1]{\For{#1}}%
}
\AtBeginDocument{\bibpunct{[}{]}{,}{n}{}{,}}
\usepackage{hyperref}
\hypersetup{colorlinks=true,citecolor=blue,urlcolor=blue,linkcolor=blue}

\newcolumntype{R}[1]{>{\raggedright\arraybackslash}p{#1}}
\newcolumntype{C}[1]{>{\centering\let\newline\\\arraybackslash\hspace{0pt}}m{#1}}

\graphicspath{{Pic/}}

\renewcommand{\topfraction}{1}
\renewcommand{\bottomfraction}{1}
\renewcommand{\textfraction}{0}
\renewcommand{\floatpagefraction}{1}

\newcommand{\proposed}{HGPRGDel\xspace}

\makeatletter
\def\@journaltitle{}%
\def\@pubyear{}%
\def\@copyrightyear{}%
\def\@vol{}%
\def\@issue{}%
\def\@volume{}%
\def\@appnotes{}%
\def\@DOI{}%
\def\@access{}%
\def\@history{}%
\def\ps@opening{%
  \let\@oddhead\@empty
  \let\@evenhead\@empty
  \def\@oddfoot{\hfill\normalsize\thepage\hfill}%
  \def\@evenfoot{\hfill\normalsize\thepage\hfill}%
}
\def\ps@headings{%
  \let\@oddhead\@empty
  \let\@evenhead\@empty
  \def\@oddfoot{\hfill\normalsize\thepage\hfill}%
  \def\@evenfoot{\hfill\normalsize\thepage\hfill}%
  \def\titlemark##1{\markboth{##1}{##1}}%
}
\makeatother

\begin{document}

\firstpage{1}

\title[HGPRGDel]{HGPRGDel: GPR-Aware Learning and Ranking for Growth-Coupled Gene Deletion Design}

\author[1]{Ziwei Yang}
\author[1]{Takeyuki Tamura}

\address[1]{\orgdiv{Bioinformatics Center}, \orgname{Institute for Chemical Research, Kyoto University}, \orgaddress{Kyoto, Japan}}

\corresp[$\ast$]{To whom correspondence should be addressed. E-mail: \href{mailto:tamura@kuicr.kyoto-u.ac.jp}{tamura@kuicr.kyoto-u.ac.jp}.}

\abstract{
Designing microbial strains for growth-coupled production, in which cellular growth is linked to the synthesis of a target compound through genetic interventions, is a fundamental goal in metabolic engineering. Constraint-based methods can identify growth-coupled deletion strategies, but design for each target typically requires a separate search across a large number of possible gene-deletion combinations. This motivates a reusable approach that learns from previously evaluated strain designs to prioritize candidate interventions before downstream constraint-based evaluation.
We present a reproducible protocol for growth-coupled gene deletion design that learns and ranks gene-deletion priorities for each target metabolite. Using computationally curated deletion strategies as learning resources, the protocol generates continuous deletion-priority rankings, constructs candidate strategies at adjustable search-space budgets, and evaluates these candidates through constraint-based metabolic analysis. Its learning component, \proposed, explicitly represents gene--protein--reaction (GPR) logic, capturing how multiple genes can jointly support a reaction or alternative genes can maintain its activity, while integrating molecular and reaction-level information.
Across \textit{e\_coli\_core}, \textit{iMM904}, and \textit{iML1515}, the protocol achieves the highest macro-F1 and AUC among the evaluated learning configurations. \proposed's rankings recovered an average of 90\% of the reference deletion genes from MetNetComp by examining fewer genes than the evaluated baseline methods.
By separating learned intervention prioritization from downstream constraint-based assessment, the protocol provides a reproducible framework for reusing strain-design knowledge while controlling the size of the candidate space and retaining mechanistic evaluation of growth-coupled production.
}


\maketitle


\begin{figure*}[h]
  \centering
  \includegraphics[width=0.97\linewidth]{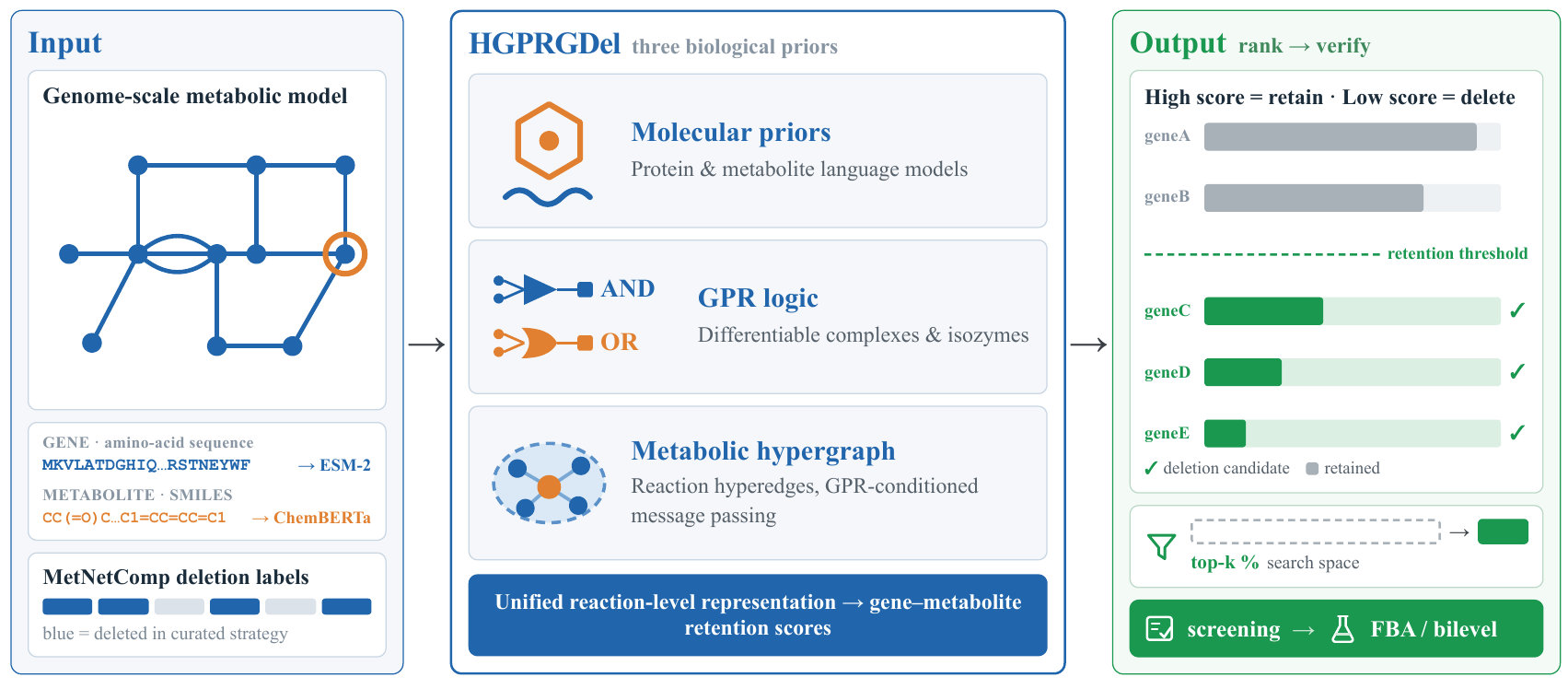}
  \caption*{\textbf{Graphical Abstract.}
\proposed\ is the growth-coupled gene deletion protocol's learning model; MetNetComp provides its reference-strategy layer. Molecular representations, differentiable GPR logic, and reaction-centric hypergraph learning produce continuous gene-retention scores and deletion-priority rankings, which define an ordered search space for candidate deletion strategies and downstream constraint-based evaluation.}
  \label{fig:ga}
\end{figure*}

\section{Introduction}\label{sec:introduction}
Cells rely on a network of biochemical reactions to support growth and produce metabolites. These reactions are catalyzed by enzymes encoded by genes, linking genetic information to metabolic activity. Genome-scale metabolic models (GEMs) are computational models that represent cellular metabolism as a network of biochemical reactions, describing how metabolites are converted into one another and how genes are associated with these reactions~\cite{reed2003ijr904,thiele2010protocol}. Gene--protein--reaction (GPR) associations specify how genes support reaction activity, including cases where multiple genes are jointly required or alternative genes can independently support the same reaction. These relationships are important for evaluating genetic interventions because deleting a gene can alter or eliminate the activity of associated reactions. Constraint-based modeling (CBM) uses mass-balance relationships and constraints on reaction fluxes to identify feasible metabolic states, while flux balance analysis (FBA) estimates metabolic flux distributions under specified constraints and objectives~\cite{orth2010flux,simeonidis2015genome,heirendt2019creation}. Together, these methods provide a mechanistic basis for analyzing how gene deletions affect cellular growth and metabolite production.

Growth-coupled production (GCP) is a strain-design objective in which cellular growth is coupled to the production of a target compound, such that maintaining growth requires the cell to produce that compound. This coupling can provide a selective advantage to productive strains and is therefore useful for designing microbial cell factories~\cite{trinh2008minimal,banerjee2020genome}. Computational strain design seeks gene-deletion strategies that establish this coupling by changing the available metabolic routes. A successful strategy must preserve sufficient metabolic activity for growth while constraining the network so that target production is linked to growth. Identifying such strategies is challenging because each gene-deletion combination can produce a different metabolic state, and the number of possible combinations grows exponentially with the number of genes.

This combinatorial challenge motivates a complementary approach: rather than replacing constraint-based strain design, previously evaluated designs can be used to prioritize candidate interventions for downstream evaluation. We develop a reusable protocol that learns gene--target deletion priorities from computationally curated strategies and uses these priorities to construct candidate deletion sets at adjustable search-space budgets. Each candidate is then evaluated independently using constraint-based metabolic analysis to determine whether it satisfies the growth-coupling objective. The protocol thus separates \textbf{learning-based candidate prioritization} from \textbf{constraint-based evaluation}, allowing the learning stage to guide the search without replacing mechanistic assessment.

The MetNetComp database~\cite{tamura2023metnetcomp} provides reference gene-deletion strategies across GEM--target combinations. These curated strategies serve as learning targets for estimating gene--target deletion priorities and as reference sets for assessing the recovery of known designs. By learning from this resource, the protocol reuses information from previously evaluated strain designs rather than treating each target-specific search as an isolated problem. However, learning useful deletion priorities requires accounting for how gene deletions affect the metabolic network, not merely which genes appear in a reference strategy.

A central consideration is the GPR structure, which encodes the functional relationships between genes and metabolic reactions. GPR rules are Boolean expressions that describe how gene states determine reaction activity~\cite{covert2001regulation,covert2003constraints}. In an AND rule, all genes encoding components of an enzyme complex are required for the associated reaction to remain active; in an OR rule, alternative isozymes can preserve reaction activity when one gene is deleted. Consequently, the effect of a deletion depends not only on gene identity but also on its role in the GPR rules of the affected reactions.

Existing learning-based predictors, including DeepGDel~\cite{yang2025deepgdel} and GraphGDel~\cite{yang2025graphgdel}, incorporate gene--reaction relationships and metabolic network context. However, they do not explicitly represent the compositional effects of GPR AND/OR logic within the learned reaction representation. This motivates a learning component that explicitly accounts for GPR structure while integrating reaction-level network context and molecular information.

\begin{figure}[t]
\centering
\includegraphics[width=0.85\linewidth]{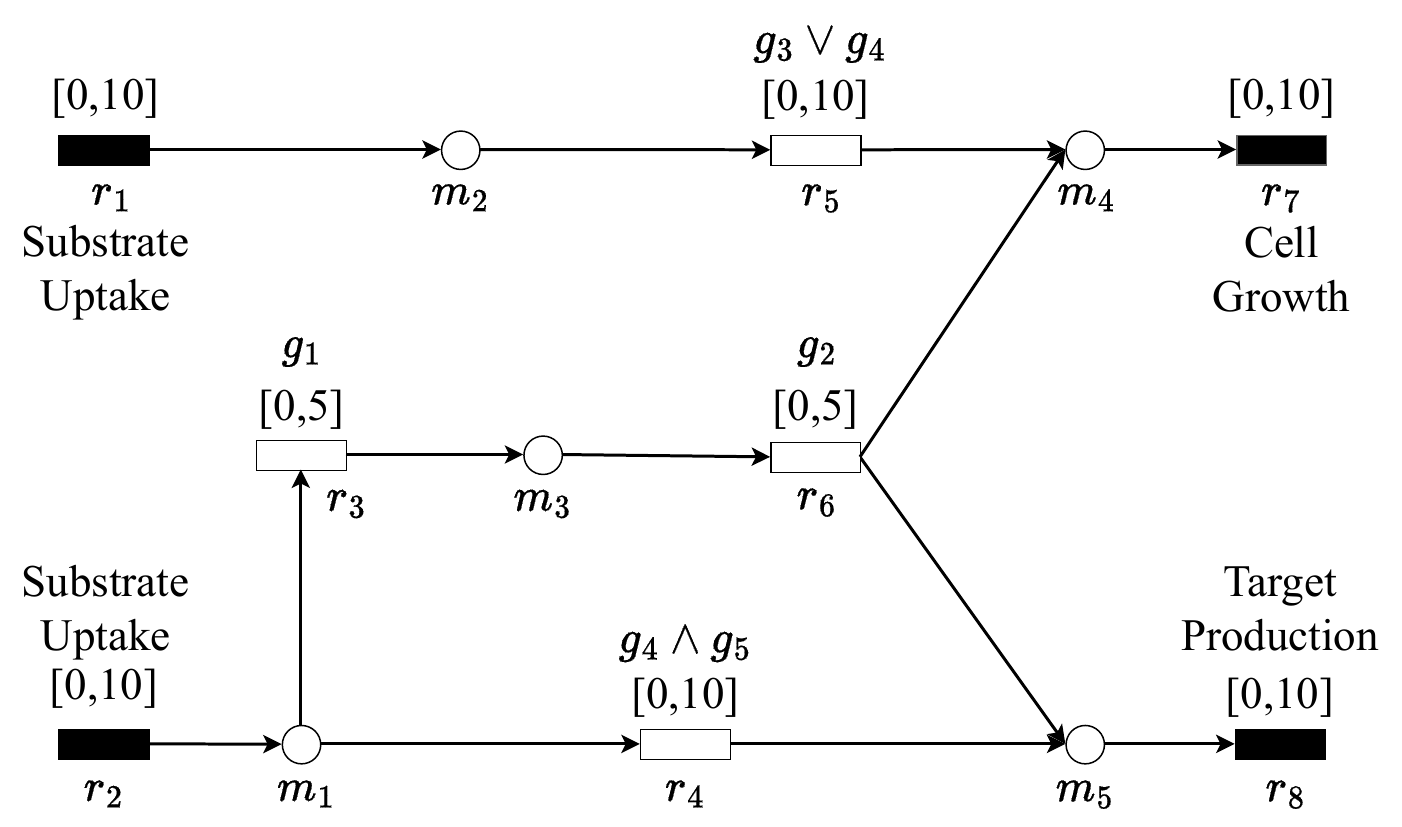}
\caption{Toy model illustrating GPR-dependent deletion patterns. Circles represent metabolites, and rectangles represent reactions ($r_7$ for growth and $r_8$ for target production). The GPR rule for $r_4$ requires both $g_4$ and $g_5$ (enzyme complex), whereas the rule for $r_5$ requires either $g_3$ or $g_4$ (alternative isozymes). Deleting $\{g_3,g_4\}$ disables both reactions, removing the alternative route and the enzyme-complex-associated reaction. Among all $2^5=32$ deletion combinations, only two achieve growth-coupled production of $m_5$, illustrating how GPR logic shapes feasible deletion strategies.}
\label{fig:toy_model}
\end{figure}

Fig.~\ref{fig:toy_model} illustrates the importance of GPR logic through a toy model. Reaction $r_4$ follows an AND rule ($g_4 \wedge g_5$), representing an enzyme complex, whereas $r_5$ follows an OR rule ($g_3 \vee g_4$), representing alternative isozymes. Among all $2^5=32$ deletion combinations, only two strategies, $\{g_3,g_4\}$ and $\{g_3,g_4,g_5\}$, achieve growth-coupled production of the target metabolite. Deleting both $g_3$ and $g_4$ disables $r_5$ by removing isozyme redundancy, while deleting $g_4$ disables $r_4$ by disrupting the enzyme complex. The example illustrates how the same gene can affect multiple reactions through different Boolean relationships, making the functional context of a gene deletion important for identifying growth-coupled strategies.

Because GPR rules determine how gene states affect reaction activity, reactions provide a natural interface between gene-level interventions and metabolic-network behavior. We therefore formulate the learning component around reaction representations (Fig.~\ref{fig:toy_model}). For each gene--target metabolite pair, \proposed aggregates reaction-level representations associated with the candidate gene and target metabolite to produce a continuous \textbf{retention score} $\hat{y}_{g,m}\in[0,1]$. The corresponding deletion priority is defined as $d_{g,m}=1-\hat{y}_{g,m}$, with higher values indicating higher deletion priority. Ranking genes by decreasing $d_{g,m}$ defines an ordered intervention space from which nested candidate deletion strategies can be constructed at different search-space budgets. These candidates are subsequently evaluated using constraint-based metabolic analysis.

The learning component, \proposed, integrates three complementary sources of biological information:

\begin{itemize}
\item \textbf{GPR-M (SoftBooleanAggregator):} Transforms each reaction's Boolean GPR rule into a differentiable reaction embedding through tree-structured AND/OR aggregation, encoding enzyme-complex and isozyme logic as learnable inductive biases.

\item \textbf{Graph-M (reaction-centric hypergraph):} Represents each reaction as a hyperedge over its participating metabolites and propagates network context through hypergraph convolution conditioned on GPR-derived attributes, preserving reaction-level multi-metabolite structure.

\item \textbf{Gene-M / Meta-M (frozen ESM-2 and ChemBERTa):} Provide pretrained representations of protein sequences and metabolite SMILES, respectively, without fine-tuning these molecular encoders on the limited per-GEM labels.

\end{itemize}

Together, these components implement the learning stage of the protocol. They prioritize candidate interventions but do not replace the downstream constraint-based evaluation used to determine whether a constructed deletion set achieves the intended growth-coupling phenotype.

This work makes three contributions:

\begin{enumerate}
\item \textbf{Protocol.}
We present a reusable protocol for growth-coupled gene deletion design that transfers information from computationally curated deletion strategies to gene--target prioritization tasks. The protocol combines learned deletion priorities, nested candidate construction, and downstream constraint-based evaluation within a reproducible workflow.

\item \textbf{Learning component.}
We develop \proposed, a GPR-aware learning component within the protocol that integrates differentiable GPR logic, reaction-centric hypergraph structure, and pretrained molecular representations to generate gene--target deletion priorities.

\item \textbf{Evaluation and evidence.}
We evaluate the protocol across three GEMs and show that \proposed achieves the highest macro-F1 and AUC among the evaluated learning methods on each GEM. The learned rankings recover computationally curated MetNetComp reference gene sets and support adjustable candidate-space restriction. Ablation and representation analyses further characterize the contributions of GPR, molecular, and topological information to the learned rankings.

\end{enumerate}

\section{Related Work}\label{sec:related}

\subsection{Constraint-Based Methods for Growth-Coupled Strain Design}\label{subsec:cbm} 
Flux balance analysis~\cite{orth2010flux} provides the steady-state metabolic flux prediction that underpins computational strain design. Bilevel programming methods such as OptKnock~\cite{burgard2003optknock} formulate growth-coupled production as an inner/outer optimization, in which the inner program maximizes growth and the outer program identifies gene knockouts that enforce production at the growth optimum. Subsequent methods improved search efficiency through local search, preprocessing, mixed-integer formulations, minimum-cut-set enumeration, and reaction-network-based restrictions~\cite{yang2011emilio,ohno2014fastpros,gu2016idealknock,schneider2020extended,tamura2023gene,yang2025dbgdel,tamura2022trimming}. These methods established constraint-based optimization as a general route for growth-coupled strain design, but the search remains target-specific. Each new GEM--target combination requires a separate optimization or search procedure.

\subsection{Machine Learning for Metabolic Gene Perturbation}\label{subsec:ml} 
Machine learning has been applied to metabolic modeling for tasks including pathway flux prediction~\cite{ML_meta_1} and enzyme kinetics~\cite{ML_meta_2}. Direct learning of gene deletion priorities from curated design strategies is a more recent direction. DeepGDel~\cite{yang2025deepgdel} introduced supervised learning of gene deletion strategies using sequence-based representations, while GraphGDel~\cite{yang2025graphgdel} added metabolic-network topology through graph-based representations. These methods established learning-based deletion prioritization as a complementary approach to target-specific optimization. An important question is how to represent the biological relationships that connect gene perturbations to reaction activity. GPR associations provide an explicit representation of this mapping through Boolean AND/OR rules. Reaction-centric and hypergraph representations have been explored for metabolic-network learning, including gene essentiality prediction~\cite{hasibi2024integration} and reaction prediction in genome-scale models~\cite{wang2026hypergraph}. However, these approaches do not formulate deletion prioritization around an explicit differentiable representation of GPR logic. This motivates a reaction-centric formulation in which GPR structure is treated as a biological prior within the learned reaction representation, while metabolic topology provides additional context. 

\subsection{Foundation Models in Systems Biology}\label{subsec:fm} 
Protein and molecular language models provide pretrained representations of sequence and chemical information. ESM-2~\cite{esm2} derives protein representations from large-scale sequence pretraining, while ChemBERTa~\cite{chemberta} provides molecular representations from SMILES strings. Given the limited supervision available within individual GEMs, freezing these pretrained backbones and training lightweight task-specific projections provides a practical way to use these representations without introducing a large number of trainable parameters. Together, these developments suggest a progression from target-specific constraint-based search toward reusable learning-based design, and from generic learned representations toward models that explicitly encode biological structure. Our framework follows this progression by combining curated computational design knowledge, GPR logic, metabolic topology, and pretrained molecular representations within a reusable learning protocol for growth-coupled gene deletion design.

\section{The Growth-Coupled Gene Deletion Protocol}\label{sec:framework}

\subsection{Protocol Overview}\label{subsec:protocol-overview}

Drawing on previous work on growth-coupled gene deletion design, we present a reproducible protocol for learning, ranking, constructing, and evaluating candidate gene-deletion strategies for growth-coupled metabolite production (Fig.~\ref{fig:framework}). The protocol builds on the MetNetComp reference database~\cite{tamura2023metnetcomp} and previous methods, including DeepGDel~\cite{yang2025deepgdel}, GraphGDel~\cite{yang2025graphgdel}, and the DBgDel optimization framework~\cite{yang2025dbgdel}.
The protocol takes a GEM, a target metabolite, and molecular sequence information as inputs. Computationally curated reference deletion strategies from MetNetComp serve as supervised learning targets and reference sets for evaluating gene-set recovery (Section~\ref{subsec:problem}). The protocol comprises five stages: (i) target-specific problem definition, (ii) \proposed\ scoring, (iii) deletion-priority ranking, (iv) candidate strategy construction, and (v) downstream constraint-based assessment. Throughout this section, \emph{protocol} refers to the complete workflow, whereas \proposed\ denotes its GPR-aware learning component.
Within this workflow, \proposed\ integrates molecular and metabolic information into a shared GPR-aware reaction representation (Sections~\ref{subsec:molecular}--\ref{subsec:graph}) and produces a continuous gene-retention score for each gene--target pair (Section~\ref{subsec:pred}). These scores are converted into deletion priorities to define an ordered search space, from which nested candidate deletion sets are constructed (Section~\ref{subsec:candidates}). The candidate sets are then assessed independently using downstream constraint-based metabolic analysis (Supplementary Section~S3.3).
 
\subsection{Problem Definition and Reference Strategies}\label{subsec:problem}

A genome-scale metabolic model (GEM) $\mathcal{C}$ encodes cellular metabolism as a stoichiometric network with metabolite, reaction, and gene sets $M$, $R$, and $G$, together with GPR associations that link gene presence to reaction activity~\cite{covert2001regulation}.
A deletion strategy is considered growth-coupled for a target metabolite when it satisfies the weak growth-coupling criterion used in our downstream FBA evaluation: biomass is first maximized and fixed at its optimum, target production is then maximized, and both growth and production must reach at least $0.001$ (Supplementary Section~S2.2). The supervised reference labels are the deletion gene sets reported by MetNetComp for each target metabolite~\cite{tamura2023metnetcomp}; their label construction and encoding are described in Supplementary Section~S1.2.

\textbf{Learning task.} \proposed\ outputs a continuous retention score $\hat{y}_{g,m}\in[0,1]$ for each gene--target pair, with lower values indicating higher deletion priority.
Defining $d_{g,m}=1-\hat{y}_{g,m}$, ranking genes by decreasing $d_{g,m}$ produces a complete deletion-priority order.
This ranking defines an ordered search space from which candidate deletion sets of different sizes can be derived: the candidate set of size $k$ consists of the $k$ genes with highest deletion priority, so candidate sets at increasing $k$ are nested.
A threshold $\theta$ may be applied to select one operating point from this ranking; the choice of operating point is not part of the model definition.
The learning task is to estimate and rank deletion priorities; exhaustive enumeration of feasible knockout sets is outside its scope.

\begin{figure*}[t]
  \centering
  \includegraphics[width=0.9\linewidth]{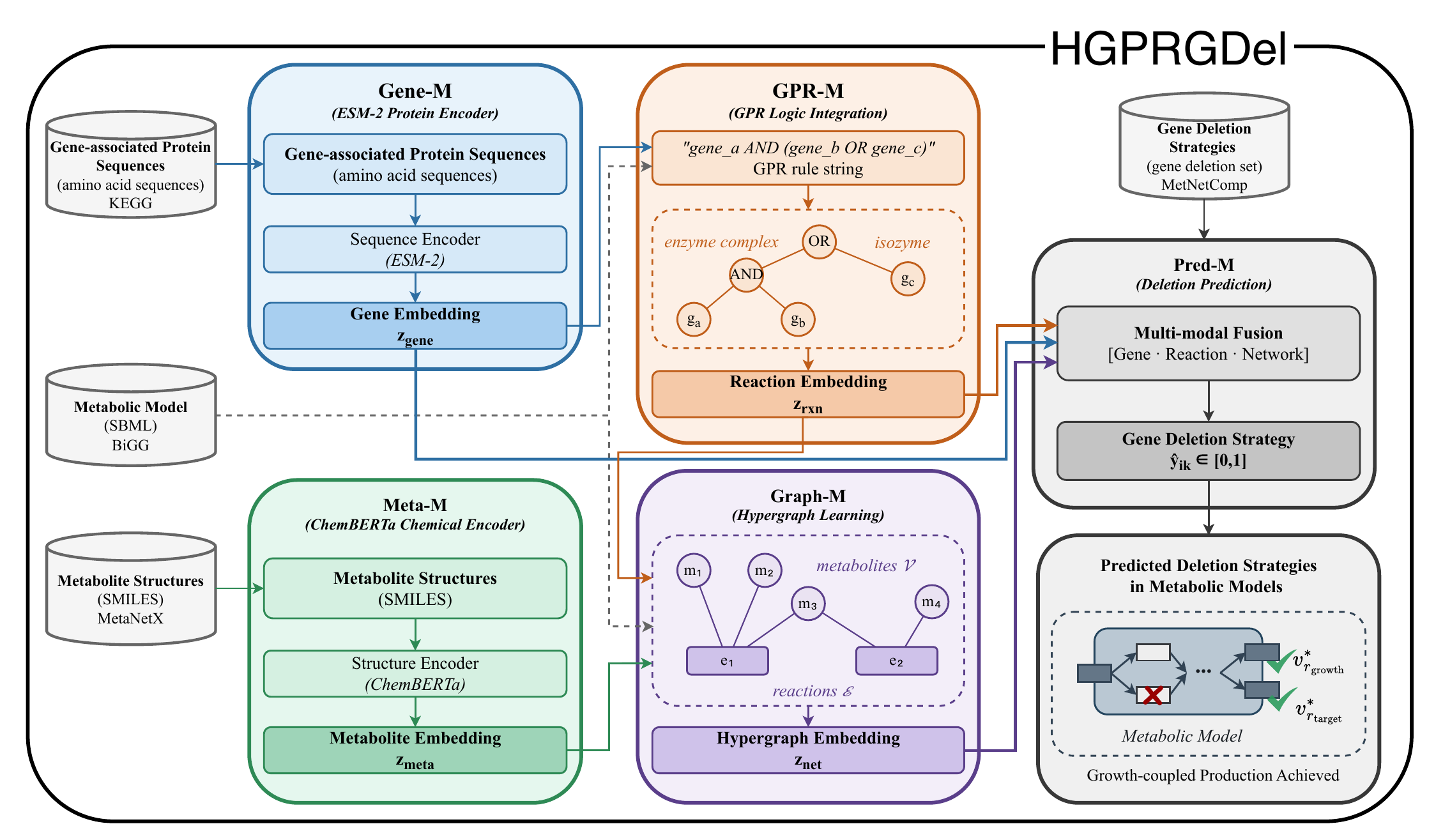}
  \caption{Overview of the growth-coupled gene deletion protocol. The protocol takes a GEM, a target metabolite, and molecular sequence information as input and uses computationally curated MetNetComp reference strategies for supervised learning. \proposed\ serves as the learning component: GPR-M converts Boolean AND/OR GPR structure into reaction embeddings; Graph-M propagates metabolic context through a reaction-centric hypergraph; Gene-M and Meta-M provide frozen molecular representations; and Pred-M converts the resulting gene, metabolite, and reaction representations into a continuous retention score $\hat{y}_{g,m}$. The resulting gene--target scores define a deletion-priority ranking, which provides an ordered search space for constructing nested candidate deletion sets and downstream constraint-based assessment.}
  \label{fig:framework}
\end{figure*}

Within the learning layer, \proposed\ is organized around three complementary biological priors introduced sequentially into a shared reaction representation (Fig.~\ref{fig:framework}):
\textbf{Molecular prior} encodes sequence-derived information via frozen pretrained language models;
\textbf{GPR prior} transforms Boolean GPR rules into differentiable reaction embeddings;
\textbf{Topology prior} propagates metabolite context through a reaction-centric hypergraph.
Five modules implement the learning component (Gene-M, Meta-M, GPR-M, Graph-M, and Pred-M); their interactions are detailed in the following subsections.

\subsection{Encoding Molecular Information}\label{subsec:molecular}

Within \proposed, compound and gene features are extracted from frozen pretrained language models to introduce the \textbf{molecular prior} into reaction representation learning, without fine-tuning on limited per-GEM labels.
The molecular prior module (Meta-M) encodes metabolite SMILES with frozen ChemBERTa~\cite{chemberta} (\texttt{DeepChem/ChemBERTa-77M-MLM}); the gene prior module (Gene-M) encodes protein sequences with frozen ESM-2~\cite{esm2} (\texttt{esm2\_t6\_8M\_UR50D}).
For each input, mean-pooled backbone outputs are mapped by a trained linear projection into the corresponding module's representation space, yielding compound features $\mathbf{X}\in\mathbb{R}^{d_m}$ (stacked over GEM metabolites) and gene embeddings $\mathbf{Z}_{\text{gene}}^{(g)}\in\mathbb{R}^{d_g}$.
The two projections are independent, so $d_g$ and $d_m$ need not coincide ($d_g{=}320$, $d_m{=}256$; Supplementary Table~S5).
Only the projection layers are trained: this limits trainable parameters under the available per-GEM supervision and enables controlled comparisons with non-sequence baselines.
Molecular priors are thus available as inputs to reaction representation learning without supervised fine-tuning of the backbone encoders.

\subsection{Differentiable GPR Logic Modeling}\label{subsec:gpr}

\subsubsection{Biological Rationale}

GPR rules determine how gene deletions propagate to reaction activity.
Enzyme complexes, encoded as AND nodes, require all subunits; deleting any single subunit is sufficient to block the reaction.
Isozymes, encoded as OR nodes, allow any alternative to sustain the reaction; blocking requires deleting all alternatives.
Because these Boolean distinctions must be represented differentiably for gradient-based learning, they require a tailored aggregation scheme rather than standard mean pooling.

\subsubsection{GPR-Specific Differentiable Aggregator}

The \textbf{GPRRuleParser} converts COBRA-style Boolean rule strings into AND/OR parse trees with gene leaves; nested rules are handled recursively.
The \textbf{SoftBooleanAggregator} evaluates each tree differentiably (Algorithm~\ref{algo:gpr-soft-bool}).

For a conjunction (enzyme complex) with child embeddings $\{\mathbf{u}_i\}_{i=1}^{k}$,
\begin{equation}
  \mathbf{z}_{\text{and}} = \phi_{\land}\bigl([\mathbf{m} \| \mathbf{s}]\bigr),
  \quad
  \mathbf{m} = \tfrac{1}{K}\sum_{i=1}^{K}\mathbf{u}_i,
  \quad
  \mathbf{s} = -\log\sum_{i=1}^{K}\exp(-\mathbf{u}_i/\tau),
\end{equation}
where $K$ is the number of children of the node, $\mathbf{s}$ is a soft bottleneck over the weakest-contributing subunit, and $\tau>0$ is the effective soft-AND temperature, obtained from a single learnable scalar $\hat\tau$ (Supplementary Section~S7.1).
The combiners $\phi_{\land}$ and $\phi_{\lor}$ are two-layer multilayer perceptrons (MLPs) with ReLU activation (hidden width $256$; Supplementary Section~S7.1).
For a disjunction (isozymes),
\begin{equation}
  \mathbf{z}_{\text{or}} = \phi_{\lor}\bigl([\mathbf{a} \| \mathbf{l}]\bigr),
  \quad
  \mathbf{a} = \mathrm{MHA}(\mathbf{q},\mathbf{U},\mathbf{U}),
  \quad
  \mathbf{l} = \log\sum_{i=1}^{K}\exp(\mathbf{u}_i),
\end{equation}
where $\mathrm{MHA}$ denotes multi-head attention, $\mathbf{q}=\tfrac{1}{K}\sum_i\mathbf{u}_i$, and $\mathbf{U}=[\mathbf{u}_1^\top;\cdots;\mathbf{u}_K^\top]$. The attention weights capture differences among isozyme alternatives, while log-sum-exp encodes the any-sufficient principle.
Both combiners map $\mathbb{R}^{2d_g}\to\mathbb{R}^{d_g}$, so GPR-M maps each reaction to a reaction embedding $\mathbf{Z}_{\text{rxn}}^{(r)}\in\mathbb{R}^{d_g}$ that preserves the gene-embedding dimension.
The reduction to the smaller reaction-context dimension $D_{\text{rxn}}$ happens downstream, in Pred-M (Section~\ref{subsec:pred}).
Gradient expressions, temperature constraints, and numerical stability details are given in Supplementary Section~S7.
The GPR prior is thus embedded into reaction representations, with Boolean AND/OR logic made differentiable and available for downstream topology-level learning. GPR-M operates on gene embeddings derived from protein sequence alone; deletion labels are used as supervised targets, and their relationship to the resulting reaction representations is learned downstream by Pred-M (Section~\ref{subsec:pred}).

\begin{algorithm}[t]
  \caption{Differentiable evaluation of GPR logic for reaction representation learning.\label{algo:gpr-soft-bool}}
  \begin{algorithmic}[1]
    \STATE \textbf{Input:} parse tree $T$; gene embeddings $\{\mathbf{z}_g\}$; temperature $\tau>0$; MLPs $\phi_{\land}$, $\phi_{\lor}$; multi-head attention $\mathrm{MHA}$.
    \STATE \textbf{Output:} reaction embedding $\mathbf{z}_{\mathrm{rxn}} \in \mathbb{R}^{d_g}$ ($d_g{=}320$; the combiners preserve the gene-embedding dimension).
    \STATE \textbf{procedure} $\mathrm{GPR\text{-}Eval}(T)$
    \IF{$T$ is empty} \RETURN $\mathbf{0}$ \ENDIF
    \IF{$T$ is leaf gene $g$} \RETURN $\mathbf{z}_g$ (or $\mathbf{0}$ if missing) \ENDIF
    \IF{$T = T_1 \,\texttt{and}\, \cdots \,\texttt{and}\, T_K$}
      \FOR{$i=1$ to $K$} \STATE $\mathbf{u}_i \leftarrow \mathrm{GPR\text{-}Eval}(T_i)$ \ENDFOR
      \STATE $\mathbf{m} \leftarrow \frac{1}{K}\sum_i\mathbf{u}_i$; \quad $\mathbf{s} \leftarrow -\log\sum_i\exp(-\mathbf{u}_i/\tau)$
      \RETURN $\phi_{\land}([\mathbf{m} \| \mathbf{s}])$
    \ENDIF
    \IF{$T = T_1 \,\texttt{or}\, \cdots \,\texttt{or}\, T_K$}
      \FOR{$i=1$ to $K$} \STATE $\mathbf{u}_i \leftarrow \mathrm{GPR\text{-}Eval}(T_i)$ \ENDFOR
      \STATE $\mathbf{U} \leftarrow [\mathbf{u}_1^\top;\cdots;\mathbf{u}_K^\top]$; \quad $\mathbf{q} \leftarrow \frac{1}{K}\sum_i\mathbf{u}_i$
      \STATE $\mathbf{a} \leftarrow \mathrm{MHA}(\mathbf{q},\mathbf{U},\mathbf{U})$; \quad $\mathbf{l} \leftarrow \log\sum_i\exp(\mathbf{u}_i)$
      \RETURN $\phi_{\lor}([\mathbf{a} \| \mathbf{l}])$
    \ENDIF
  \end{algorithmic}
\end{algorithm}

\subsection{Reaction-Centric Hypergraph Learning}\label{subsec:graph}

\subsubsection{Hypergraph Formulation}

We represent the metabolic network as a hypergraph $\mathcal{H} = (\mathcal{V}, \mathcal{E})$ where compound nodes $\mathcal{V}$ are metabolites and each reaction $r \in R$ is a hyperedge connecting all its substrates and products.
The incidence matrix $\mathbf{H} \in \{0,1\}^{|\mathcal{V}| \times |R|}$ marks metabolites that participate in each reaction.
Currency metabolites (ATP, NADH, H$_2$O, etc.) are filtered to suppress topological hubs that would otherwise dominate message passing.
Unlike pairwise metabolite graphs, each reaction remains an explicit computational unit throughout message passing, preserving reaction identity and multi-metabolite structure rather than decomposing multi-participant reactions into pairwise edges.

\subsubsection{GPR-Conditioned Message Passing (Graph-M)}

Graph-M takes compound features $\mathbf{X}$ from Meta-M, the GEM incidence matrix $\mathbf{H}$ (fixed across folds; Supplementary Section~S2.6.1), and GPR-derived reaction embeddings $\mathbf{Z}_{\text{rxn}}^{(r)}$.
Each hyperedge attribute combines aggregated compound features with the GPR projection:
\begin{equation}
  \mathbf{A}_{\text{he}}^{(r)}
  = \mathbf{W}_{\text{he}} \Bigl[
      \bigl(\mathbf{H}^\top \mathbf{X}\bigr)_r
      \;\Big\|\;
      \mathbf{W}_{\text{gpr}} \mathbf{Z}_{\text{rxn}}^{(r)}
    \Bigr].
\end{equation}
Two stacked \texttt{HypergraphConv} layers~\cite{hypergraph_conv} (\texttt{torch\_geometric}) with node-level attention produce graph-refined compound representations $\mathbf{Z}_{\text{metaG}}$; the attention weights $\alpha_{v,r}^{(\ell)}$ and $\beta_{v,r}^{(\ell)}$ follow this layer's built-in attention mechanism.
At each layer $\ell\in\{0,1\}$, message passing proceeds in two steps.
First, each hyperedge (reaction) aggregates its incident node features, conditioned on the GPR-derived attribute $\mathbf{A}_{\text{he}}^{(r)}$:
\begin{equation}
  \mathbf{m}^{(r,\ell)}
  = \sigma\!\Bigl(
      \mathbf{W}_{1}^{(\ell)}
      \sum_{v \in \mathcal{N}(r)} \alpha_{v,r}^{(\ell)}\,\mathbf{h}_{v}^{(\ell)}
      + \mathbf{A}_{\text{he}}^{(r)}
    \Bigr),
\end{equation}
where $\mathbf{h}_{v}^{(0)} = \mathbf{X}_v$ (Meta-M compound features), $\mathcal{N}(r)$ denotes the metabolites participating in reaction $r$ (i.e.\ the non-zero rows of column $r$ of $\mathbf{H}$), and $\alpha_{v,r}^{(\ell)}$ are learned node-level attention weights.
Second, each node (metabolite) updates its representation by aggregating messages from all reactions it participates in:
\begin{equation}
  \mathbf{h}_{v}^{(\ell+1)}
  = \sigma\!\Bigl(
      \mathbf{W}_{2}^{(\ell)}
      \sum_{r \in \mathcal{R}(v)} \beta_{v,r}^{(\ell)}\,\mathbf{m}^{(r,\ell)}
    \Bigr),
\end{equation}
where $\mathcal{R}(v)$ is the set of reactions in which metabolite $v$ participates and $\beta_{v,r}^{(\ell)}$ are a second set of learned attention weights.
The final node representations after two layers are collected as $\mathbf{Z}_{\text{metaG}} = \{\mathbf{h}_{v}^{(2)}\}_{v \in \mathcal{V}}$.
GPR-derived hyperedge attributes make gene-level Boolean structure available during message passing, conditioning metabolite context on the GPR-derived reaction attributes of the reactions in which each metabolite participates.

\subsection{Deletion-Priority Ranking: Continuous Gene-Retention Prediction and Training}\label{subsec:pred}

For gene--target pair $(g, m)$, Pred-M pools GPR-M reaction embeddings over $R_{g \cup m} = R_g \cup R_m$, where $R_g$ denotes the set of reactions associated with gene $g$ and $R_m$ denotes the set of reactions involving target metabolite $m$; this union pools reaction-level evidence relevant to either the candidate gene or the target metabolite, so Pred-M retains context even when the two do not co-occur in a shared reaction:
\begin{equation}
  \mathbf{c}_{\text{rxn}}^{(g,m)} = \frac{1}{|R_{g \cup m}|}\!\!\sum_{r \in R_{g \cup m}}\!\! \mathbf{Z}_{\text{rxn}}^{(r)}.
\end{equation}
The pooled context $\mathbf{c}_{\text{rxn}}^{(g,m)}\in\mathbb{R}^{d_g}$ is reduced by the association MLP $\mathbb{R}^{d_g}\to\mathbb{R}^{D_{\text{rxn}}}$ ($D_{\text{rxn}}{=}128$; Supplementary Table~S5). The reduced reaction context is concatenated with the gene and target-metabolite representations to form the joint representation $\mathbf{Z}^{(g,m)} = [\mathbf{Z}_{\text{gene}}^{(g)} \| \mathbf{h}_{m}^{(2)} \| \mathrm{MLP}(\mathbf{c}_{\text{rxn}}^{(g,m)})] \in \mathbb{R}^{d_g + d_m + D_{\text{rxn}}}$, where $\mathbf{h}_{m}^{(2)} \in \mathbf{Z}_{\text{metaG}}$ is the embedding of the compound node corresponding to target metabolite $m$. An MLP followed by a sigmoid maps this joint representation to the gene-retention score $\hat{y}_{g,m}\in[0,1]$, where lower values indicate higher deletion priority.
Model selection and operating-point selection are decoupled: training is early-stopped on the threshold-free AUC of an inner validation split ($20\%$ of each fold's training partition), and only afterwards is $\theta$ chosen on that same inner split by maximizing macro-F1 over an $81$-point grid on $[0.1,0.9]$ ($\theta$ is distinct from the learnable GPR-M temperature $\tau$).
The ranking induced by $\hat{y}_{g,m}$ is retained independent of threshold selection.

Training minimizes a per-gene entropy-weighted binary cross-entropy over supervised pairs:
\begin{equation}
  \mathcal{L}(\mathcal{B}) = \frac{\sum_{(g,m)\in\mathcal{B}} w_g\,\ell(y_{g,m},\hat{y}_{g,m})}{\sum_{(g,m)\in\mathcal{B}} w_g},
  \qquad
  w_g = \epsilon + (1-\epsilon)\,H_g^{\gamma},
\end{equation}
where $H_g$ is the binary Shannon entropy of gene $g$'s label distribution across supervised targets ($\gamma{=}1.0$, $\epsilon{=}0.05$; Supplementary Section~S2.5). This weighting reduces the influence of genes with nearly constant labels across targets and emphasizes genes whose deletion status varies with the target metabolite.
Molecular encoders (ESM-2, ChemBERTa) remain frozen while all other modules are trained jointly.
Full implementation details and hyperparameters are listed in Supplementary Table~S5.
The molecular, GPR-derived, and topology-conditioned representations are thus jointly translated into gene-retention scores, completing the learning layer of the protocol and yielding the continuous deletion-priority ranking used in downstream candidate construction.

\section{Results}\label{sec:experiments}

\subsection{Experimental Setup}\label{subsec:setup}

\textbf{Datasets.}
We evaluate \proposed\ on three GEMs:
\textit{e\_coli\_core}~\cite{orth2010reconstruction} (95 reactions, compact \textit{E.\ coli});
\textit{iMM904}~\cite{mo2009connecting} (1,577 reactions, genome-scale \textit{S.\ cerevisiae}); and
\textit{iML1515}~\cite{monk2017ml1515} (2,712 reactions, large \textit{E.\ coli} reconstruction).
Growth-coupled deletion labels are obtained from MetNetComp~\cite{tamura2023metnetcomp}, using the curated maximal deletion strategy available for each target metabolite.
Gene amino acid sequences are from KEGG~\cite{KEGG}; metabolite SMILES from MetaNetX~\cite{MetaNetX}.
Table~\ref{table:model_sum} summarizes model and dataset statistics.

\textbf{Evaluation protocol.}
We use metabolite-based five-fold cross-validation (seed 42): the split unit is the target metabolite, so all gene labels for a metabolite remain in the same fold.
Genes with near-constant labels across metabolites are excluded from the primary reported metrics because their labels provide limited discrimination for target-dependent ranking.
The remaining target-dependent genes, whose deletion labels vary across supervised targets, number 66 on \textit{e\_coli\_core}, 55 on \textit{iMM904}, and 33 on \textit{iML1515} (Table~\ref{table:model_sum}).
This filtering is performed independently of model predictions and applied identically to all methods and baselines under the same CV folds.
A full-gene supplementary evaluation and an all-target extension of the top-10 recovery diagnostic are provided in Supplementary Sections~S3.4 and~S3.5.
Tables~\ref{table:main_exp} and~\ref{table:ablation_exp} report macro-F1 and area under the ROC curve (AUC) as mean $\pm$ s.d.\ over five held-out test folds on target-dependent pairs.

\textbf{Model selection and thresholding.}
Each outer fold reserves $20\%$ of its training partition as an inner validation split, used for model selection and threshold selection only; neither the inner validation metabolites nor the outer test metabolites contribute to the loss.
For each fold, the model state is selected by early stopping on the inner-validation AUC, a threshold-free criterion over target-dependent gene--metabolite pairs (full early-stopping protocol in Supplementary Section~S2.5).
Only after the model is fixed is the decision threshold $\theta$ determined, by maximizing unweighted macro-F1 on the same inner validation split over a grid of $81$ values on $[0.1,0.9]$; $\theta$ is then applied unchanged to the held-out test metabolites.

\textbf{Baselines.}
We compare with: \textbf{DNN} (sequence encoders, no graph); \textbf{DeepGDel}~\cite{yang2025deepgdel} (LSTM-based encoders with supervised prediction); \textbf{NMA}~\cite{yang2025graphgdel} (fixed one-hop metabolite neighborhood aggregation); and \textbf{GraphGDel}~\cite{yang2025graphgdel} (learnable GCN on pairwise metabolite graph).
All baselines are trained and evaluated under identical CV folds, graph inputs, and threshold selection procedure.
Thresholded scores are evaluated as candidate deletion sets against the curated reference gene sets, following the convention that genes outside the reference are labeled retained; the continuous ranking is evaluated separately as a deletion-priority order (Section~\ref{subsec:recovery}).

\begin{table}[h]
  \centering
  \caption{Metabolic model and dataset statistics. ``\#GCP target metabolites'': growth-coupled production (GCP) annotated targets; ``\#Deletion strategies'': targets with a usable curated maximal deletion set. ``\#Target-dependent genes'' and ``\#Target-dependent pairs'' exclude pre-determined genes (constant labels for $\geq 90\%$ of targets; defined above), and, on \textit{iML1515}, gene--target slots without an explicit supervised label in the curated reference (Section~S1.2).}
  \label{table:model_sum}
  \resizebox{0.99\linewidth}{!}{%
  \begin{tabular}{lccc}
  \toprule
  & \textbf{\textit{e\_coli\_core}} & \textbf{\textit{iMM904}} & \textbf{\textit{iML1515}} \\
  \midrule
  \#Genes                             & 137         & 905         & 1516        \\
  \#Pre-determined genes              & 71          & 850         & 1483        \\
  \#Target-dependent genes            & 66          & 55          & 33          \\
  \#Reactions                         & 95          & 1577        & 2712        \\
  \#GCP target metabolites            & 48          & 782         & 1085        \\
  \#Deletion strategies               & 41          & 105         & 443         \\
  \#Target-dependent pairs            & 2706        & 5775        & 13464       \\
  \bottomrule
  \end{tabular}
  }
\end{table}

\subsection{The Core Learning Component Performs Consistently Across Three GEMs}\label{subsec:benchmark}

Across the three GEMs, \proposed\ achieves higher macro-F1 than all four evaluated baselines and outperforms GraphGDel on every individual fold (Table~\ref{table:main_exp}; Fig.~\ref{fig:benchmark_ablation}A; Supplementary Section~S3.2).
The evaluated methods show incremental gains from sequence-based encoders (DNN, DeepGDel) to graph-augmented representations (NMA, GraphGDel), while \proposed\ additionally models GPR logic at the reaction level.

Relative to GraphGDel, \proposed\ combines GPR-aware reaction embeddings with reaction-centric hypergraph propagation, with macro-F1 gains of $14.9$, $7.1$, and $18.4$ percentage points on \textit{e\_coli\_core}, \textit{iMM904}, and \textit{iML1515}, respectively.
The margin is largest on the genome-scale \textit{iML1515} reconstruction, which has the largest reaction count and the largest absolute number of multi-gene (AND/OR) rules.
Cross-fold spread is lowest on the genome-scale \textit{iML1515} ($90.38 \pm 0.74$) and widest on \textit{e\_coli\_core} ($89.11 \pm 3.81$), the GEM with the fewest supervised pairs per fold.

\proposed\ also shows a consistently wider AUC--macro-F1 gap than any baseline ($7.3$, $7.6$, and $6.7$ percentage points across the three GEMs, versus $0.3$--$3.7$ for the baselines).
Because the operating threshold is selected within each training fold, fold-level macro-F1 reflects both ranking quality and the fold-specific operating point; we therefore use the continuous-score and ranking metrics below to assess prioritization independently of any fixed threshold.

\begin{table}[t]
\centering
\caption{Benchmark performance (mean\,$\pm$\,s.d.\ over five held-out metabolite test folds on target-dependent pairs). Bold: best per panel.
The \proposed\ rows report the five-fold protocol of Section~\ref{subsec:setup}, in which each fold uses its own threshold chosen on that fold's inner validation split.
The same scores, extended to all five metrics, are tabulated in Supplementary Table~S7.}
\label{table:main_exp}

\smallskip
\textbf{(A)}\enspace\textit{e\_coli\_core}\\[3pt]
\begin{tabular}{lcc}
\toprule
Method & Macro-F1 (\%) & AUC (\%) \\
\midrule
DNN        & $58.94 \pm 3.53$ & $62.18 \pm 3.96$ \\
DeepGDel   & $70.16 \pm 3.73$ & $70.47 \pm 3.90$ \\
NMA        & $70.91 \pm 3.78$ & $71.25 \pm 3.94$ \\
GraphGDel  & $74.22 \pm 3.27$ & $76.33 \pm 3.34$ \\
\proposed  & $\mathbf{89.11 \pm 3.81}$ & $\mathbf{96.42 \pm 1.57}$ \\
\bottomrule
\end{tabular}

\medskip
\textbf{(B)}\enspace\textit{iMM904}\\[3pt]
\begin{tabular}{lcc}
\toprule
Method & Macro-F1 (\%) & AUC (\%) \\
\midrule
DNN        & $65.43 \pm 1.63$ & $67.02 \pm 1.58$ \\
DeepGDel   & $75.98 \pm 1.51$ & $79.71 \pm 1.80$ \\
NMA        & $76.83 \pm 1.55$ & $80.46 \pm 1.86$ \\
GraphGDel  & $81.54 \pm 1.38$ & $84.09 \pm 1.36$ \\
\proposed  & $\mathbf{88.61 \pm 1.42}$ & $\mathbf{96.22 \pm 1.12}$ \\
\bottomrule
\end{tabular}

\medskip
\textbf{(C)}\enspace\textit{iML1515}\\[3pt]
\begin{tabular}{lcc}
\toprule
Method & Macro-F1 (\%) & AUC (\%) \\
\midrule
DNN        & $58.02 \pm 1.21$ & $59.53 \pm 1.18$ \\
DeepGDel   & $62.74 \pm 1.38$ & $65.08 \pm 1.20$ \\
NMA        & $63.62 \pm 1.40$ & $65.89 \pm 1.14$ \\
GraphGDel  & $72.03 \pm 1.05$ & $73.02 \pm 1.09$ \\
\proposed  & $\mathbf{90.38 \pm 0.74}$ & $\mathbf{97.12 \pm 1.20}$ \\
\bottomrule
\end{tabular}
\end{table}

Performance gains persist across GEM scales.
\proposed\ was applied to three GEMs spanning a 30-fold range in reaction count without GEM-specific hyperparameter tuning.
The largest gain appears on \textit{iML1515}, which contains the largest absolute number of multi-gene (flat AND/OR or nested) reactions (24.0\% flat OR, 7.6\% flat AND, 3.9\% nested over 2{,}712 reactions; Supplementary Table~S4).
Performance on the eukaryotic \textit{iMM904} reconstruction is comparable to the genome-scale \textit{E.~coli} model \textit{iML1515} ($88.61\%$ vs.\ $90.38\%$ macro-F1).
Cross-GEM absolute scores should be interpreted cautiously because target-dependent-pair numbers and label distributions differ across GEMs; comparisons therefore focus on relative improvements over matched baselines under identical CV folds.

Supplementary Table~S4 summarizes GPR rule distributions across GEMs, showing that \textit{iML1515} contains substantially more AND/OR-associated reactions in absolute numbers.
The same continuous scores also support gene-set recovery on held-out targets (Section~\ref{subsec:recovery}); the resulting rankings require fewer genes than the evaluated baselines to reach 90\% mean cumulative recall on each GEM.

\subsection{Molecular, GPR, and Topology Priors Provide Complementary Signal}\label{subsec:priors}

\begin{table}[t]
\centering
\caption{Component ablation (mean\,$\pm$\,s.d.\ over five held-out folds). Bold: best per panel.
The \textbf{Full} rows report the five-fold protocol of Section~\ref{subsec:setup}, in which each fold uses its own threshold chosen on that fold's inner validation split.}
\label{table:ablation_exp}

\smallskip
\textbf{(A)}\enspace\textit{e\_coli\_core}\\[3pt]
\begin{tabular}{lcc}
\toprule
Configuration & Macro-F1 (\%) & AUC (\%) \\
\midrule
Full           & $\mathbf{89.11 \pm 3.81}$ & $\mathbf{96.42 \pm 1.57}$ \\
  Mean Pooling   & $78.80 \pm 3.21$ & $88.10 \pm 3.24$ \\
  No Graph-M     & $77.58 \pm 3.29$ & $86.55 \pm 3.32$ \\
  No pretrained  & $76.50 \pm 3.30$ & $87.00 \pm 3.20$ \\
  No GPR-M       & $75.30 \pm 3.29$ & $84.60 \pm 3.34$ \\
\bottomrule
\end{tabular}

\medskip
\textbf{(B)}\enspace\textit{iMM904}\\[3pt]
\begin{tabular}{lcc}
\toprule
Configuration & Macro-F1 (\%) & AUC (\%) \\
\midrule
Full           & $\mathbf{88.61 \pm 1.42}$ & $\mathbf{96.22 \pm 1.12}$ \\
  Mean Pooling   & $85.30 \pm 1.19$ & $93.50 \pm 1.11$ \\
  No Graph-M     & $83.30 \pm 1.25$ & $92.10 \pm 1.14$ \\
  No pretrained  & $82.00 \pm 1.20$ & $91.00 \pm 1.10$ \\
  No GPR-M       & $81.30 \pm 1.27$ & $89.50 \pm 1.15$ \\
\bottomrule
\end{tabular}

\medskip
\textbf{(C)}\enspace\textit{iML1515}\\[3pt]
\begin{tabular}{lcc}
\toprule
Configuration & Macro-F1 (\%) & AUC (\%) \\
\midrule
Full           & $\mathbf{90.38 \pm 0.74}$ & $\mathbf{97.12 \pm 1.20}$ \\
  Mean Pooling   & $86.40 \pm 0.82$ & $95.20 \pm 1.29$ \\
  No Graph-M     & $84.40 \pm 0.92$ & $91.50 \pm 1.36$ \\
  No pretrained  & $83.35 \pm 0.80$ & $92.50 \pm 1.20$ \\
  No GPR-M       & $82.40 \pm 0.96$ & $89.30 \pm 1.41$ \\
\bottomrule
\end{tabular}
\end{table}

\begin{figure*}[t]
  \centering
  \includegraphics[width=0.97\linewidth]{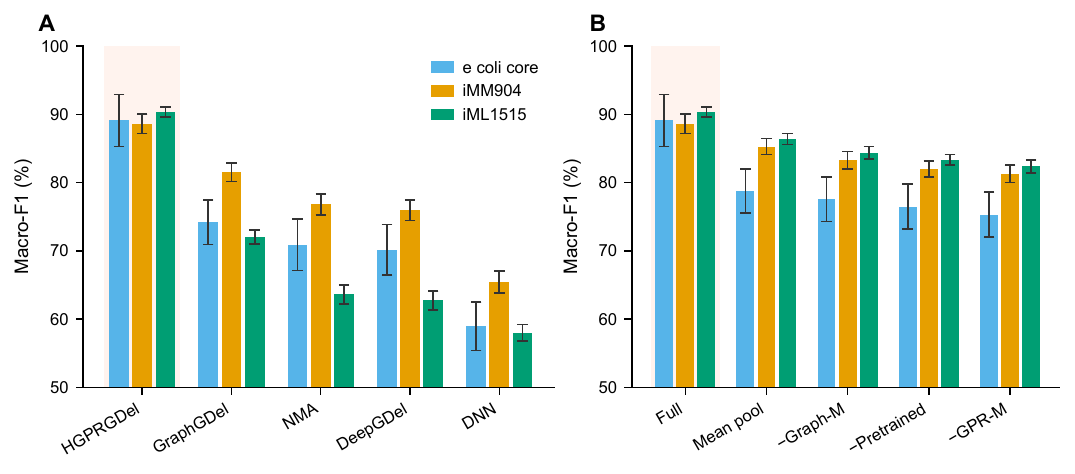}
  \caption{\textbf{Benchmark and component ablation across three GEMs.}
    (A) Macro-F1 across evaluated methods.
    (B) Macro-F1 for \proposed\ and its ablated variants.
    Values are mean $\pm$ s.d.\ over five held-out metabolite folds.}
  \label{fig:benchmark_ablation}
\end{figure*}

Table~\ref{table:ablation_exp} and Fig.~\ref{fig:benchmark_ablation}B examine the contribution of each biological prior within \proposed.
We next asked whether the three representation sources contribute complementary information or whether performance is dominated by a single component.
Variant definitions are given in Supplementary Section~S4.1.

\textbf{Molecular prior.}
Removing the pretrained weights (\textbf{No pretrained}; Supplementary Section~S2.3) costs ${\approx}6.6$--$12.6$\,pp macro-F1 across GEMs.
This variant retains the ESM-2 and ChemBERTa architectures and trains them end-to-end from random initialization, so the drop isolates the contribution of the pretrained weights themselves rather than of model capacity or input information.
Its cost is the most GEM-dependent of the four ablations ($12.6$, $6.6$, and $7.0$\,pp on \textit{e\_coli\_core}, \textit{iMM904}, and \textit{iML1515}, respectively), and is largest on the smallest GEM, where supervision per fold is scarcest.

\textbf{GPR prior.}
The ordering \textbf{Full} $>$ \textbf{Mean Pooling} $>$ \textbf{No GPR-M} decomposes the GPR-M benefit: \textbf{Mean Pooling} preserves gene--reaction connectivity but collapses AND/OR distinctions, while \textbf{No GPR-M} removes both.
On \textit{e\_coli\_core}, the AND/OR distinction contributes ${\approx}10.3$\,pp and connectivity ${\approx}3.5$\,pp of the total ${\approx}13.8$\,pp GPR-M gain; disabling GPR-M entirely removes the largest single-component contribution in all three GEMs (${\approx}7.3$--$13.8$\,pp macro-F1).
To further dissect the SoftBooleanAggregator design, we compared three AND-node variants on \textit{e\_coli\_core}: using only the mean branch ($\mathbf{m}$), only the soft-min branch ($\mathbf{s}$), or both combined (default).
Isolating either branch alone drops macro-F1 by approximately 20 percentage points relative to the combined configuration ($89.11\%$ vs.\ $68.64\%$ / $69.38\%$), supporting complementary roles for the two branches: $\mathbf{m}$ captures overall complex representation, while $\mathbf{s}$ emphasizes the lower-valued subunit representations.
Detailed AND-branch ablation results are provided in Supplementary Table~S17.

\textbf{Topology prior.}
Disabling Graph-M drops performance by ${\approx}5$--$12$\,pp across GEMs.
To further justify the hypergraph design, we evaluated three Graph-M variants on \textit{e\_coli\_core} while holding GPR-M and Pred-M fixed (Supplementary Table~S18): replacing the hypergraph with a reaction--metabolite bipartite network reduces macro-F1 by 9.7\,pp ($79.44\%$); a pairwise metabolite GCN reduces it by 13.1\,pp ($75.99\%$), approaching the GraphGDel baseline ($74.22\%$) that uses no GPR information.
The hypergraph variant retains reaction-level groupings of all participating metabolites and accepts GPR-derived hyperedge attributes without structural approximation, consistent with its stronger performance than both alternatives.

\subsection{Learned Reaction Embeddings Recover GPR Logic and Metabolic Organization}\label{subsec:mechanism}

\begin{figure*}[t]
  \centering
  \includegraphics[width=0.98\linewidth]{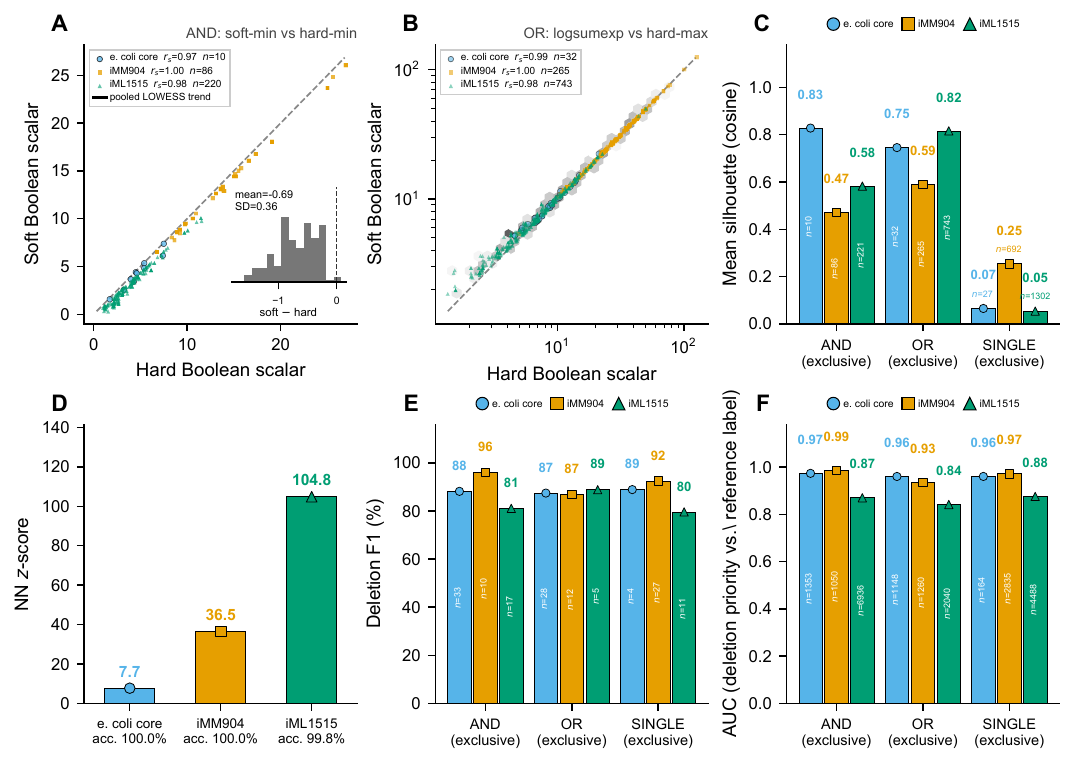}
  \caption{\textbf{Mechanistic analysis of GPR-aware reaction representations and their association with deletion prioritization.}
    (\textbf{A}) SoftBoolean AND fidelity: soft-min versus hard-min leaf $L_2$ scalars across GEMs (linear axes; dashed $y{=}x$; black line, pooled LOWESS trend). Inset shows the distribution of soft-min$-$hard-min residuals; mean $\pm$ SD is annotated. Spearman $r_s$ and $n$ are reported for each GEM.
    (\textbf{B}) SoftBoolean OR fidelity: log-sum-exp versus hard-max leaf $L_2$ scalars (log--log axes; dashed $y{=}x$). Grey hexbin density summarizes the pooled distribution, with GEM-specific points overlaid. Spearman $r_s$ and $n$ are reported for each GEM.
    (\textbf{C}) GPR-rule representation separation: mean silhouette score (cosine distance) for AND, OR, and single-gene reactions, computed jointly with empty reactions in the reference representation space. The number of reactions ($n$) is annotated for each GEM.
    (\textbf{D}) Embedding diagnostics: nearest-neighbor subsystem-enrichment $z$-scores across three GEMs, shown as bars; linear-probe accuracy for AND/OR rule classification (\%) is annotated for each GEM.
    (\textbf{E}) Deletion F1 on target-dependent gene--metabolite pairs, stratified by genes whose associated reactions are exclusively AND, OR, or single-gene (MIXED-rule genes excluded). The number of genes ($n$) is annotated for each GEM and GPR role.
    (\textbf{F}) Representation--output association: AUC of the continuous deletion-priority score ($1-\hat y$) for distinguishing reference-set membership, stratified by exclusive GPR role. The number of target-dependent gene--metabolite pairs ($n$) is annotated for each GEM and GPR role.
    UMAP embeddings and OR-branch attention entropy are shown in Supplementary Fig.~S3.} 
  \label{fig:mechanism}
\end{figure*}

The architecture explicitly models GPR logic and metabolic topology at the reaction level.
To dissect whether the learned reaction embeddings capture these structures and are associated with downstream deletion prioritization, we performed a six-panel mechanistic analysis (Fig.~\ref{fig:mechanism}): Panels A--B assess operator fidelity; C and D quantify GPR-related and biological properties of the learned representations; E and F examine downstream deletion behavior.
 
GPR-M reaction embeddings (Fig.~\ref{fig:mechanism}; Supplementary Section~S8) encode complementary structure without explicit pathway supervision.
First, SoftBoolean AND and OR aggregations track hard-min / hard-max scalars with Spearman $r_s \geq 0.86$ across GEMs (Fig.~\ref{fig:mechanism}A,B), indicating that the differentiable gates preserve the expected ordering properties of Boolean aggregation.
Second, AND/OR type is linearly decodable at near-ceiling probe accuracy and nearest-neighbor subsystem enrichment strengthens with GEM scale ($z{=}7.7$--$104.8$; Fig.~\ref{fig:mechanism}D), indicating that subsystem organization is retained without explicit pathway supervision; GPR-rule representation separation is consistently weaker for single-gene reactions than for AND/OR reactions across all three GEMs (mean silhouette $0.05$--$0.25$ versus $0.47$--$0.83$; Fig.~\ref{fig:mechanism}C), suggesting that AND/OR combinatorial structure is more distinctly represented than single-gene structure.
Third, deletion F1 varies systematically by exclusive gene--GPR role (AND / OR / SINGLE; Fig.~\ref{fig:mechanism}E). On \textit{iML1515}, the OR-exclusive stratum has the highest deletion F1 ($88.8\%$), while AND- and SINGLE-exclusive genes are closer ($81.1\%$ and $79.6\%$), so AND-exclusive genes are not uniformly the easiest stratum. The continuous deletion-priority score discriminates reference-set membership consistently across GPR roles and GEMs (AUC $0.84$--$0.99$; Fig.~\ref{fig:mechanism}F), indicating that this representation--output association is not confined to a single GPR role.
Together, these analyses provide multi-level evidence that the learned reaction representations retain GPR-related structure and that this structure is reflected in downstream deletion prioritization.
Reaction-type UMAP and OR-branch attention entropy (Supplementary Fig.~S3) provide additional supporting evidence: partial GPR-type co-localization in embedding space and non-uniform isozyme attention weighting, respectively.
Coarse-category UMAP is provided in Supplementary Fig.~S4.

\subsection{Continuous Retention Scores Recover Curated Reference Gene Sets}\label{subsec:recovery}

We next systematically evaluated the model's primary output, the continuous retention score, across held-out targets and GEMs, asking whether the continuous retention scores recover curated reference gene sets and whether the complete ranking provides prioritization value beyond a single threshold. Because the intended downstream use is deletion prioritization, ranking metrics ($\mathrm{AP}$, cumulative recall, $k_{90}$) provide the primary evidence for ordering quality, while threshold-dependent classification and reference-set recovery provide complementary diagnostics.
These analyses evaluate recovery of known metabolic-design patterns rather than claim de novo discovery, since the MetNetComp reference provides the supervised annotation~\cite{tamura2023metnetcomp}.
Method-level predictive comparisons (DNN, DeepGDel, NMA, GraphGDel) are reported in Table~\ref{table:main_exp}; here we focus on reference gene-set recovery and the additional value of ranked gene prioritization.

The continuous retention score encodes deletion priority at every operating point, enabling both candidate-set recovery through thresholding and a complete ranking that retains prioritization information beyond any single operating point.

\begin{figure*}[t]
  \centering
  \includegraphics[width=0.95\linewidth]{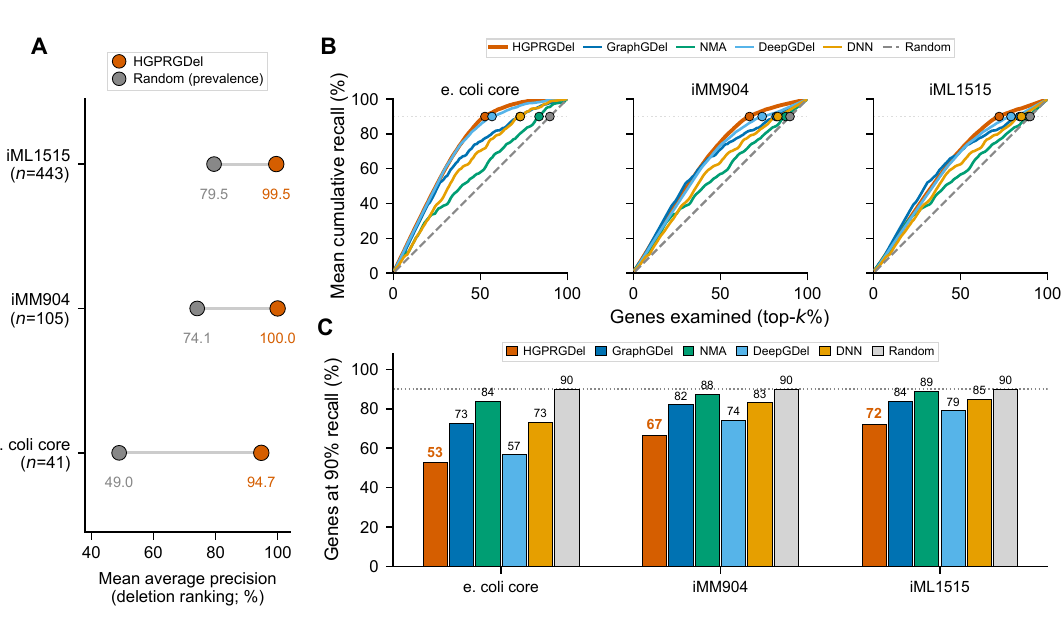}
  \caption{\textbf{Deletion-priority ranking recovers curated reference gene sets.}
  All methods are evaluated on the same held-out metabolites, gene roster, and curated labels; scores are continuous retention probabilities ranked by $1-\hat{y}_{g,m}$ (higher $=$ earlier deletion).
  (\textbf{A}) Mean average precision.
  (\textbf{B}) Cumulative recall as the top fraction of the ranked gene list increases, for \proposed, GraphGDel, NMA, DeepGDel, DNN, and a random ranking. Circles mark each method's $k_{90}$.
  (\textbf{C}) $k_{90}$, the fraction of genes required to recover 90\% of reference genes; lower is better.
  Reference gene sets are computationally curated MetNetComp annotations and are not experimentally validated.}
  \label{fig:ranking_recovery}
\end{figure*}

\textbf{Recovery.}
Average precision evaluates the ordering of reference genes independently of any fixed operating threshold, and therefore matches the model's primary output.
Deletion rankings measured by mean average precision exceed a random prevalence baseline (Fig.~\ref{fig:ranking_recovery}A), with the largest margin on \textit{e\_coli\_core} (mean AP $94.7\%$ vs.\ $49\%$ prevalence); genome-scale AP values ($99.5$--$100\%$) should be interpreted alongside the elevated prevalence baseline ($74$--$80\%$) and read together with the cumulative-recall curves rather than in isolation.
At one operating point, thresholded \proposed\ predictions additionally achieve high gene-set overlap with the curated reference across all three GEMs (recall $92.0$--$99.0\%$; precision $92.5$--$99.3\%$; F1 $92.0$--$99.2\%$; Table~\ref{tab:ranking_recovery}).

\textbf{Ranking.}
\proposed\ produces a continuous retention score for every gene--target pair, providing a complete deletion-priority ordering.
The same held-out metabolites, gene roster, and curated labels used for \proposed\ were scored with GraphGDel, NMA, DeepGDel, and DNN checkpoints.
Ordering genes by deletion priority ($1-\hat{y}_{g,m}$) concentrates curated deletions toward the top-ranked fraction (Fig.~\ref{fig:ranking_recovery}B).

Across all three GEMs, \proposed\ yields the lowest $k_{90}$ among evaluated methods, defined as the fraction of genes that must be examined to reach ${\approx}90\%$ mean cumulative recall (Fig.~\ref{fig:ranking_recovery}C).
On \textit{e\_coli\_core}, $k_{90}{=}53\%$ vs.\ DeepGDel $57\%$, GraphGDel/DNN $73\%$, NMA $84\%$, and random $90\%$.
On genome-scale GEMs, the margin widens: $k_{90}{=}67\%$ on \textit{iMM904} and $72\%$ on \textit{iML1515}, compared with DeepGDel ($74\%$ / $79\%$), GraphGDel ($82\%$ / $84\%$), DNN ($83\%$ / $85\%$), and NMA ($88\%$ / $89\%$).
Notably, GraphGDel, despite strong macro-F1 (Table~\ref{table:main_exp}), produces genome-scale rankings only modestly better than a random order ($k_{90}{\approx}82$--$84\%$ against $90\%$ for random), showing that binary classification strength does not necessarily transfer to gene prioritization.

\begin{table}[t]
  \centering
  \caption{\textbf{Recovery of curated MetNetComp reference gene sets.}
Gene-level recovery metrics are computed for targets with explicit supervised labels; definitions are provided in Methods and Supplementary Section~S1.2.}
  \label{tab:ranking_recovery}
  \footnotesize
  \setlength{\tabcolsep}{3.5pt}
  \resizebox{\columnwidth}{!}{%
  \begin{tabular}{lrrrrrrrr}
    \toprule
    GEM & \#Held-out & \#Scored & Avg.\ $|GT|$ & Avg.\ $|Pred|$ & Rec.\ (\%) & Prec.\ (\%) & F1 (\%) & AP (\%) \\
    \midrule
    \textit{e\_coli\_core} & 41  & 41  & 67.1   & 66.4   & 92.03 & 92.53 & 91.98 & 94.71 \\
    \textit{iMM904}        & 105 & 105 & 670.4  & 672.5  & 98.91 & 98.62 & 98.76 & 99.97 \\
    \textit{iML1515}       & 443 & 408 & 1205.3 & 1201.6 & 99.02 & 99.34 & 99.18 & 99.53 \\
    \bottomrule
  \end{tabular}
  }

  \vspace{2pt}
  {\footnotesize
  \raggedright
  \#Held-out: number of held-out target metabolites.
  \#Scored: number of targets with explicit supervised labels in the curated reference; on \textit{iML1515}, 35 of the 443 held-out targets have no label rows in the curated reference (Section~S1.2), and all metric columns and the mean AP are computed over the \#Scored targets only.
  ``Avg.\ $|GT|$'': mean number of genes deleted in the reference gene set; ``Avg.\ $|Pred|$'': mean number of genes predicted deleted at one operating point.
  Both are counted over the full curated maximal gene roster (including pre-determined genes), distinct from the target-dependent-gene filter used in CV metrics.
  Recall / Precision / F1: gene-set overlap with the reference under thresholded \proposed\ predictions; mean AP: deletion-ranking quality of \proposed\ on the same \#Scored targets, with baseline rankings compared in Fig.~\ref{fig:ranking_recovery}.
  Set-level recovery metrics and ranking quality (mean AP) are reported as complementary views of the same retention scores, and recovery denotes overlap with reference gene sets rather than experimental validation.\par}
\end{table}

\subsection{A Target-Level Case Study of GPR-Aware Deletion Prioritization}\label{subsec:case}

\begin{figure*}[t]
  \centering
  \includegraphics[width=0.98\linewidth]{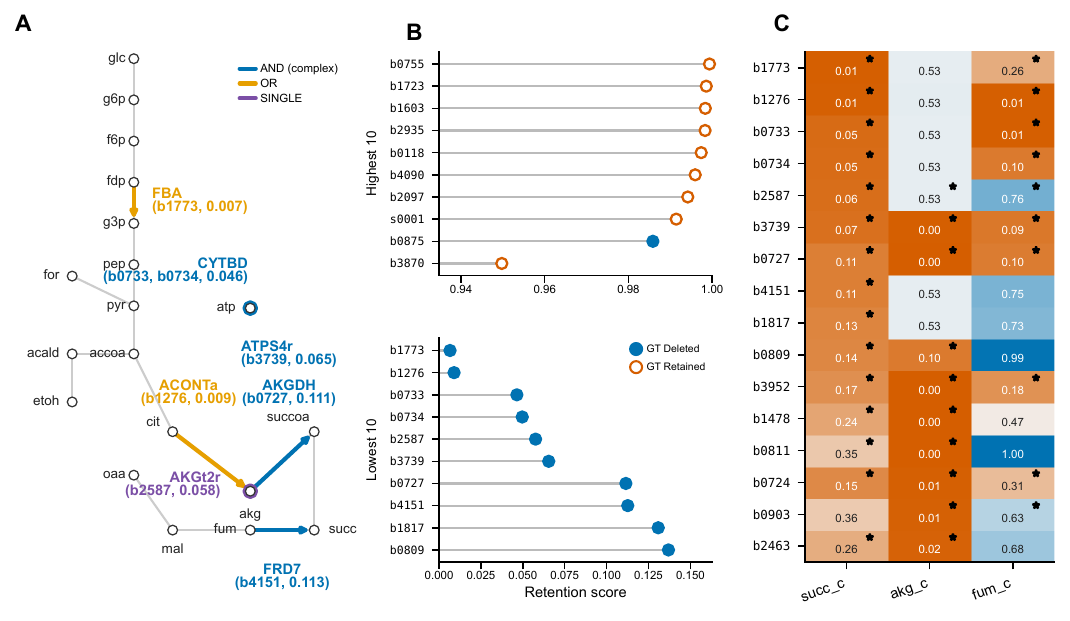}
  \caption{\textbf{Target-level case study of GPR-aware deletion prioritization for succinate} (\texttt{succ\_c}) on \textit{e\_coli\_core} (target-dependent genes; Supplementary Section~S1).
  (\textbf{A}) Central-carbon schematic highlighting reactions hit by the ten lowest-retention predicted deletions.
  Colors denote GPR type: blue, AND complex; gold, OR isozymes; purple, SINGLE.
  Labels show reaction, lowest-retention gene(s), and retention score (lower $=$ higher deletion priority); bold labels mark genes in the curated reference.
  (\textbf{B}) Ten highest- and ten lowest-retention scores; filled/open markers denote GT deleted/retained.
  (\textbf{C}) Retention heatmap for top-ranked genes across \texttt{succ\_c}, \texttt{akg\_c}, and \texttt{fum\_c}; $*$ denotes GT deleted.
  \texttt{akg\_c} ($\alpha$-ketoglutarate) and \texttt{fum\_c} (fumarate) are TCA-neighbor targets detailed in Supplementary Section~S5.}
  \label{fig:case_succinate}
\end{figure*}

To illustrate how the learned ranking behaves at the target level, we examined succinate (\texttt{succ\_c}) in \textit{e\_coli\_core} as a representative case study, among target-dependent genes (Fig.~\ref{fig:case_succinate}).
Succinate was selected because it represents a central-carbon target with diverse GPR configurations, including AND-type enzyme complexes and OR-type isozymes.
Among 66 target-dependent genes, the curated reference annotates 44 deletions; the continuous ranking attains average precision $0.91$ with all ten of the lowest-retention calls matching the curated reference (Supplementary Table~S19).
This AP is a case-study metric computed on the 66 target-dependent genes of a single target and is not directly comparable to the full-roster mean AP of Table~\ref{tab:ranking_recovery} (Supplementary Section~S5).
The ranking places genes associated with distinct GPR roles at different priorities, illustrating how reaction-level structure can shape the predicted deletion order.

Among the lowest-retention predicted deletions, OR-type isozyme calls (e.g., \texttt{b1773} on \texttt{FBA}; \texttt{b1276} on \texttt{ACONTa}) co-occur with AND-type complex subunits (e.g., \texttt{CYTBD}, \texttt{ATPS4r}, \texttt{AKGDH}, \texttt{FRD7}) and a single-gene transport call (\texttt{AKGt2r}), illustrating different deletion priorities associated with AND, OR, and single-gene GPR structures.
No false positives appear among the ten lowest-retention calls.
Cytosolic $\alpha$-ketoglutarate (\texttt{akg\_c}) and fumarate (\texttt{fum\_c}), two additional TCA-neighbor targets with reordered GPR-typed priorities, are reported in Supplementary Section~S5.

\subsection{Deletion-Priority Rankings Provide an Ordered Search Space for Candidate Strategy Construction}\label{subsec:candidates}

The complete deletion-priority ranking provides a natural way to construct candidate deletion sets without fixing a common strategy size across targets.
For a given target, progressively larger candidate sets are obtained by taking the first $q$ genes in the ranking, so strategy size is an operating-point choice applied to the ranking rather than a model hyperparameter.

The curated reference sets also vary substantially in size across GEMs, with mean sizes of $67.1$, $670.4$, and $1205.3$ genes, respectively (Table~\ref{tab:ranking_recovery}).

These candidate sets are intended for downstream constraint-based evaluation, and growth-coupling feasibility is not determined by the ranking alone. To connect the ranking to this assessment, genes were added to a candidate deletion set one at a time in decreasing deletion priority, and the weak growth-coupling criterion was applied after each addition (Supplementary Section~S3.3). This localized a weakly growth-coupled deletion set for 26 of 41 held-out \textit{e\_coli\_core} targets, or $63.4\%$, and for 44 of 105 \textit{iMM904} targets, or $41.9\%$. These values report whether the ranked candidate sets contained a deletion set that satisfied the downstream criterion. They are not prediction rates, because the ranking does not determine growth-coupling feasibility. Every unsolved target ended at a lethal prefix before a feasible one was reached, whereas placing the curated reference genes first reached weak growth coupling for all $41$ and $105$ targets, so these failures do not reflect infeasibility of the reference sets. The reference-informed ordering is therefore used as a coverage control rather than as a predictive baseline.

A separate controlled analysis on \textit{e\_coli\_core} examined how much of the ranked gene space the downstream search needs. When gDel-minRN was restricted to the top-ranked fraction of genes, the fraction of targets with a feasible growth-coupled solution rose from $4.9\%$ at $5\%$ and $10\%$ of genes to $9.8\%$ at $20\%$, $24.4\%$ at $30\%$, and $78.0\%$ at $50\%$, and reached $100\%$ when all genes were available (Supplementary Section~S3.3.3, which also compares alternative rankings). The candidate-space budget therefore trades the scope of the downstream search against coverage of feasible interventions, and targets missed at a smaller budget were recovered by the unrestricted search.

\section{Discussion}\label{sec:discussion}

\subsection{Reusing Computational Design Knowledge}\label{subsec:practical}

The central contribution of this work is a reusable formulation for growth-coupled gene deletion design. Constraint-based strain-design methods can identify feasible deletion strategies, but related combinatorial searches must often be repeated across gene--target combinations. Here, previously obtained design strategies are used as a learning resource to derive gene--target deletion priorities, while constraint-based analysis remains responsible for evaluating the resulting interventions.

This separation gives learning a specific role in strain design: rather than replacing mechanistic optimization, it narrows and structures the intervention space explored downstream. The learned ranking can therefore be reused across targets while the candidate-space budget remains an explicit design parameter.

\subsection{Structured Biological Knowledge in Learning}\label{subsec:bio-impl}

GPR associations provide a natural biological prior because they determine how gene perturbations propagate to reaction activity. Incorporating this structure into the learning representation allows the model to capture dependencies arising from enzyme complexes and alternative isozymes, while molecular and network information provide complementary context. The ablation and representation analyses support the contribution of these information sources (Section~\ref{subsec:priors}; Fig.~\ref{fig:mechanism}).

More broadly, this illustrates a useful division of roles between biological knowledge and learning: mechanistic structure can shape the learning problem without requiring the learning component to reproduce the full constraint-based model.

\subsection{Search-Space Control and Scope}\label{subsec:bridge}

The learned ranking is most naturally interpreted as an ordered intervention space rather than a final strain design. Candidate sets can therefore be constructed at different budgets and evaluated independently by constraint-based analysis. Not every target reached a feasible solution at the evaluated budgets because the corresponding feasible strategy may lie outside the selected candidate space; increasing the budget progressively restored access to feasible strategies, with the unrestricted search recovering all evaluated \textit{e\_coli\_core} targets. The reference-informed ordering provides a coverage control under the same mechanistic criterion rather than a predictive baseline.

The framework is suited to settings where prior computational design knowledge can be reused to guide subsequent search. Its scope is limited by the reference strategies used for supervision, the target-dependent gene definition, and the validation design.

\subsection{Limitations and Future Work}\label{subsec:limitations}

The reference strategies do not exhaustively represent all feasible or experimentally observed deletion strategies. Moreover, the available evaluation is limited by the target-dependent gene definition and the limited number of cross-validation folds (Supplementary Sections~S3.2, S3.6, and S3.7). The reference strategies and downstream assessment are both based on constraint-based metabolic modeling, so this study evaluates the reuse and prioritization of computational design knowledge rather than independent biological ground truth.

Future work can incorporate multiple feasible strategies per target and experimental strain-engineering outcomes, and can integrate learned priorities more adaptively with constraint-based search. The same framework could also be extended to reaction-, pathway-, or metabolic-state interventions.

\subsection{Conclusion}\label{subsec:conclusion}

This work establishes a reusable approach for growth-coupled gene deletion design in which computational design knowledge is converted into gene--target priorities and used to structure downstream constraint-based search. By separating intervention prioritization from mechanistic assessment, the framework retains the strengths of constraint-based modeling while making previously obtained design knowledge reusable across related design problems.

\section*{Key Points}

\begin{itemize}

  \item We present a reproducible framework that reuses previously obtained computational design knowledge to prioritize gene deletions for growth-coupled production across gene--target combinations.

  \item The framework formulates deletion design as a gene--target learning-and-ranking problem, with \proposed\ providing a GPR-aware learning component that integrates GPR logic, molecular information, and metabolic-network context.

  \item Across three GEMs, \proposed\ achieves the highest macro-F1 and AUC among the evaluated methods, with ablation and representation analyses supporting the contribution of these biological information sources.

  \item The learned deletion-priority rankings recover computationally curated MetNetComp reference gene sets and reach 90\% mean cumulative recall after examining fewer genes than the evaluated baseline methods, providing an ordered intervention space for downstream constraint-based assessment.

  \item By separating intervention prioritization from mechanistic feasibility assessment, the framework makes previously obtained design knowledge reusable while retaining explicit control over the downstream candidate space.

\end{itemize}

\section*{Data and Code Availability}
Processed MetNetComp gene-deletion labels, COBRA-style GEM JSON models, metabolite SMILES tables, amino acid sequences, source code, and pretrained checkpoints are available at \url{https://github.com/MetNetComp/HGPRGDel}.

\section*{Funding}
This work was supported by JSPS, KAKENHI under Grant 26K02972.

\section*{Conflict of Interest}
The authors declare that they have no competing interests.

\section*{Author Contributions}
Z.Y. and T.T. conceived and designed the study. Z.Y. implemented the model and performed the computational experiments. T.T. supervised the project and acquired funding. Z.Y. drafted the manuscript, and both authors revised and approved the final version.

\section*{Acknowledgements}
The authors thank colleagues at BIC, Kyoto University for helpful discussions and computational support.

\bibliographystyle{oup-unsrt}
\bibliography{Bibliography}

@article{cobrapy,
  title={{COBRApy}: constraint-based reconstruction and analysis for python},
  author={Ebrahim, Ali and Lerman, Joshua A and Palsson, Bernhard O and Hyduke, Daniel R},
  journal={BMC Systems Biology},
  volume={7},
  number={1},
  pages={74},
  year={2013},
  publisher={BioMed Central},
  doi={10.1186/1752-0509-7-74}
}

@article{covert2001regulation,
  title={Regulation of gene expression in flux balance models of metabolism},
  author={Covert, Markus W and Schilling, Christophe H and Palsson, Bernhard},
  journal={Journal of Theoretical Biology},
  volume={213},
  number={1},
  pages={73--88},
  year={2001},
  publisher={Elsevier},
  doi={10.1006/jtbi.2001.2405}
}

@article{covert2003constraints,
  title={Constraints-based models: regulation of gene expression reduces the steady-state solution space},
  author={Covert, Markus W and Palsson, Bernhard O},
  journal={Journal of Theoretical Biology},
  volume={221},
  number={3},
  pages={309--325},
  year={2003},
  publisher={Elsevier},
  doi={10.1006/jtbi.2003.3071}
}

@article{reed2003ijr904,
  title={An expanded genome-scale model of {Escherichia coli} {K-12} ({iJR904 GSM/GPR})},
  author={Reed, Jennifer L and Vo, Thuy D and Schilling, Christophe H and Palsson, Bernhard O},
  journal={Genome Biology},
  volume={4},
  number={9},
  pages={R54},
  year={2003},
  publisher={BioMed Central},
  doi={10.1186/gb-2003-4-9-r54}
}

@article{thiele2010protocol,
  title={A protocol for generating a high-quality genome-scale metabolic reconstruction},
  author={Thiele, Ines and Palsson, Bernhard {\O}},
  journal={Nature Protocols},
  volume={5},
  number={1},
  pages={93--121},
  year={2010},
  publisher={Nature Publishing Group},
  doi={10.1038/nprot.2009.203}
}

@article{KEGG,
  title={KEGG: kyoto encyclopedia of genes and genomes},
  author={Kanehisa, Minoru and Goto, Susumu},
  journal={Nucleic acids research},
  volume={28},
  number={1},
  pages={27--30},
  year={2000},
  publisher={Oxford University Press}
}

@article{MetaNetX,
  title={MetaNetX/MNXref: unified namespace for metabolites and biochemical reactions in the context of metabolic models},
  author={Moretti, S{\'e}bastien and Tran, Van Du T and Mehl, Florence and Ibberson, Mark and Pagni, Marco},
  journal={Nucleic acids research},
  volume={49},
  number={D1},
  pages={D570--D574},
  year={2021},
  publisher={Oxford University Press}
}

@article{simeonidis2015genome,
  title={Genome-scale modeling for metabolic engineering},
  author={Simeonidis, Evangelos and Price, Nathan D},
  journal={Journal of Industrial Microbiology and Biotechnology},
  volume={42},
  number={3},
  pages={327--338},
  year={2015},
  publisher={Oxford University Press}
}

@article{yang2025deepgdel,
  title={{DeepGDel}: Deep Learning-Based Gene Deletion Prediction Framework for Growth-Coupled Production in Genome-Scale Metabolic Models},
  author={Yang, Ziwei and Tamura, Takeyuki},
  journal={IEEE Transactions on Computational Biology and Bioinformatics},
  volume={22},
  number={5},
  pages={2252--2266},
  year={2025},
  month={sep},
  doi={10.1109/TCBBIO.2025.3590141},
  publisher={IEEE}
}

@article{yang2025graphgdel,
  title={{GraphGDel}: Constructing and Learning Graph Representations of Genome-Scale Metabolic Models for Growth-Coupled Gene Deletion Prediction},
  author={Yang, Ziwei and Tamura, Takeyuki},
  journal={arXiv preprint arXiv:2504.06316},
  year={2025}
}

@article{esm2,
  title={Evolutionary-scale prediction of atomic-level protein structure with a language model},
  author={Lin, Zeming and Akin, Halil and Rao, Roshan and Hie, Brian and Zhu, Zhongkai and Lu, Wenting and Smetanin, Nikita and Verkuil, Robert and Kabeli, Ori and Shmueli, Yair and others},
  journal={Science},
  volume={379},
  number={6637},
  pages={1123--1130},
  year={2023},
  publisher={American Association for the Advancement of Science}
}

@article{chemberta,
  title={ChemBERTa: Large-scale self-supervised pretraining for molecular property prediction},
  author={Chithrananda, Seyone and Grand, Gabriel and Ramsundar, Bharath},
  journal={arXiv preprint arXiv:2010.09885},
  year={2020}
}

@inproceedings{hypergraph_conv,
  title={Hypergraph neural networks},
  author={Feng, Yifan and You, Haoxuan and Zhang, Zizhao and Ji, Rongrong and Gao, Yue},
  booktitle={Proceedings of the AAAI conference on artificial intelligence},
  volume={33},
  pages={3558--3565},
  year={2019}
}

@article{DNN,
  title={Pytorch: An imperative style, high-performance deep learning library},
  author={Paszke, A},
  journal={arXiv preprint arXiv:1912.01703},
  year={2019}
}

@article{trinh2008minimal,
  title={Minimal Escherichia coli cell for the most efficient production of ethanol from hexoses and pentoses},
  author={Trinh, Cong T and Unrean, Pornkamol and Srienc, Friedrich},
  journal={Applied and environmental microbiology},
  volume={74},
  number={12},
  pages={3634--3643},
  year={2008},
  publisher={Am Soc Microbiol}
}

@article{yang2025dbgdel,
  title={{DBgDel}: Database-Enhanced Gene Deletion Framework for Growth-Coupled Production in Genome-Scale Metabolic Models},
  author={Yang, Ziwei and Tamura, Takeyuki},
  journal={IEEE Transactions on Computational Biology and Bioinformatics},
  volume={22},
  number={4},
  pages={1415--1427},
  year={2025},
  month={jul},
  doi={10.1109/TCBBIO.2025.3555999},
  publisher={IEEE}
}

@article{ML_meta_2,
  title={Deep learning-based k cat prediction enables improved enzyme-constrained model reconstruction},
  author={Li, Feiran and Yuan, Le and Lu, Hongzhong and Li, Gang and Chen, Yu and Engqvist, Martin KM and Kerkhoven, Eduard J and Nielsen, Jens},
  journal={Nature Catalysis},
  volume={5},
  number={8},
  pages={662--672},
  year={2022},
  publisher={Nature Publishing Group UK London}
}

@article{ML_meta_1,
  title={A machine learning approach to predict metabolic pathway dynamics from time-series multiomics data},
  author={Costello, Zak and Martin, Hector Garcia},
  journal={NPJ systems biology and applications},
  volume={4},
  number={1},
  pages={1--14},
  year={2018},
  publisher={Nature Publishing Group}
}

@article{tamura2022trimming,
  title={Trimming gene deletion strategies for growth-coupled production in constraint-based metabolic networks: TrimGdel},
  author={Tamura, Takeyuki},
  journal={IEEE/ACM Transactions on Computational Biology and Bioinformatics},
  volume={20},
  number={2},
  pages={1540--1549},
  year={2023},
  publisher={IEEE}
}

@article{burgard2003optknock,
  title={Optknock: a bilevel programming framework for identifying gene knockout strategies for microbial strain optimization},
  author={Burgard, Anthony P and Pharkya, Priti and Maranas, Costas D},
  journal={Biotechnology and bioengineering},
  volume={84},
  number={6},
  pages={647--657},
  year={2003},
  publisher={Wiley Online Library}
}

@article{orth2010flux,
  title={What is flux balance analysis?},
  author={Orth, Jeffrey D and Thiele, Ines and Palsson, Bernhard {\O}},
  journal={Nature biotechnology},
  volume={28},
  number={3},
  pages={245--248},
  year={2010},
  publisher={Nature Publishing Group US New York}
}

@article{yang2011emilio,
  title={EMILiO: a fast algorithm for genome-scale strain design},
  author={Yang, Laurence and Cluett, William R and Mahadevan, Radhakrishnan},
  journal={Metabolic engineering},
  volume={13},
  number={3},
  pages={272--281},
  year={2011},
  publisher={Elsevier}
}

@article{gu2016idealknock,
  title={IdealKnock: a framework for efficiently identifying knockout strategies leading to targeted overproduction},
  author={Gu, Deqing and Zhang, Cheng and Zhou, Shengguo and Wei, Liujing and Hua, Qiang},
  journal={Computational biology and chemistry},
  volume={61},
  pages={229--237},
  year={2016},
  publisher={Elsevier}
}

@article{ohno2014fastpros,
  title={FastPros: screening of reaction knockout strategies for metabolic engineering},
  author={Ohno, Satoshi and Shimizu, Hiroshi and Furusawa, Chikara},
  journal={Bioinformatics},
  volume={30},
  number={7},
  pages={981--987},
  year={2014},
  publisher={Oxford University Press}
}

@article{schneider2020extended,
  title={An extended and generalized framework for the calculation of metabolic intervention strategies based on minimal cut sets},
  author={Schneider, Philipp and von Kamp, Axel and Klamt, Steffen},
  journal={PLoS computational biology},
  volume={16},
  number={7},
  pages={e1008110},
  year={2020},
  publisher={Public Library of Science San Francisco, CA USA}
}

@article{banerjee2020genome,
  title={Genome-scale metabolic rewiring improves titers rates and yields of the non-native product indigoidine at scale},
  author={Banerjee, Deepanwita and Eng, Thomas and Lau, Andrew K and Sasaki, Yusuke and Wang, Brenda and Chen, Yan and Prahl, Jan-Philip and Singan, Vasanth R and Herbert, Robin A and Liu, Yuzhong and others},
  journal={Nature communications},
  volume={11},
  number={1},
  pages={5385},
  year={2020},
  publisher={Nature Publishing Group UK London}
}

@article{tamura2023gene,
  title={Gene deletion algorithms for minimum reaction network design by mixed-integer linear programming for metabolite production in constraint-based models: gDel\_minRN},
  author={Tamura, Takeyuki and Muto-Fujita, Ai and Tohsato, Yukako and Kosaka, Tomoyuki},
  journal={Journal of Computational Biology},
  volume={30},
  number={5},
  pages={553--568},
  year={2023},
  publisher={Mary Ann Liebert, Inc., publishers 140 Huguenot Street, 3rd Floor New~…}
}

@article{tamura2023metnetcomp,
  title={MetNetComp: Database for minimal and maximal gene-deletion strategies for growth-coupled production of genome-scale metabolic networks},
  author={Tamura, Takeyuki},
  journal={IEEE/ACM Transactions on Computational Biology and Bioinformatics},
  year={2023},
  publisher={IEEE}
}

@article{nadeau2003inference,
  title={Inference for the generalization error},
  author={Nadeau, Claude and Bengio, Yoshua},
  journal={Machine Learning},
  volume={52},
  number={3},
  pages={239--281},
  year={2003},
  publisher={Springer}
}

@article{heirendt2019creation,
  title={Creation and analysis of biochemical constraint-based models using the COBRA Toolbox v. 3.0},
  author={Heirendt, Laurent and Arreckx, Sylvain and Pfau, Thomas and Mendoza, Sebasti{\'a}n N and Richelle, Anne and Heinken, Almut and Haraldsd{\'o}ttir, Hulda S and Wachowiak, Jacek and Keating, Sarah M and Vlasov, Vanja and others},
  journal={Nature protocols},
  volume={14},
  number={3},
  pages={639--702},
  year={2019},
  publisher={Nature Publishing Group UK London}
}

@article{orth2010reconstruction,
  title={Reconstruction and use of microbial metabolic networks: the core Escherichia coli metabolic model as an educational guide},
  author={Orth, Jeffrey D and Fleming, Ronan MT and Palsson, Bernhard {\O}},
  journal={EcoSal plus},
  volume={4},
  number={1},
  pages={10--1128},
  year={2010},
  publisher={Am Soc Microbiol}
}

@article{mo2009connecting,
  title={Connecting extracellular metabolomic measurements to intracellular flux states in yeast},
  author={Mo, Monica L and Palsson, Bernhard {\O} and Herrg{\aa}rd, Markus J},
  journal={BMC systems biology},
  volume={3},
  pages={1--17},
  year={2009},
  publisher={Springer}
}

@article{monk2017ml1515,
  title={iML1515, a knowledgebase that computes Escherichia coli traits},
  author={Monk, Jonathan M and Lloyd, Colton J and Brunk, Elizabeth and Mih, Nathan and Sastry, Anand and King, Zachary and Takeuchi, Rikiya and Nomura, Wataru and Zhang, Zhen and Mori, Hirotada and others},
  journal={Nature biotechnology},
  volume={35},
  number={10},
  pages={904--908},
  year={2017},
  publisher={Nature Publishing Group US New York}
}

@article{hasibi2024integration,
  title={Integration of graph neural networks and genome-scale metabolic models for predicting gene essentiality},
  author={Hasibi, Ramin and Michoel, Tom and Oyarz{\'u}n, Diego A},
  journal={npj Systems Biology and Applications},
  volume={10},
  pages={24},
  year={2024},
  publisher={Nature Publishing Group}
}

@article{wang2026hypergraph,
  title={Hypergraph learning with multi-dimensional metabolite feature extractions and static--dynamic attention mechanisms to fill missing reactions in metabolic networks},
  author={Wang, Kai and Qu, Jiajun and Liu, Fei and Luan, Xiaoli and Zhou, Jingwen},
  journal={Briefings in Bioinformatics},
  volume={27},
  number={3},
  pages={bbag314},
  year={2026}
}

\clearpage
\onecolumn
\makeatletter
\@unnumsecfalse
\renewcommand{\@seccntformat}[1]{\csname the#1\endcsname\quad}
\makeatother
\setcounter{section}{0}
\setcounter{subsection}{0}
\setcounter{subsubsection}{0}
\setcounter{figure}{0}
\setcounter{table}{0}
\setcounter{equation}{0}
\renewcommand{\thesection}{S\arabic{section}}
\renewcommand{\thesubsection}{\thesection.\arabic{subsection}}
\renewcommand{\thesubsubsection}{\thesubsection.\arabic{subsubsection}}
\renewcommand{\thefigure}{S\arabic{figure}}
\renewcommand{\thetable}{S\arabic{table}}
\renewcommand{\theequation}{S\arabic{equation}}
\renewcommand{\topfraction}{0.92}
\renewcommand{\bottomfraction}{0.85}
\renewcommand{\textfraction}{0.06}
\renewcommand{\floatpagefraction}{0.6}

\section*{Supplementary Information}
\addcontentsline{toc}{section}{Supplementary Information}
\FloatBarrier

This supplement provides implementation details and extended analyses that complement the main text. Cross-references indicate where supplementary material provides additional detail or extends a main-text result.

\section{Dataset Details}
\label{si:dataset}

\subsection{Genome-Scale Metabolic Model Overview}

We evaluate \proposed\ on three GEMs of increasing complexity: \textit{e\_coli\_core} (a curated core reconstruction of \textit{Escherichia coli} K-12 MG1655), \textit{iMM904} (a genome-scale reconstruction of \textit{Saccharomyces cerevisiae}), and \textit{iML1515} (the most comprehensive \textit{E.~coli} K-12 MG1655 reconstruction to date).
Table~\ref{tab:si-gem-overview} summarizes key network statistics.

\begin{table}[h]
  \centering
  \caption{Genome-scale metabolic model overview (extends main-text Table~1). ``GCP targets'': GCP-annotated targets in the GEM. ``Deletion strategies'': targets with a usable curated maximal deletion set used for supervision (the supervised-target count). ``Target-dep.\ pairs'' excludes pre-determined genes (Section~S1.2).}
  \label{tab:si-gem-overview}
  \resizebox{0.98\linewidth}{!}{%
  \begin{tabular}{lrrrrrrr}
    \toprule
    GEM & $|M|$ & $|R|$ & $|G|$ & GCP targets & Deletion strategies & Target-dep.\ pairs & Del.\ rate (\%) \\
    \midrule
    \textit{e\_coli\_core} & 72   & 95    & 137   & 48   & 41  & 2{,}706   & 43.4 \\
    \textit{iMM904}        & 1226 & 1577  & 905   & 782  & 105 & 5{,}775   & 64.6 \\
    \textit{iML1515}       & 1877 & 2712  & 1516  & 1085 & 443 & 13{,}464  & 52.7 \\
    \bottomrule
  \end{tabular}
  }
\end{table}

All three GEMs are accessed via COBRApy~\cite{cobrapy} from COBRA-style JSON model files.
GPR associations are stored as infix Boolean strings (e.g., \texttt{b0114 and b0115 and b0116}) and parsed by the \proposed\ GPRRuleParser (Section~S6.1).

\subsection{MetNetComp Label Construction}
\label{si:labels}

For each GEM and target metabolite, one MetNetComp maximal strategy table is used with columns \texttt{Gene} and \texttt{Deleted}. MetNetComp provides the reference strategies used as supervised learning targets and recovery references in the protocol.
We set $y_{g,m}=0$ for deleted genes and $y_{g,m}=1$ for retained genes.
In the processed \texttt{Label/} CSV files, the \texttt{Deleted} column stores this encoding directly ($0{=}\text{deleted}$, $1{=}\text{retained}$), consistent with the \proposed\ release format.

Reported macro-averaged precision, recall, and F1 average the per-class metrics for $y{=}0$ and $y{=}1$ with equal weight; AUC treats $y{=}1$ (retained) as the positive class.

Each GEM uses one curated reference gene set per target metabolite.
We refer to these computationally curated sets as reference gene sets throughout.
On \textit{e\_coli\_core} and \textit{iMM904}, every gene in the GEM roster appears in each strategy table, so supervised pairs equal the full roster ($|\mathit{G}|\times|\mathcal{M}_{\mathrm{target}}|$).
On \textit{iML1515}, $618{,}528$ gene--metabolite pairs have explicit MetNetComp rows ($y_{g,m}\in\{0,1\}$); the remaining $53{,}060$ roster slots have no strategy-table row and are omitted from supervised-pair counts; these slots correspond to $35$ \textit{iML1515} targets that have no supervised labels in the curated reference; these $35$ targets contribute no supervised gene--metabolite pairs and remain in the held-out roster only for cross-validation bookkeeping.
The target-dependent-pair count follows from the same accounting: the $1516-1483=33$ target-dependent genes give $33 \times 443 = 14{,}619$ candidate pairs, of which $1{,}155$ fall in the unlabeled remainder above, leaving $14{,}619-1{,}155 = 13{,}464$ reported target-dependent pairs.
  
  Pre-determined genes (those whose majority deletion label holds for a fraction $\geq\rho_{\mathrm{label}}$ of their supervised metabolites, including growth-essential genes) remain in the training matrix but are excluded from all reported metrics.
  Here $\rho_{\mathrm{label}}{=}0.9$ is the \emph{label-consistency threshold}, a data-preparation constant distinct from the GPR-M soft-AND temperature $\tau$ of Section~\ref{si:sba} and from the decision threshold $\theta$ of Section~\ref{si:training}.
The complement is termed the target-dependent set and forms the basis of all reported metrics. It consists of genes whose labels vary across supervised targets.

\subsection{Label Composition and Class Imbalance}
\label{si:label-composition}

\begin{table}[t]
  \centering
  \caption{Label composition across the three GEMs (curated supervision). ``Roster pairs'' is $|\mathit{G}|\times|\mathcal{M}_{\mathrm{target}}|$ (on \textit{iML1515}, $53{,}060$ roster slots lack explicit curated labels and are excluded from supervised pair counts). ``Target-dep.\ pairs'' and ``Del.\ pairs (target-dep.)'' exclude pre-determined genes (majority-label consistency $\geq 0.9$; Section~\ref{si:labels}). ``Del.\ rate'' is the fraction of target-dependent pairs with $y_{g,m}=0$.}
  \label{tab:si-label-composition}
  \resizebox{0.98\linewidth}{!}{%
  \begin{tabular}{lrrrrr}
    \toprule
    GEM & Roster pairs & Target-dep.\ pairs & Del.\ pairs (target-dep.) & Pre-det.\ genes & Del.\ rate (\%) \\
    \midrule
    \textit{e\_coli\_core} & 5{,}617  & 2{,}706  & 1{,}175 & 71   & 43.4 \\
    \textit{iMM904}        & 95{,}025 & 5{,}775  & 3{,}730 & 850  & 64.6 \\
    \textit{iML1515}       & 671{,}588 & 13{,}464 & 7{,}097 & 1{,}483 & 52.7 \\
    \bottomrule
  \end{tabular}
  }
\end{table}

Gene deletion strategies exhibit strong class imbalance: across GEMs, 43--65\% of target-dependent pairs carry a deletion label.
All reported main-text and ablation tables use no class balancing during training (Section~S2.4).

\subsection{Cross-Validation Partitions}
\label{si:cv}

All reported results use metabolite-based five-fold cross-validation with fixed random seed~42.
Target metabolites are shuffled once and partitioned evenly; all gene labels for a metabolite remain in the same fold.
Across the five rounds, every supervised target metabolite is evaluated exactly once (complete out-of-fold test coverage).
The target-dependent gene set is fixed across folds; only the target metabolites change between rounds.

\begin{table}[t]
  \centering
  \caption{Metabolite-based five-fold cross-validation partitions (seed~42).
  Entries are the numbers of roster gene--metabolite pairs in the held-out
  metabolite set for each fold. Counts use the full gene roster, including
  pre-determined genes. Target-dependent pair counts are reported in
  Table~\ref{tab:si-label-composition}. On \textit{iML1515}, counts include
  unlabeled roster slots.}
  \label{tab:si-cv-partitions}

  \begin{tabular}{lrrrrr}
    \toprule
    GEM & Fold 1 & Fold 2 & Fold 3 & Fold 4 & Fold 5 \\
    \midrule
    \textit{e\_coli\_core}
      & 1,096 & 1,096 & 1,096 & 1,096 & 1,233 \\
    \textit{iMM904}
      & 19,005 & 19,005 & 19,005 & 19,005 & 19,005 \\
    \textit{iML1515}
      & 133,408 & 133,408 & 133,408 & 133,408 & 137,956 \\
    \bottomrule
  \end{tabular}
\end{table}

\subsection{GPR Rule Complexity}
\label{si:gpr-complexity}

Table~\ref{tab:si-gpr-complexity} reports the distribution of GPR rule types across the three evaluated GEMs.

\begin{table}[t]
  \centering
  \caption{GPR rule complexity across the three GEMs. Depth is the root-to-leaf length of the parsed Boolean tree. Percentages are over all reactions; the largest absolute number of multi-gene rules occurs on \textit{iML1515} (2{,}712 reactions), whereas multi-gene proportions are highest on \textit{e\_coli\_core}.}
  \label{tab:si-gpr-complexity}
  \resizebox{0.9\linewidth}{!}{%
  \begin{tabular}{lrrrrrrrr}
    \toprule
    GEM & $|R|$ & Med.\ depth & Max depth & Single gene (\%) & Flat AND (\%) & Flat OR (\%) & Nested (\%) & Empty (\%) \\
    \midrule
    \textit{e\_coli\_core} & 95   & 1 & 4 & 28.4 & 10.5 & 28.4 & 5.3 & 27.4 \\
    \textit{iMM904}        & 1577 & 1 & 3 & 43.9 & 4.5  & 13.6 & 4.2 & 33.9 \\
    \textit{iML1515}       & 2712 & 1 & 4 & 48.0 & 7.6  & 24.0 & 3.9 & 16.5 \\
    \bottomrule
  \end{tabular}
  }
\end{table}

\section{Implementation Details}
\label{si:implementation}

\subsection{Hardware and Software Environment}
\label{si:hardware}

We run single-GPU training on an NVIDIA A100-PCIE-40GB GPU with a Xeon Gold 6258R CPU.
All experiments use PyTorch with the PyG (PyTorch Geometric) hypergraph convolution implementation.
Software dependencies and version information are provided with the code release.

\subsection{Downstream FBA Evaluation Protocol}
\label{si:fba_protocol}

COBRApy~\cite{cobrapy} is used for FBA re-simulation and GEM parsing.
MetNetComp curated deletion sets are taken as the reference labels and are not re-verified~\cite{tamura2023metnetcomp}.
Weak growth-coupling checks are applied only to predicted (and matched-size random) knockout sets, using the biomass-then-production criteria of prior constraint-based studies~\cite{yang2025dbgdel,tamura2023metnetcomp}: production flux uses \texttt{EX\_}$\langle$\textit{targetMet}$\rangle$ if present, else \texttt{DM\_}$\langle$\textit{targetMet}$\rangle$, else a newly created demand on that exact metabolite; growth is maximised, then biomass is fixed at the optimum while production is maximised; success requires $\mathrm{GR},\mathrm{PR}\geq 0.001$.

\subsection{Pre-trained Models and Sequence Processing}
\label{si:pretrained}

We use ESM-2 (\texttt{esm2\_t6\_8M\_UR50D}) for gene amino acid sequences and ChemBERTa (\texttt{DeepChem/\allowbreak ChemBERTa-77M-MLM}) for metabolite SMILES.
Both encoders are frozen; only the projection layers and downstream modules are trained end-to-end.
Amino acid sequences and SMILES strings are tokenized with the respective model tokenizers; long sequences are truncated to 256 tokens to fit available GPU memory.

In the \textbf{No pretrained} ablation variant (main-text Table~3), the published checkpoints are not loaded: the ESM-2 and ChemBERTa architectures are instantiated from their configurations with randomly initialized weights and trained end-to-end together with the projection layers and downstream modules.
This holds the architecture, parameter count, and information flow of the frozen-encoder configuration fixed, so the variant isolates the contribution of the pretrained weights themselves.
Because the backbones are trainable in this variant, the frozen pooled states are neither precomputed nor reused: its trainable parameter count necessarily exceeds that of the full model, in which both language-model backbones are frozen.

Because the language-model backbones are frozen, their pooled hidden states are precomputed and cached from sequences and SMILES alone, without access to labels, before any fold-specific supervised fitting under cross-validation.
This caching protocol ensures that sequence representations are computed identically across folds and eliminates redundant forward passes.

SMILES strings for metabolites were obtained from MetaNetX~\cite{MetaNetX}.
Compound-node coverage is $100.0\%$ on \textit{e\_coli\_core}, $86.3\%$ on \textit{iMM904} and $89.9\%$ on \textit{iML1515}.
The uncovered remainder consists of compounds for which no structure is available, chiefly acyl-carrier proteins, charged tRNAs and polymers whose model formula is a generic repeat unit (\texttt{C11H21N2O7PRS}, \texttt{C1920H3622O700P100}) rather than a single molecule.
A metabolite without a SMILES string is passed to ChemBERTa as an empty sequence, which tokenizes to \texttt{[CLS]\,[SEP]} and yields a single fixed embedding shared by all such metabolites; it is projected to $d_m{=}256$ dimensions by the same linear layer as every other metabolite.
These nodes are therefore mutually indistinguishable to the model rather than zero-valued, and are reachable through the hypergraph only via their reaction neighborhood.

\subsection{Full Hyperparameter Settings}
\label{si:hyperparams}

Table~\ref{tab:si-hyperparams} provides the complete hyperparameter configuration used for all reported experiments.
No per-GEM retuning was performed; the same configuration is applied to all three GEMs.

\begin{table}[h]
  \centering
  \caption{Full hyperparameter configuration for \proposed. All reported experiments use these settings without per-GEM retuning.}
  \label{tab:si-hyperparams}
  \resizebox{0.9\linewidth}{!}{%
  \begin{tabular}{ll}
    \toprule
    \textbf{Parameter} & \textbf{Value} \\
    \midrule
    Gene representation dim ($d_g$)          & 320 \\
    Metabolite representation dim ($d_m$)     & 256 \\
    Hypergraph conv dim                       & 128 \\
    Hypergraph conv layers                    & 2 \\
    GPR-M reaction-embedding dim              & $d_g$ ($=320$; \textbf{GPR-M} output) \\
    Reaction-context dim ($D_{\text{rxn}}$)  & 128 (association-MLP output, $d_g \to D_{\text{rxn}}$) \\
    Classifier input dim                      & $d_g + d_m + D_{\text{rxn}} = 704$ \\
    Prediction MLP hidden dims               & $256 \to 128 \to 1$ \\
    GPR-M MHA heads ($H$)                    & 4 \\
    Dropout                                   & 0 \\
    Optimizer                                 & Adam ($\beta_1{=}0.9$, $\beta_2{=}0.999$) \\
    Learning rate                             & $10^{-3}$ \\
    Weight decay                              & none ($0$) \\
    Batch size                                & 16 \\
    Class balancing                           & none: no per-class weight term \\
    Gene-pair loss weighting                  & per-gene label-entropy weight \\
                                              & ($\gamma{=}1.0$, $\epsilon{=}0.05$; Section~\ref{si:training}) \\
    Inner validation split                    & $20\%$ of each fold's training partition \\
    Model selection                           & best inner-validation AUC \\
    Decision threshold $\theta$               & macro-F1-optimal on the inner validation split, \\
                                              & grid $[0.1, 0.9]$ (81 values), fixed after early stopping \\
    GPR-M soft-AND temperature $\tau$        & $\mathrm{softplus}(\hat\tau)+10^{-6}$, \\
                                              & $\hat\tau$ init.\ so $\tau \approx 1$ \\
    GPR-M OR-branch temperature               & none (log-sum-exp is untempered) \\
    Max epochs                                & 50 \\
    Early stopping patience                   & 5 \\
    \bottomrule
  \end{tabular}
  }
\end{table}

\subsection{Training Procedure}
\label{si:training}

Each of the five outer folds reserves $20\%$ of its training partition as an \emph{inner validation split}; the remaining $80\%$ is the inner training set, and every supervised gene--metabolite pair of a fold is drawn from that inner training set.
Training runs for up to $50$ epochs (Adam, learning rate $10^{-3}$, weight decay $0$, batch size $16$) with early stopping driven by the \textbf{inner-validation AUC} computed over target-dependent gene--metabolite pairs.
AUC is threshold-free, which is what decouples the stopping decision from the choice of operating point.
If the inner-validation AUC does not improve for $5$ consecutive epochs, training is halted and the best-AUC state is restored; the outer test fold is never consulted.
The per-fold states behind the results reported in the manuscript are the per-fold model states, i.e.\ the best inner-validation states this rule selects, so no additional training was run for them.

Only after the model state is fixed is the decision threshold $\theta$ selected, by maximizing macro-F1 on the same inner validation split over target-dependent pairs on a grid of $81$ values in $[0.1, 0.9]$.
This is the unweighted macro-F1 that the tables report, deliberately not the gene-pair-weighted training objective.
The resulting fold-specific $\theta$ is applied unchanged to the held-out test metabolites.
Each batch consists of gene--metabolite pairs sampled from the inner training set.
The training objective is a \emph{per-gene weighted} binary cross-entropy.
For a batch $\mathcal{B}$ of gene--metabolite pairs with logits $z_{g,m}$ and labels $y_{g,m}$,
\begin{align}
  \mathcal{L}(\mathcal{B}) &\;=\; \frac{\sum_{(g,m)\in\mathcal{B}} w_g \,\ell\big(y_{g,m},\,\hat y_{g,m}\big)}{\sum_{(g,m)\in\mathcal{B}} w_g}, \label{eq:loss} \\
  \ell(y,\hat y) &\;=\; -\big[y\log \hat y + (1-y)\log(1-\hat y)\big],
  \qquad
  \hat y_{g,m} \;=\; \mathrm{sigmoid}(z_{g,m}), \notag
\end{align}
where the weight depends on the gene alone,
\begin{equation}
  w_g \;=\; \epsilon + (1-\epsilon)\,H_g^{\gamma},
  \qquad
  H_g \;=\; -\big[p_g\log_2 p_g + (1-p_g)\log_2(1-p_g)\big],
  \qquad
  p_g \;=\; \frac{\max\!\big(n^{\mathrm{del}}_g,\, n^{\mathrm{ret}}_g\big)}{n^{\mathrm{del}}_g + n^{\mathrm{ret}}_g},
  \label{eq:entropy-weight}
\end{equation}
with $\gamma{=}1.0$, $\epsilon{=}0.05$, and $n^{\mathrm{del}}_g$, $n^{\mathrm{ret}}_g$ the numbers of deleted and retained supervised metabolites for gene $g$.
$H_g\in[0,1]$ is the binary label entropy in bits, so $w_g$ ranges from $\epsilon$ for a gene whose label is constant across its supervised metabolites to $1$ for a perfectly balanced gene.

Equations~\eqref{eq:loss}--\eqref{eq:entropy-weight} clarify how this weighting relates to the ``Class balancing: none'' entry of Table~\ref{tab:si-hyperparams}.
\textbf{(i)} $w_g$ is a function of the gene only and never of the label, so within a gene the deleted and retained pairs receive the \emph{same} weight.
The scheme therefore contains no per-class weight parameter, which is what ``no class balancing'' reports; the weighting is over gene--metabolite pairs, not over classes.
\textbf{(ii)} It is nevertheless \emph{not} equivalent to unweighted BCE.
Unweighted BCE is recovered exactly only when $\gamma{=}0$ (then $w_g{\equiv}1$) or when every gene is balanced ($H_g{\equiv}1$); at $\gamma{=}1.0$ the objective down-weights near-constant genes, reducing the effective pair count to $\sum_g w_g = 0.52$, $0.11$ and $0.09$ of the supervised pair count of Section~\ref{si:label-composition} on \textit{e\_coli\_core}, \textit{iMM904} and \textit{iML1515}, respectively.
\textbf{(iii)} Because genes differ in class composition, the aggregate effective class ratio shifts as a side effect: the effective share of deleted pairs falls from $49.0\%$, $74.1\%$ and $79.5\%$ (unweighted) to $45.2\%$, $63.8\%$ and $70.8\%$ (entropy-weighted) on the three GEMs, redistributing weight toward retained labels.
This rebalancing is a by-product of down-weighting low-entropy genes rather than an explicit class-balancing term; all reported metrics are computed unweighted throughout (Section~\ref{si:training}).

The evaluation split is defined over \textit{target metabolites}, not over sequence clusters: the same gene may appear in both training and test folds with different metabolite partners because deletion labels are metabolite-conditioned.
Fold composition and pair counts are reported in Table~\ref{tab:si-cv-partitions}.
This nested protocol is the source of every classification result in the manuscript (main-text Tables~2 and 3, Tables~S6, S7 and S9--S12, and main-text Fig.~3).

\subsection{Baseline Comparison Protocol}
\label{si:baselines}

All five methods (DNN, DeepGDel, NMA, GraphGDel, \proposed) use metabolite-based five-fold cross-validation (seed~42, Table~\ref{tab:si-cv-partitions}) with no class balancing during training.
Held-out target metabolites and the complete, unmasked graph inputs are identical across methods.
All baselines are trained under the same inner-validation early stopping and the same inner-validation macro-F1 threshold rule described in Section~\ref{si:training}, so every method's reported macro-F1 and AUC come from an operating point chosen without reference to the test fold.
NMA follows the GraphGDel definition: DeepGDel LSTM encoders with fixed neighborhood-mean aggregation on the GraphGDel metabolite pairing graph $\mathcal{G}_{\mathrm{final}}$ (not the \proposed\ compound hypergraph).
Within-model comparisons (ablations in main-text Table~3) share the same folds and seeds as the \proposed\ full model; for \textbf{No pretrained} this means the two differ only in backbone weight initialization and trainability.

Table~\ref{tab:si-training-configs} summarizes the training configuration of each evaluated method.

\begin{table}[!htbp]
  \centering
  \caption{Training configurations of the evaluated methods.}
  \label{tab:si-training-configs}
  \newcolumntype{L}[1]{>{\raggedright\arraybackslash}p{#1}}
  \begin{tabular}{l L{4.0cm} L{3.6cm} c L{2.4cm}}
    \toprule
    Method & Input representation & Pretrained backbone & Optimization & Training \\
    \midrule
    DNN
      & Sequence encoders (from scratch), no graph
      & None
      & Adam, 0.001
      & 50 epochs, patience 5 \\

    DeepGDel
      & LSTM sequence encoders (from scratch)
      & None
      & Adam, 0.001
      & 50 epochs, patience 5 \\

    NMA
      & DeepGDel encoders + fixed one-hop neighborhood-mean graph feature
      & None
      & Adam, 0.001
      & 50 epochs, patience 5 \\

    GraphGDel
      & LSTM sequence encoders + GCN (from scratch) on pairwise metabolite graph
      & None
      & Adam, 0.001
      & 50 epochs, patience 5 \\

    \proposed
      & ESM-2 (gene) + ChemBERTa (metabolite)
      & ESM-2 / ChemBERTa (frozen)
      & Adam, 0.001
      & 50 epochs, patience 5 \\
    \bottomrule
  \end{tabular}

  \vspace{2pt}
  \begin{minipage}{0.95\linewidth}
    \footnotesize
    Configuration values are taken directly from the shared experiment protocol; per-method architecture sizes otherwise follow the released baseline implementations.
    Trainable parameter counts are not reported here, as they were not recorded consistently across the evaluated implementations.
  \end{minipage}
\end{table}

\subsubsection{Graph Convention and the Declared Inductive Sensitivity}
\label{si:graph-convention}

Every reported number in the manuscript comes from one five-fold protocol and one graph
convention: the model is forwarded through the complete GEM, and the folds differ only in which gene--metabolite pairs are supervised, never in the graph
the model sees.

Main-text Tables~2 and 3, Tables~S6, S7 and S9--S12 and main-text Fig.~3 report
per-fold metrics with each fold's own frozen $\theta$; the mechanism, deletion-recovery,
operating-point and case-study analyses (main-text Tables~4, Figs.~4--6 and
Tables~S8, S13--S19) average the five folds' scores first and then binarize with the
mean of the five folds' $\theta$.

\clearpage
\section{Additional Results}
\label{si:results}

\subsection{Full Performance Comparison (All Metrics)}
\label{si:full-main-table}

Main-text Table~2 reports macro-F1 and AUC.
Table~\ref{tab:si-main-full} extends this to all five metrics.

\begin{table*}[h]
  \centering
  \caption{Full baseline comparison: all five metrics across three GEMs (mean\,$\pm$\,s.d.\ over five held-out metabolite test folds on target-dependent pairs). Bold: best per GEM block. Every row reports per-fold metrics with each fold's own frozen $\theta$, under the convention of Section~\ref{si:graph-convention}. Tabulated metrics are computed on the target-dependent pairs (evaluable pairs: 2706, 5775 and 13464 for \textit{e\_coli\_core}, \textit{iMM904} and \textit{iML1515}). The target-dependent gene sets are defined with the label-consistency threshold $\rho_{\mathrm{label}}{=}0.9$ (Section~\ref{si:labels}) in all three GEMs.}
  \label{tab:si-main-full}
  \resizebox{0.99\linewidth}{!}{%
  \begin{tabular}{llccccc}
    \toprule
    GEM & Method & Accuracy (\%) & M-Prec.\ (\%) & M-Recall (\%) & Macro-F1 (\%) & AUC (\%) \\
    \multicolumn{7}{l}{\footnotesize M-Prec.\ $=$ macro-precision; M-Recall $=$ macro-recall; AUC $=$ area under the ROC curve.} \\
    \midrule
    \multirow{5}{*}{\textit{e\_coli\_core}}
      & DNN       & $62.77\pm3.84$ & $64.31\pm3.66$ & $63.58\pm3.38$ & $58.94\pm3.53$ & $62.18\pm3.96$ \\
      & DeepGDel  & $70.83\pm3.57$ & $71.92\pm3.62$ & $72.41\pm3.60$ & $70.16\pm3.73$ & $70.47\pm3.90$ \\
      & NMA       & $71.54\pm3.63$ & $72.65\pm3.67$ & $73.03\pm3.64$ & $70.91\pm3.78$ & $71.25\pm3.94$ \\
      & GraphGDel & $77.16\pm3.39$ & $76.12\pm3.31$ & $75.91\pm3.42$ & $74.22\pm3.27$ & $76.33\pm3.34$ \\
      & \proposed & $\mathbf{89.34\pm3.76}$ & $\mathbf{89.33\pm3.75}$ & $\mathbf{89.27\pm3.82}$ & $\mathbf{89.11\pm3.81}$ & $\mathbf{96.42\pm1.57}$ \\
    \midrule
    \multirow{5}{*}{\textit{iMM904}}
      & DNN       & $69.55\pm1.69$ & $71.64\pm1.51$ & $68.12\pm1.45$ & $65.43\pm1.63$ & $67.02\pm1.58$ \\
      & DeepGDel  & $80.03\pm1.49$ & $77.08\pm1.46$ & $77.36\pm1.48$ & $75.98\pm1.51$ & $79.71\pm1.80$ \\
      & NMA       & $80.86\pm1.55$ & $77.92\pm1.51$ & $78.02\pm1.52$ & $76.83\pm1.55$ & $80.46\pm1.86$ \\
      & GraphGDel & $85.27\pm1.39$ & $83.31\pm1.30$ & $82.61\pm1.37$ & $81.54\pm1.38$ & $84.09\pm1.36$ \\
      & \proposed & $\mathbf{89.65\pm1.21}$ & $\mathbf{88.64\pm1.27}$ & $\mathbf{88.61\pm1.59}$ & $\mathbf{88.61\pm1.42}$ & $\mathbf{96.22\pm1.12}$ \\
    \midrule
    \multirow{5}{*}{\textit{iML1515}}
      & DNN       & $59.21\pm1.15$ & $63.95\pm1.20$ & $61.13\pm1.16$ & $58.02\pm1.21$ & $59.53\pm1.18$ \\
      & DeepGDel  & $67.12\pm0.70$ & $68.33\pm1.39$ & $63.71\pm1.29$ & $62.74\pm1.38$ & $65.08\pm1.20$ \\
      & NMA       & $67.88\pm0.73$ & $69.06\pm1.42$ & $64.58\pm1.21$ & $63.62\pm1.40$ & $65.89\pm1.14$ \\
      & GraphGDel & $72.95\pm1.08$ & $74.83\pm1.12$ & $71.92\pm1.10$ & $72.03\pm1.05$ & $73.02\pm1.09$ \\
      & \proposed & $\mathbf{90.43\pm0.71}$ & $\mathbf{90.39\pm1.02}$ & $\mathbf{90.43\pm0.95}$ & $\mathbf{90.38\pm0.74}$ & $\mathbf{97.12\pm1.20}$ \\
    \bottomrule
  \end{tabular}
  }
\end{table*}

\subsection{Fold-Level Paired Significance Tests}
\label{si:wilcoxon}

For each GEM, we compare \proposed\ and GraphGDel on the five paired held-out test macro-F1 scores (per-fold values in Table~\ref{tab:si-fold-f1}) using a two-sided Wilcoxon signed-rank test, its one-sided counterpart, and a paired $t$-test with the Nadeau--Bengio variance correction~\cite{nadeau2003inference}.
Because \proposed\ extends the same graph-based line of work as GraphGDel~\cite{yang2025graphgdel} with explicit GPR-aware reaction modeling, the direction of the effect is expected a priori to be non-negative, which is what makes the one-sided test interpretable in addition to the two-sided one.

\begin{table}[!htbp]
  \centering
  \caption{Paired tests of \proposed\ vs.\ GraphGDel on held-out test macro-F1 (five folds per GEM). ``Mean $\Delta$'' is the mean macro-F1 difference (pp); the $95\%$ CI is the paired $t$ interval on the mean difference. ``NB $t$-test'' is the Nadeau--Bengio corrected resampled $t$-test, in which the paired standard error $s/\sqrt{n}$ is replaced by $s\sqrt{1/n + n_{\mathrm{test}}/n_{\mathrm{train}}}$; with $n{=}5$ folds and $n_{\mathrm{test}}/n_{\mathrm{train}}{=}1/4$ this inflates the standard error by exactly $1.5\times$, to account for the dependence between the training sets of different folds.}
  \label{tab:si-wilcoxon}
  \resizebox{0.98\linewidth}{!}{%
  \begin{tabular}{lccccc}
    \toprule
    GEM & Mean $\Delta$ (pp) & $95\%$ CI & Wilcoxon (two-sided) & Wilcoxon (one-sided) & NB $t$-test \\
    \midrule
    \textit{e\_coli\_core} & $14.89$  & $[10.47, 19.32]$  & $0.0625$ & $0.0312$ & $3.4\times10^{-3}$ \\
    \textit{iMM904}        & $7.07$  & $[5.21, 8.94]$   & $0.0625$ & $0.0312$ & $2.2\times10^{-3}$ \\
    \textit{iML1515}       & $18.38$ & $[16.95, 19.82]$ & $0.0625$ & $0.0312$ & $1.9\times10^{-5}$ \\
    \bottomrule
  \end{tabular}
  }
\end{table}

On all three GEMs, \proposed\ achieves higher macro-F1 than GraphGDel on every individual fold, and the mean differences are large ($14.9$, $7.1$, and $18.4$\,pp; Table~\ref{tab:si-wilcoxon}).
With $n{=}5$ paired folds and all five differences in the same direction, the two-sided Wilcoxon signed-rank test attains its minimum attainable $p$-value ($p = 2/2^5 = 0.0625$).
This floor is a property of assumption-free testing at this fold count rather than evidence of a small effect: the sign test and an exact permutation test over the $2^5$ sign flips attain the same $0.0625$ on all three GEMs.
The one-sided counterpart reaches $p = 1/2^5 = 0.0312$, its own minimum attainable value, which is below $\alpha{=}0.05$ under the pre-specified direction motivated above.
Two caveats apply to that figure: it sits at the resolution floor of the test, and a Bonferroni correction across the three GEMs raises it to $0.094$, above $\alpha{=}0.05$.
We therefore report the one-sided result alongside the two-sided one rather than in place of it.

The Nadeau--Bengio correction addresses a separate concern.
The five cross-validation folds share four fifths of their training data, so fold-level scores are not independent and the standard error of the mean difference is underestimated.
Replacing the paired standard error $s/\sqrt{n}$ by $s\sqrt{1/n + n_{\mathrm{test}}/n_{\mathrm{train}}}$ (with $n{=}5$ folds and $n_{\mathrm{test}}/n_{\mathrm{train}}{=}1/4$, so the standard error is inflated by $1.5\times$) still leaves $p \leq 3.4\times10^{-3}$ on all three GEMs, and the corresponding corrected $95\%$ CIs ($[8.25, 21.53]$, $[4.28, 9.87]$, and $[16.23, 20.54]$) remain well clear of zero.
Taken together, these comparisons indicate a large and directionally consistent effect that is robust to the fold-dependence correction, while the assumption-free tests alone cannot attain conventional significance at $n{=}5$ folds.

\subsection{Operating-Point Analysis from Deletion-Priority Rankings}
\label{si:fba-operating-points}

\subsubsection{Operating-Point Analysis on \textit{e\_coli\_core}: Weak Growth Coupling from Continuous Rankings}
\label{si:fba-forward-selection}

The continuous ranking defines a sequence of candidate deletion sets at increasing intervention sizes.
We therefore evaluated whether the ranking can directly localize operating points that satisfy the same weak growth-coupling criterion used throughout this analysis (Section~\ref{si:fba_protocol}), rather than treating gene deletion prediction as a fixed-size set prediction problem; the analysis assesses how the ranking can be used for downstream constraint-based evaluation, not FBA feasibility as a separate prediction endpoint.

This analysis was first carried out on \textit{e\_coli\_core}, the smallest GEM evaluated; Section~\ref{si:fba-forward-selection-imm904} repeats it on the genome-scale \textit{iMM904} reconstruction.
For each of the $41$ evaluated targets we sorted all $137$ roster genes by deletion priority $1-\hat{y}_{g,m}$ in descending order, started from an empty deletion set, and added genes one at a time.
After each addition we re-checked weak growth coupling using the same protocol as Section~\ref{si:fba_protocol} ($\mathrm{GR},\mathrm{PR}\geq0.001$) and stopped at the first successful prefix $D^*$.
By monotonicity of biomass under knockout accumulation, a prefix whose maximum biomass falls below $0.001$ is lethal for every longer prefix; such a target was therefore classified as a \emph{lethal break}, while a target that traversed all $137$ genes without reaching weak growth coupling was classified as \emph{exhausted}.

\begin{figure}[!p]
  \centering
  \includegraphics[width=0.56\linewidth]{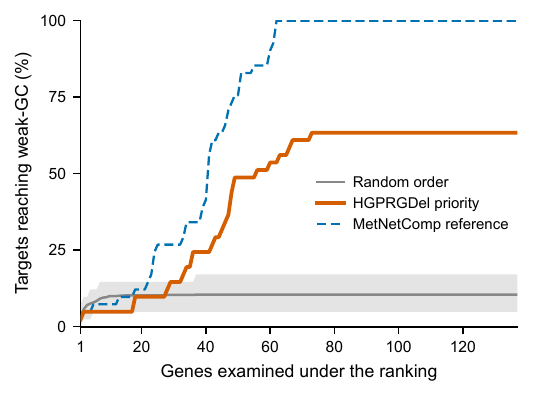}
  \caption{Localization of weakly growth-coupled operating points from continuous deletion rankings on \textit{e\_coli\_core} ($n{=}41$ targets): cumulative fraction reaching weak growth coupling versus genes examined.
  The $x$-axis counts every gene examined, \emph{including} those added before a lethal break, so such targets advance the $x$-axis without ever contributing to the cumulative count and a plateau reflects targets that ordering cannot solve.
  \proposed{} priority (solid); random order (mean over $20$ shuffles, band $=$ min--max); curated reference genes placed first (reference-informed ordering, dashed).}
  \label{fig:si-fba-forward-selection}
\end{figure}

The \proposed-guided ranking localized a weakly growth-coupled operating point for $26/41$ targets ($63.4\%$) after examining a mean of $41.7$ genes (Table~\ref{tab:si-fba-forward-selection}, Fig.~\ref{fig:si-fba-forward-selection}).
The first weakly growth-coupled operating point therefore occurred well before the full ranking was exhausted: across the $26$ solved targets the stopping rank was $41.7$ genes (median $45.5$), corresponding to a relatively compact deletion set.
As a control, the identical procedure was repeated with $20$ randomly shuffled orders per target.
Random order reached weak growth coupling for a mean of only $4.3/41$ targets (range $2$--$7$ across the $20$ orders); its successes were confined to nine targets, of which seven were also solved by the \proposed\ ranking and two (\texttt{ru5p\_\_D\_c}, \texttt{xu5p\_\_D\_c}) were not.
Within those nine targets, the random sets recovered far fewer reference genes than the ranking did (mean GT recall $0.077$ versus $0.617$; mean Jaccard $0.069$ versus $0.616$), while stopping much earlier (mean stopping rank $9.2$ versus $41.7$).
These early random solutions therefore generally did not reproduce the curated reference gene sets.
On the $32$ targets that no random order solved in any of the $640$ trials, the \proposed\ ranking reached weak growth coupling for $19$.
Across the $15$ lethal breaks, biomass was lost only after a mean of $51.5$ genes had been examined, compared with $9.5$ for the random control on the same targets, indicating that the ranking tends to place growth-essential genes later in the deletion order. For all 15 lethal-break targets, the computationally curated reference set itself reached the weak growth-coupling criterion under reference-informed ordering. Thus, these failures were not attributable to infeasibility of the reference set; the learned ranking reached a lethal prefix before reaching the reference-informed feasible ordering.
Every failure ended in a lethal break at a rank between $34$ and $81$; once such a break occurs, no later prefix can satisfy the growth criterion.
Placing the curated reference genes first (reference-informed ordering) reached weak growth coupling for $41/41$ targets at a mean of $38.7$ genes, providing an upper reference for the achievable operating-point localization. Here, reference-informed ordering denotes the ordering obtained by placing the computationally curated reference deletions first; it is used only as a reference ordering for the forward-selection analysis.

Together, these results show that the continuous ranking can be used directly as a deletion-priority order for downstream constraint-based evaluation.
On \textit{e\_coli\_core}, it localized weakly growth-coupled operating points for $26/41$ targets, substantially exceeding randomized ordering while retaining substantially greater agreement with the curated reference gene sets.

\begin{table}[!p]
  \centering
  \caption{Localization of weakly growth-coupled operating points from the continuous deletion ranking on \textit{e\_coli\_core} ($n{=}41$ targets, $137$ roster genes). Genes are added one at a time in decreasing deletion priority $1-\hat{y}_{g,m}$; the search stops at the first prefix that is weakly growth coupled, and ``Genes examined'' is that stopping rank (equal to $|D^*|$ by construction). Recall and Jaccard are averaged over each row's solved targets. Random order is repeated $20$ times per target (mean over orders; min--max in brackets); reference-informed ordering places the curated reference genes first.}
  \label{tab:si-fba-forward-selection}
  \resizebox{0.9\linewidth}{!}{%
  \begin{tabular}{lrrrr}
    \toprule
    Ranking source & Successes & Genes examined & Mean GT recall & Mean Jaccard \\
    \midrule
    \proposed{} deletion priority & 26          & 41.7 & 0.617 & 0.616 \\
    Random order (mean of $20$)   & 4.3\,[2--7] & 9.2  & 0.077 & 0.069 \\
    Curated reference (reference-informed) & 41          & 38.7 & 0.591 & 0.591 \\
    \bottomrule
  \end{tabular}
  }\\[2pt]
  \begin{minipage}{\linewidth}
    \footnotesize
    Successes are out of $41$; \proposed{} terminations were $26$ weak-GC, $15$ lethal break, and $0$ exhausted.
    Mean lethal-break position is $51.5$ for \proposed{} (over its $15$ lethal breaks) and $9.5$ for random restricted to those same $15$ targets, so the two values are paired on identical targets.
    Random's ``Genes examined'' averages over its nine solved targets.
    Of the $32$ targets unsolved in any of the $640$ random trials, \proposed{} solved $19$.
    The reference-informed ordering's mean recall is below $1$ because it stops at the first weakly growth-coupled prefix, shorter than the full reference set (mean $38.7$ genes examined versus mean $|\mathrm{GT}|=67.1$); as reference-informed ordering ranks reference genes first, it is the mean of the per-target ratios $|D^*|/|\mathrm{GT}|$, not the ratio of means.
  \end{minipage}
\end{table}

\subsubsection{Operating-Point Analysis on \textit{iMM904}: Weak Growth Coupling at Genome Scale}
\label{si:fba-forward-selection-imm904}

We repeated the identical forward-selection procedure on \textit{iMM904}, the genome-scale yeast reconstruction, to test whether operating-point localization extends beyond the smallest evaluated GEM.
For each of the $105$ evaluated targets we sorted all $905$ roster genes by deletion priority $1-\hat{y}_{g,m}$ in descending order, started from an empty deletion set, and added genes one at a time, using the same termination rule (weak growth coupling, lethal break, or exhausted) and the same $20$-random-order control as Section~\ref{si:fba-forward-selection}.

\begin{figure}[!p]
  \centering
  \includegraphics[width=0.56\linewidth]{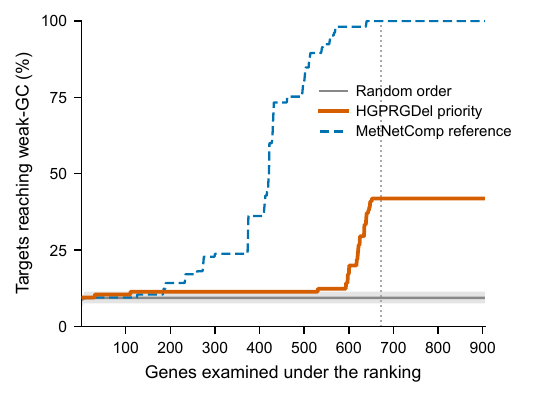}
  \caption{Localization of weakly growth-coupled operating points from continuous deletion rankings on \textit{iMM904} ($n{=}105$ targets): cumulative fraction reaching weak growth coupling versus genes examined.
  The $x$-axis counts every gene examined, \emph{including} those added before a lethal break, so such targets advance the $x$-axis without ever contributing to the cumulative count and a plateau reflects targets that ordering cannot solve.
  \proposed{} priority (solid); random order (mean over $20$ shuffles, band $=$ min--max); curated reference genes placed first (reference-informed ordering, dashed).}
  \label{fig:si-fba-forward-selection-imm904}
\end{figure}

The \proposed-guided ranking localized a weakly growth-coupled operating point for $44/105$ targets ($41.9\%$) after examining a mean of $455.0$ genes (median $618.0$ among solved targets; Table~\ref{tab:si-fba-forward-selection-imm904}, Fig.~\ref{fig:si-fba-forward-selection-imm904}).
Random order reached weak growth coupling for a mean of only $9.9/105$ targets (range $8$--$12$ across the $20$ orders); its successes were confined to $18$ targets, of which $14$ were also solved by the \proposed\ ranking and four (\texttt{2mppal\_e}, \texttt{glyc\_e}, \texttt{ppi\_m}, \texttt{ru5p\_\_D\_c}) were not.
The random sets recovered far fewer reference genes than the ranking did (mean GT recall $0.005$ versus $0.681$; mean Jaccard $0.005$ versus $0.680$), consistent with the pattern on \textit{e\_coli\_core}.
On the $87$ targets that no random order solved in any of the $1{,}740$ trials, the \proposed\ ranking reached weak growth coupling for $30$.
Across the $61$ lethal breaks, biomass was lost only after a mean of $646.8$ genes had been examined, compared with $8.0$ for the random control on the same targets; every failure ended in a lethal break at a rank between $576$ and $660$.
As on \textit{e\_coli\_core}, placing the curated reference genes first (reference-informed ordering) reached weak growth coupling for $105/105$ targets, here at a mean of $377.4$ genes against a mean reference size of $670.4$ genes, indicating that the reference sets remain FBA-feasible at genome scale and that operating-point localization from the continuous ranking is not restricted to the smallest evaluated GEM. Thus, the unsolved targets under the learned ordering reflect failure to reach a feasible prefix before the lethal break rather than infeasibility of the corresponding reference deletion sets under the same criterion.

\begin{table}[!p]
  \centering
  \caption{Localization of weakly growth-coupled operating points from the continuous deletion ranking on \textit{iMM904} ($n{=}105$ targets, $905$ roster genes), computed identically to Table~\ref{tab:si-fba-forward-selection}.}
  \label{tab:si-fba-forward-selection-imm904}
  \resizebox{0.9\linewidth}{!}{%
  \begin{tabular}{lrrrr}
    \toprule
    Ranking source & Successes & Genes examined & Mean GT recall & Mean Jaccard \\
    \midrule
    \proposed{} deletion priority & 44          & 455.0 & 0.681 & 0.680 \\
    Random order (mean of $20$)   & 9.9\,[8--12] & 3.8  & 0.005 & 0.005 \\
    Curated reference (reference-informed) & 105          & 377.4 & 0.563 & 0.563 \\
    \bottomrule
  \end{tabular}
  }\\[2pt]
  \begin{minipage}{\linewidth}
    \footnotesize
    Successes are out of $105$; \proposed{} terminations were $44$ weak-GC, $61$ lethal break, and $0$ exhausted.
    Mean lethal-break position is $646.8$ for \proposed{} (over its $61$ lethal breaks) and $8.0$ for random restricted to those same $61$ targets; random's ``Genes examined'' averages over its $18$ solved targets; of the $87$ targets unsolved in any of the $1{,}740$ random trials, \proposed{} solved $30$.
    As in Table~\ref{tab:si-fba-forward-selection} (see that table's note for why this is the mean of per-target ratios, not the ratio of means), the reference-informed ordering's mean recall is below $1$ because it stops at the first weakly growth-coupled prefix (mean $377.4$ genes examined versus mean $|\mathrm{GT}|=670.4$).
  \end{minipage}
\end{table}

\subsubsection{Search-Space Restriction under Alternative Gene Rankings}
\label{si:fba-search-cost}

Unlike the forward-selection analyses above, which evaluate whether prefixes of a ranking already contain a feasible strategy, this analysis asks whether a ranking can be used to restrict the candidate intervention space supplied to a constraint-based search algorithm. gDel-minRN does not simply test the restricted gene pool for a feasible subset; it searches that pool with its own MILP objective, which minimizes the number of repressed reactions subject to feasibility. Success rates between the two analyses are therefore not directly comparable.

To test whether this effect is specific to HGPRGDel's ranking or would arise under any reasonably constructed ranking, we repeated the analysis using gDel-minRN~\cite{tamura2023gene}, the MILP-based search algorithm used to construct the MetNetComp reference strategies, with three gene rankings: HGPRGDel's deletion-priority ranking, GraphGDel's retention scores (obtained from its existing trained checkpoint without retraining), and five independent random gene orderings per target. All three rankings used the same gDel-minRN search procedure, budgets ($B \in \{5,10,20,30,50,100\}\%$, i.e.\ $7$ to $137$ of the $137$ roster genes), and feasibility thresholds ($\mathrm{GR},\mathrm{PR}\geq0.001$, maxLoop $=10$) described above; no $B=100\%$ run was truncated at this limit, and an internal consistency check confirmed that restricting the HGPRGDel-ranked candidate pool to all $137$ genes reproduced the unrestricted search exactly for all $41$ targets.

\begin{figure}[!p]
  \centering
  \includegraphics[width=0.85\linewidth]{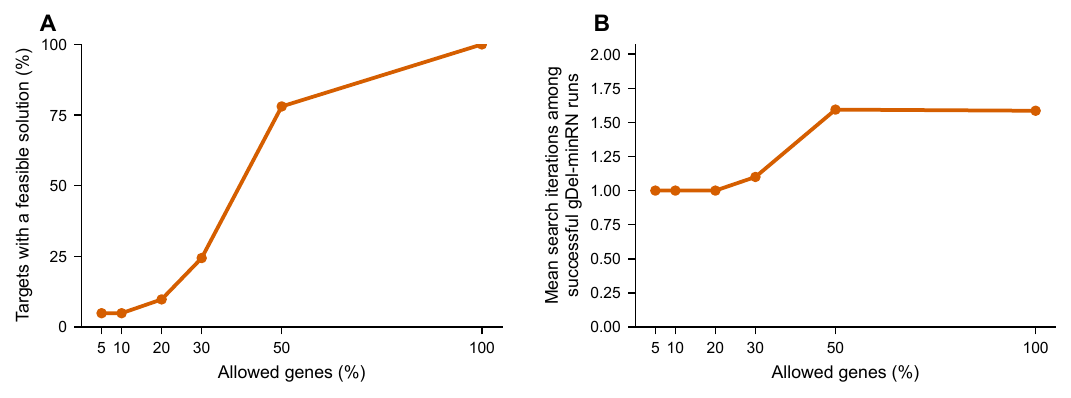}
  \caption{Restricting gDel-minRN's candidate gene pool to the top-$B\%$ of a gene ranking, on held-out \textit{e\_coli\_core} targets ($n{=}41$), under three ranking sources: HGPRGDel's deletion-priority ranking, GraphGDel's retention scores, and random gene order (mean over $5$ independent shuffles per target). (\textbf{A}) Fraction of targets for which a feasible growth-coupled solution exists within the allowed gene pool, as a function of $B$. (\textbf{B}) Mean number of search iterations required by gDel-minRN among the runs that found a feasible solution, as a function of $B$.}
  \label{fig:si-search-cost}
\end{figure}

At $5$--$20\%$, random ranking yielded higher success rates ($7.8$--$14.1\%$) than HGPRGDel ($4.9$--$9.8\%$) or GraphGDel ($2.4$--$9.8\%$). From $30\%$ onward, HGPRGDel achieved a higher success rate than both alternatives, reaching $78.0\%$ at $50\%$, compared with $70.7\%$ for GraphGDel and $35.1\%$ for random ranking (Fig.~\ref{fig:si-search-cost}).

The reversal at very small budgets indicates that restricting the candidate pool to the highest-ranked genes is not uniformly beneficial when the allowed pool is extremely small. One possible explanation is that a highly concentrated ranking may provide less pathway diversity within a small prefix than a random sample, although this possibility was not directly tested here.

Mean search iterations among successful runs remained within a narrow range ($1.00$--$1.90$) across ranking sources and budgets, with no consistent ranking-dependent pattern (Fig.~\ref{fig:si-search-cost}). Thus, the primary effect of the ranking was on the probability that a feasible solution was contained in the restricted intervention space, rather than on the number of search iterations once a solution was reachable.

Detailed per-budget values for HGPRGDel, including recall and Jaccard overlap with the MetNetComp reference set, are reported in Table~\ref{tab:si-search-cost} for completeness rather than as the primary endpoint of this analysis: gDel-minRN is formulated to minimize the number of repressed reactions subject to feasibility, not to reproduce a specific curated reference strategy, so enlarging the allowed gene pool does not necessarily increase its overlap with the reference set.

This experiment therefore supports a ranking- and budget-dependent search-space coverage--feasibility trade-off, rather than a uniform claim that learned ranking improves constraint-based search at every budget. This analysis was performed as a controlled case study on \textit{e\_coli\_core}, chosen because its small gene roster allowed the unrestricted ($B=100\%$) search to be run to completion rather than approximated; it does not constitute a claim about computational savings on larger GEMs.

\begin{table}[!p]
  \centering
  \caption{Search-space restriction and downstream gDel-minRN outcomes on held-out \textit{e\_coli\_core} targets ($n{=}41$). Mean recall and Jaccard are computed against the MetNetComp reference strategy and are reported for completeness rather than as the primary endpoint.}
  \label{tab:si-search-cost}
  \resizebox{0.9\linewidth}{!}{%
  \begin{tabular}{rrrrrr}
    \toprule
    Budget (\%) & Genes allowed & Successes & Mean search iterations & Mean GT recall & Mean Jaccard \\
    \midrule
    5   & 7   & 2/41  & 1.00 & 0.074 & 0.074 \\
    10  & 14  & 2/41  & 1.00 & 0.181 & 0.181 \\
    20  & 28  & 4/41  & 1.00 & 0.404 & 0.403 \\
    30  & 42  & 10/41 & 1.10 & 0.607 & 0.601 \\
    50  & 69  & 32/41 & 1.59 & 0.885 & 0.822 \\
    100 & 137 & 41/41 & 1.59 & 0.874 & 0.653 \\
    \bottomrule
  \end{tabular}
  }\\[2pt]
  \begin{minipage}{\linewidth}
    \footnotesize
    Mean search iterations are averaged over successful runs only. Budget $100\%$ corresponds to the unrestricted gDel-minRN search; an internal consistency check (restricting the candidate pool to all $137$ genes explicitly) reproduced this result exactly for all $41$ targets.
  \end{minipage}
\end{table}

\subsection{Full-Gene Supplementary Evaluation}
\label{si:full-gene}

The primary evaluation (main-text Experimental Setup) is restricted to target-dependent genes, since this is the scientific question of interest: deletion behavior that varies with the target metabolite.
Pre-determined genes remain part of the supervised training signal (Methods) but are excluded from the reported metrics.
As a supplementary check, Table~\ref{tab:si-full-gene} recomputes macro-F1 and AUC over the complete gene roster of each GEM, using the same fold-averaged predictions that underlie the primary evaluation.

\begin{table}[!htbp]
  \centering
  \caption{Full-gene supplementary evaluation. ``Target-dependent'' reproduces the primary evaluation subset (pooled over the fold-averaged predictions); ``Full gene roster'' additionally includes pre-determined genes.}
  \label{tab:si-full-gene}
  \resizebox{0.9\linewidth}{!}{%
  \begin{tabular}{lrrrr}
    \toprule
    GEM & Target-dependent Macro-F1 & Target-dependent AUC & Full-gene Macro-F1 & Full-gene AUC \\
    \midrule
    \textit{e\_coli\_core} & $89.04\%$ & $96.48\%$ & $92.20\%$ & $97.93\%$ \\
    \textit{iMM904}        & $88.81\%$ & $96.25\%$ & $97.60\%$ & $99.76\%$ \\
    \textit{iML1515}       & $90.40\%$ & $97.10\%$ & $97.98\%$ & $99.67\%$ \\
    \bottomrule
  \end{tabular}
  }\\[2pt]
  \begin{minipage}{\linewidth}
    \footnotesize
    Values are pooled over the fold-averaged predictions rather than averaged per-fold, so they differ slightly from the per-fold mean\,$\pm$\,s.d.\ reported in main-text Tables~2 and 3.
    Full-gene macro-F1/AUC are higher than the target-dependent subset on all three GEMs, consistent with the primary-evaluation rationale (main-text Experimental Setup): pre-determined genes are, by construction, easier to classify from their near-constant label alone, so including them raises the aggregate score without adding condition-dependent evidence.
  \end{minipage}
\end{table}

\subsection{All-Target Top-10 Recovery}
\label{si:all-target-top10}

The main-text case study (``A Target-Level Case Study of GPR-Aware Deletion Prioritization'') reports top-10 recovery for a single representative target (succinate).
Table~\ref{tab:si-all-target-top10} extends the same diagnostic to every target with at least ten evaluable target-dependent genes. For each target, we count how many of the ten lowest-retention target-dependent genes match the curated reference.

\begin{table}[!htbp]
  \centering
  \caption{All-target top-10 recovery, computed identically to the main-text case study for every evaluated target.}
  \label{tab:si-all-target-top10}
  \begin{tabular}{lrrr}
    \toprule
    GEM & Targets evaluated & Mean top-10 hit rate & Targets with $10/10$ hits \\
    \midrule
    \textit{e\_coli\_core} & $41$  & $98.0\%$  & $36/41$ ($87.8\%$) \\
    \textit{iMM904}        & $105$ & $100.0\%$ & $105/105$ ($100.0\%$) \\
    \textit{iML1515}       & $408$ & $99.5\%$  & $400/408$ ($98.0\%$) \\
    \bottomrule
  \end{tabular}\\[2pt]
  \begin{minipage}{\linewidth}
    \footnotesize
    On \textit{iML1515}, the target-dependent gene set contains only $33$ genes, so the ten lowest-retention calls represent a substantial fraction of the evaluable genes; the hit rate should therefore be interpreted as a diagnostic of ranking concentration rather than as an operating-point performance metric.
  \end{minipage}
\end{table}

\subsection{Sensitivity to the Target-Dependent Gene Definition}
\label{si:rho-sensitivity}

The primary evaluation treats a gene as target-dependent as defined in Section~\ref{si:labels}.
Table~\ref{tab:si-rho-sensitivity} recomputes macro-F1 and AUC under two alternative label-consistency thresholds, $\rho\in\{0.80,0.95\}$, re-partitioning the same fold-averaged predictions into a target-dependent subset and reusing the existing per-fold decision threshold $\theta$ (i.e., this is a sensitivity of the reported evaluation to how the target-dependent gene set is defined, not a re-run of training or threshold selection).

\begin{table}[!htbp]
  \centering
  \caption{Sensitivity to the target-dependent gene definition. $\rho{=}0.90$ reproduces the primary evaluation.}
  \label{tab:si-rho-sensitivity}
  \begin{tabular}{lrrrr}
    \toprule
    GEM & $\rho$ & Target-dependent genes & Macro-F1 & AUC \\
    \midrule
    \textit{e\_coli\_core} & $0.80$ & $41$ & $87.98\%$ & $96.19\%$ \\
                            & $0.90$ & $66$ & $89.04\%$ & $96.48\%$ \\
                            & $0.95$ & $80$ & $89.54\%$ & $96.83\%$ \\
    \addlinespace
    \textit{iMM904}         & $0.80$ & $21$ & $77.39\%$ & $84.95\%$ \\
                            & $0.90$ & $55$ & $88.81\%$ & $96.25\%$ \\
                            & $0.95$ & $96$ & $91.73\%$ & $97.86\%$ \\
    \addlinespace
    \textit{iML1515}        & $0.80$ & $21$ & $81.22\%$ & $92.34\%$ \\
                            & $0.90$ & $33$ & $90.40\%$ & $97.10\%$ \\
                            & $0.95$ & $68$ & $92.14\%$ & $98.23\%$ \\
    \bottomrule
  \end{tabular}\\[2pt]
  \begin{minipage}{\linewidth}
    \footnotesize
    Macro-F1/AUC are pooled as in Table~\ref{tab:si-full-gene}, so the $\rho{=}0.90$ rows differ slightly from main-text Tables~2 and 3.
    On \textit{iMM904} and \textit{iML1515}, the stricter $\rho{=}0.80$ threshold retains a smaller and more label-diverse gene subset, on which macro-F1 is correspondingly lower; the qualitative pattern is consistent across all three GEMs. Higher $\rho$ retains more genes and increases the aggregate score, and macro-F1 increases monotonically with $\rho$ on every GEM.
  \end{minipage}
\end{table}

\subsection{Target-Level Bootstrap Uncertainty}
\label{si:target-bootstrap}

As a complement to the fold-level Nadeau--Bengio analysis (Section~\ref{si:wilcoxon}), Table~\ref{tab:si-target-bootstrap} reports a $95\%$ confidence interval on \proposed's own macro-F1 and AUC obtained by resampling target metabolites with replacement ($2{,}000$ bootstrap resamples, seed $42$) from the fold-averaged predictions.
This characterizes uncertainty in \proposed's reported metrics under the target-based evaluation unit; it is not a paired comparison against other methods, since no per-target baseline predictions under the matched cross-validation protocol are available (Section~\ref{si:baselines}).

\begin{table}[!htbp]
  \centering
  \caption{Target-level bootstrap $95\%$ confidence intervals for \proposed\ (target-dependent genes, $2{,}000$ resamples over target metabolites, seed $42$).}
  \label{tab:si-target-bootstrap}
  \begin{tabular}{lrrr}
    \toprule
    GEM & Targets & Macro-F1 ($95\%$ CI) & AUC ($95\%$ CI) \\
    \midrule
    \textit{e\_coli\_core} & $41$  & $89.04\%$ $[85.55,\,92.35]$ & $96.48\%$ $[94.61,\,98.00]$ \\
    \textit{iMM904}        & $105$ & $88.81\%$ $[87.44,\,90.05]$ & $96.25\%$ $[95.44,\,96.94]$ \\
    \textit{iML1515}       & $443$ & $90.40\%$ $[89.60,\,91.20]$ & $97.10\%$ $[96.70,\,97.50]$ \\
    \bottomrule
  \end{tabular}
\end{table}

\clearpage

\section{Additional Ablation}
\label{si:ablation}

\subsection{Ablation Variant Definitions}
\label{si:ablation-variants}

Each component-ablation variant (main-text Table~3; Table~\ref{tab:si-ablation-full}) alters exactly one architectural component of the full model.
All variants retain the same gene and metabolite encoder architectures, the same prediction MLP ($256 \to 128 \to 1$), the same classifier input dimension ($d_g + d_m + D_{\mathrm{rxn}} = 320 + 256 + 128 = 704$), and the same cross-validation folds, seeds, and threshold protocol; only the component named below is changed.
For every variant except \textbf{No pretrained} the encoders are the frozen ESM-2 and ChemBERTa backbones of the full model; \textbf{No pretrained} keeps the same architectures but replaces the frozen pretrained weights with random initialization, as detailed below.

\begin{itemize}
  \item \textbf{Mean Pooling.} GPR-M stays enabled, but the \textbf{SoftBooleanAggregator} is replaced by a parameter-free aggregator that returns the arithmetic mean of the embeddings of all leaf genes of the reaction's GPR rule (unique gene names, collected recursively). The parse tree is never evaluated, so AND/OR structure, the OR-branch attention module, and the learnable soft-AND temperature parameter $\hat\tau$ are all discarded. The output dimension remains $d_g$, and the resulting reaction embeddings still feed Graph-M and Pred-M exactly as in the full model, so downstream modules and the classifier input are unchanged.
  \item \textbf{No Graph-M.} Hypergraph propagation is disabled (\texttt{use\_hypergraph=False}). The target metabolite representation degenerates to the raw Meta-M output, i.e.\ the frozen ChemBERTa embedding after its linear projection, with no propagation over reaction--incidence structure. GPR-M and Pred-M are fully retained. Because the graph encoder acts on the metabolite side only, gene representations are unaffected both in the full model and in this variant.
  \item \textbf{No GPR-M.} GPR-M is disabled (\texttt{use\_gpr=False}) and no reaction embeddings are computed. Graph-M stays enabled, but its GPR-conditioned hyperedge channel is replaced by a zero block at every layer; in Pred-M, the reaction-context term becomes a zero vector passed through the association MLP, which returns a learned, pair-independent constant. GPR-derived information is therefore removed from both the topology and association pathways, while the classifier input dimension is unchanged.
  \item \textbf{No pretrained.} Both language-model encoders keep the ESM-2 and ChemBERTa architectures but are instantiated from their configurations with randomly initialized weights, and are trained end-to-end instead of being frozen (Supplementary Section~S2.3). This variant shares the folds and seeds of the full model; the only differences are the backbone weight initialization and the fact that both backbones are trainable rather than frozen.
\end{itemize}

\textbf{No Graph-M} is not the same as the \textbf{NMA} baseline of main-text Table~2: NMA disables both Graph-M and GPR-M and substitutes a fixed one-hop neighborhood-mean aggregation, whereas \textbf{No Graph-M} keeps GPR-M intact and applies no aggregation fallback.

\subsection{Per-Fold Held-Out Results}
\label{si:fold-f1}

Table~\ref{tab:si-fold-f1} lists per-fold held-out test macro-F1 for all methods across three GEMs.
Sample means and standard deviations over the five folds reproduce main-text Table~2.

\begin{table}[!htbp]
  \centering
  \caption{Per-fold held-out test macro-F1 (\%) on target-dependent gene--metabolite pairs for all methods. Each fold evaluates target metabolites excluded from training in that round (Section~S1.4).}
  \label{tab:si-fold-f1}
  \resizebox{0.8\linewidth}{!}{%
  \begin{tabular}{llccccc}
    \toprule
    GEM & Method & Fold 1 & Fold 2 & Fold 3 & Fold 4 & Fold 5 \\
    \midrule
    \multirow{5}{*}{\textit{e\_coli\_core}}
      & DNN       & 54.67 & 62.23 & 57.09 & 62.94 & 57.77 \\
      & DeepGDel  & 65.27 & 73.43 & 68.22 & 74.28 & 69.60 \\
      & NMA       & 67.03 & 74.19 & 68.71 & 75.67 & 68.95 \\
      & GraphGDel & 70.05 & 76.10 & 75.07 & 71.77 & 78.11 \\
      & \proposed & \textbf{88.30} & \textbf{92.90} & \textbf{92.37} & \textbf{83.45} & \textbf{88.55} \\
    \midrule
    \multirow{5}{*}{\textit{iMM904}}
      & DNN       & 63.48 & 67.01 & 64.54 & 67.23 & 64.89 \\
      & DeepGDel  & 74.04 & 77.36 & 75.14 & 77.62 & 75.74 \\
      & NMA       & 75.25 & 78.24 & 75.87 & 78.73 & 76.06 \\
      & GraphGDel & 79.71 & 82.45 & 81.81 & 80.61 & 83.12 \\
      & \proposed & \textbf{88.22} & \textbf{87.08} & \textbf{88.58} & \textbf{88.24} & \textbf{90.95} \\
    \midrule
    \multirow{5}{*}{\textit{iML1515}}
      & DNN       & 56.61 & 59.17 & 57.33 & 59.39 & 57.59 \\
      & DeepGDel  & 61.03 & 63.96 & 61.92 & 64.31 & 62.48 \\
      & NMA       & 62.27 & 64.86 & 62.69 & 65.38 & 62.90 \\
      & GraphGDel & 70.65 & 72.71 & 72.22 & 71.31 & 73.25 \\
      & \proposed & \textbf{89.66} & \textbf{90.91} & \textbf{89.69} & \textbf{91.34} & \textbf{90.46} \\
    \bottomrule
  \end{tabular}
  }
\end{table}

\subsection{Full Ablation Study (All Metrics)}
\label{si:ablation-full}

Main-text Table~3 reports macro-F1 and AUC for the component ablation study.
Table~\ref{tab:si-ablation-full} extends this to all five metrics across all three GEMs.

\begin{table*}[h]
  \centering
  \caption{Full ablation study: all five metrics across three GEMs (mean\,$\pm$\,s.d.\ over five held-out metabolite test folds on target-dependent pairs). Bold: best per GEM block.}
  \label{tab:si-ablation-full}
  \resizebox{0.99\linewidth}{!}{%
  \begin{tabular}{llccccc}
    \toprule
    GEM & Configuration & Accuracy (\%) & M-Prec.\ (\%) & M-Recall (\%) & Macro-F1 (\%) & AUC (\%) \\
    \midrule
    \multirow{5}{*}{\textit{e\_coli\_core}}
      & Full           & $\mathbf{89.34\pm3.76}$ & $\mathbf{89.33\pm3.75}$ & $\mathbf{89.27\pm3.82}$ & $\mathbf{89.11\pm3.81}$ & $\mathbf{96.42\pm1.57}$ \\
      & Mean Pooling   & $79.24\pm3.29$ & $78.96\pm3.18$ & $79.51\pm3.01$ & $78.80\pm3.21$ & $88.10\pm3.24$ \\
      & No Graph-M     & $78.31\pm3.26$ & $78.28\pm3.09$ & $77.28\pm3.03$ & $77.58\pm3.29$ & $86.55\pm3.32$ \\
      & No pretrained  & $76.80\pm3.30$ & $76.50\pm3.20$ & $76.90\pm3.10$ & $76.50\pm3.30$ & $87.00\pm3.20$ \\
      & No GPR-M       & $76.68\pm3.28$ & $76.28\pm3.12$ & $76.43\pm3.05$ & $75.30\pm3.29$ & $84.60\pm3.34$ \\
    \midrule
    \multirow{5}{*}{\textit{iMM904}}
      & Full           & $\mathbf{89.65\pm1.21}$ & $\mathbf{88.64\pm1.27}$ & $\mathbf{88.61\pm1.59}$ & $\mathbf{88.61\pm1.42}$ & $\mathbf{96.22\pm1.12}$ \\
      & Mean Pooling   & $87.40\pm1.21$ & $86.00\pm1.12$ & $85.50\pm1.29$ & $85.30\pm1.19$ & $93.50\pm1.11$ \\
      & No Graph-M     & $85.00\pm1.24$ & $83.80\pm1.21$ & $83.50\pm1.30$ & $83.30\pm1.25$ & $92.10\pm1.14$ \\
      & No pretrained  & $82.80\pm1.20$ & $82.10\pm1.20$ & $81.90\pm1.30$ & $82.00\pm1.20$ & $91.00\pm1.10$ \\
      & No GPR-M       & $81.90\pm1.27$ & $81.50\pm1.23$ & $81.20\pm1.32$ & $81.30\pm1.27$ & $89.50\pm1.15$ \\
    \midrule
    \multirow{5}{*}{\textit{iML1515}}
      & Full           & $\mathbf{90.43\pm0.71}$ & $\mathbf{90.39\pm1.02}$ & $\mathbf{90.43\pm0.95}$ & $\mathbf{90.38\pm0.74}$ & $\mathbf{97.12\pm1.20}$ \\
      & Mean Pooling   & $88.00\pm0.82$ & $87.20\pm1.05$ & $86.50\pm1.01$ & $86.40\pm0.82$ & $95.20\pm1.29$ \\
      & No Graph-M     & $85.01\pm0.88$ & $84.77\pm1.08$ & $84.80\pm1.01$ & $84.40\pm0.92$ & $91.50\pm1.36$ \\
      & No pretrained  & $84.00\pm0.80$ & $83.40\pm1.00$ & $83.80\pm0.90$ & $83.35\pm0.80$ & $92.50\pm1.20$ \\
      & No GPR-M       & $83.24\pm0.91$ & $82.60\pm1.13$ & $83.87\pm1.04$ & $82.40\pm0.96$ & $89.30\pm1.41$ \\
    \bottomrule
  \end{tabular}
  }
\end{table*}

\subsection{AND-Branch Ablation of GPR-M}
\label{si:and-branch-ablation}

Table~\ref{tab:si-and-branch-ablation} compares three AND-node variants of the SoftBooleanAggregator on \textit{e\_coli\_core}.

\begin{table}[h]
  \centering
  \caption{AND-branch ablation of GPR-M on \textit{e\_coli\_core} (mean\,$\pm$\,s.d.\ over five held-out folds). ``Mean-only'' uses $\mathbf{z}_{\text{and}}{=}\mathbf{m}$; ``Soft-min-only'' uses $\mathbf{z}_{\text{and}}{=}\mathbf{s}$; ``Combined'' is the default. Bold: best.}
  \label{tab:si-and-branch-ablation}
  \resizebox{0.7\linewidth}{!}{%
  \begin{tabular}{lcc}
    \toprule
    AND-node configuration & Macro-F1 (\%) & AUC (\%) \\
    \midrule
    Combined (default) & $\mathbf{89.11 \pm 3.81}$ & $\mathbf{96.42 \pm 1.57}$ \\
    Mean-only          & $68.64 \pm 4.08$           & $78.08 \pm 4.23$           \\
    Soft-min-only      & $69.38 \pm 2.22$           & $81.43 \pm 1.87$           \\
    \bottomrule
  \end{tabular}
  }
\end{table}

\subsection{Graph-Encoding Comparison}
\label{si:graph-enc}

Table~\ref{tab:si-graph-enc} compares three Graph-M encoding variants on \textit{e\_coli\_core} while holding GPR-M and Pred-M fixed.

\begin{table}[h]
  \centering
  \caption{Graph-encoding comparison on \textit{e\_coli\_core} (macro-F1, mean\,$\pm$\,s.d.\ over five folds). GPR-M and Pred-M are fixed. Hypergraph row matches main-text Table~3 Full.}
  \label{tab:si-graph-enc}
  \begin{tabular}{lc}
    \toprule
    Graph encoder & Macro-F1 (\%) \\
    \midrule
    Hypergraph (default) & $\mathbf{89.11 \pm 3.81}$ \\
    Bipartite MPNN       & $79.44 \pm 2.77$ \\
    Pairwise GCN         & $75.99 \pm 2.91$ \\
    \bottomrule
  \end{tabular}
\end{table}

\section{Additional Case Studies}
\label{si:case-studies}

Case studies use a trained \proposed\ model on \textit{e\_coli\_core} restricted to target-dependent genes ($n{=}66$).
They illustrate how GPR type shapes the continuous gene-retention ranking for three TCA-neighbor targets: succinate (\texttt{succ\_c}; main text), cytosolic $\alpha$-ketoglutarate (\texttt{akg\_c}), and fumarate (\texttt{fum\_c}).
The primary objects are ranking average precision (AP; positive class $=$ curated reference label; score $=1-\hat{y}_{g,m}$) and the ten lowest-retention genes; the main-text succinate figure additionally shows the ten highest-retention genes as a bidirectional contrast (panel~B).
Binary overlap after the global macro-F1 threshold, full-roster Jaccard, and a single-target weak-GC check of that one thresholded set~\cite{yang2025dbgdel,tamura2023metnetcomp} are reported only as secondary operating-point descriptors (Section~S3).
The overlap precision and recall reported in this section are computed on the $66$-gene target-dependent subset and are therefore \emph{not} directly comparable to the full-roster recall and precision of main-text Table~4, which include pre-determined genes.
As in Section~\ref{si:full-gene} and Section~\ref{si:labels}, restricting to target-dependent genes removes the near-constant genes that are easiest to predict, so overlap metrics on this subset are systematically lower than full-roster ones; the two should not be read as competing estimates of the same quantity.

\begin{table}[h]
  \centering
  \caption{Ranking and threshold summary for three TCA-neighbor targets on \textit{e\_coli\_core} (target-dependent genes, $n{=}66$). Top-10 hit: fraction of the ten lowest-retention calls labeled Deleted in the curated reference. Precision / Recall / F1: target-dependent-subset overlap after the global macro-F1 threshold. Jaccard: full-roster gene-set overlap versus the curated reference. All three targets have top-10 hit $10/10$ (no false positives in the lowest-retention ten).}
  \label{tab:si-case-overview}
  \resizebox{0.8\linewidth}{!}{%
  \begin{tabular}{lccccccc}
    \toprule
    Target & AP & Top-10 & Precision & Recall & F1 & $|$Pred$|$ & Jaccard \\
    \midrule
    \texttt{succ\_c} & 0.91 & $10/10$ & 0.84 & 0.61 & 0.71 & 32 & 0.72 \\
    \texttt{akg\_c} & 0.96 & $10/10$ & 0.96 & 0.87 & 0.92 & 28 & 0.92 \\
    \texttt{fum\_c} & 0.94 & $10/10$ & 0.94 & 0.80 & 0.86 & 17 & 0.86 \\
    \bottomrule
  \end{tabular}
  }
\end{table}

\subsection{Succinate Production}
\label{si:case-succinate}

Succinate (\texttt{succ\_c}) is a dicarboxylic acid TCA intermediate with industrial relevance as a platform chemical and polymer precursor.
Growth-coupled succinate production in \textit{e\_coli\_core} requires blocking competing TCA exit routes while maintaining anaplerotic and growth-supporting flux.
Table~\ref{tab:si-case-overview} summarizes ranking and threshold metrics; Table~\ref{tab:si-case-succinate} lists the ten lowest-retention calls (main-text case-study figure).
We read the ordering as the deletion-preference signal, independent of whether any one thresholded set is flux-feasible.

\begin{table}[h]
  \centering
  \caption{Top-10 lowest-retention \proposed\ predicted deletions for \texttt{succ\_c} on \textit{e\_coli\_core} (target-dependent genes). Retention score: model output $\hat{y}_{g,m}$ (lower $=$ stronger deletion preference). The curated reference label is $0$ for deleted genes.}
  \label{tab:si-case-succinate}
  \resizebox{0.85\linewidth}{!}{%
  \begin{tabular}{lllll}
    \toprule
    Gene & Retention score & Curated & Example reaction & GPR type \\
    \midrule
    \texttt{b1773} & 0.007 & Deleted & \texttt{FBA} & OR \\
    \texttt{b1276} & 0.009 & Deleted & \texttt{ACONTa} & OR \\
    \texttt{b0733} & 0.046 & Deleted & \texttt{CYTBD} & AND \\
    \texttt{b0734} & 0.050 & Deleted & \texttt{CYTBD} & AND \\
    \texttt{b2587} & 0.058 & Deleted & \texttt{AKGt2r} & SINGLE \\
    \texttt{b3739} & 0.065 & Deleted & \texttt{ATPS4r} & AND \\
    \texttt{b0727} & 0.111 & Deleted & \texttt{AKGDH} & AND \\
    \texttt{b4151} & 0.113 & Deleted & \texttt{FRD7} & AND \\
    \texttt{b1817} & 0.131 & Deleted & \texttt{FRUpts2} & AND \\
    \texttt{b0809} & 0.137 & Deleted & \texttt{GLNabc} & AND \\
    \bottomrule
  \end{tabular}
  }
\end{table}

\textbf{Pathway interpretation.}
The highest-confidence deletions mix OR-type central-carbon isozyme calls (\texttt{FBA}, \texttt{ACONTa}) with AND-type respiratory and TCA complex subunits (\texttt{CYTBD}, \texttt{ATPS4r}, \texttt{AKGDH}, \texttt{FRD7}) and a SINGLE transport call (\texttt{AKGt2r}), indicating that GPR-aware ranking does not collapse to a single reaction class.
No false positives appear in the top-10 target-dependent genes.

\subsection{$\alpha$-Ketoglutarate Production}
\label{si:case-akg}

\textbf{Biological background.}
Cytosolic $\alpha$-ketoglutarate (\texttt{akg\_c}) is a TCA-cycle node and industrial platform precursor (amino-acid and biopolymer routes).
Growth-coupled accumulation requires limiting downstream consumption of $\alpha$-ketoglutarate while preserving anaplerosis and growth-supporting flux.

\textbf{Ranking.}
Metrics are summarized in Table~\ref{tab:si-case-overview}; Table~\ref{tab:si-case-akg} lists the ten lowest-retention calls.
Under the Section~S3 single-target FBA check, the default-threshold predicted deletion set is not weakly growth-coupled.

\begin{table}[h]
  \centering
  \caption{Top-10 lowest-retention \proposed\ predicted deletions for \texttt{akg\_c} on \textit{e\_coli\_core} (target-dependent genes).}
  \label{tab:si-case-akg}
  \resizebox{0.85\linewidth}{!}{%
  \begin{tabular}{lllll}
    \toprule
    Gene & Retention score & Curated & Example reaction & GPR type \\
    \midrule
    \texttt{b3952} & 0.000 & Deleted & \texttt{PFL} & AND \\
    \texttt{b1478} & 0.000 & Deleted & \texttt{ALCD2x} & OR \\
    \texttt{b0727} & 0.001 & Deleted & \texttt{AKGDH} & AND \\
    \texttt{b0811} & 0.002 & Deleted & \texttt{GLNabc} & AND \\
    \texttt{b3739} & 0.004 & Deleted & \texttt{ATPS4r} & AND \\
    \texttt{b0724} & 0.011 & Deleted & \texttt{SUCDi} & AND \\
    \texttt{b0903} & 0.013 & Deleted & \texttt{PFL} & AND \\
    \texttt{b2463} & 0.019 & Deleted & \texttt{ME2} & SINGLE \\
    \texttt{b2925} & 0.024 & Deleted & \texttt{FBA} & OR \\
    \texttt{b0729} & 0.026 & Deleted & \texttt{SUCOAS} & AND \\
    \bottomrule
  \end{tabular}
  }
\end{table}

\textbf{Pathway interpretation.}
The lowest-retention calls elevate pyruvate-formate lyase (\texttt{PFL}/\texttt{b3952}), a fermentation OR isozyme (\texttt{ALCD2x}/\texttt{b1478}), and the immediate downstream TCA consumer \texttt{AKGDH}/\texttt{b0727}, followed by energy and TCA AND blocks (\texttt{ATPS4r}, \texttt{SUCDi}, \texttt{SUCOAS}) with a SINGLE anaplerotic call (\texttt{ME2}).
Relative to succinate, cytosolic $\alpha$-ketoglutarate elevates \texttt{PFL} and fermentation OR calls ahead of several TCA consumers. In the succinate ranking, \texttt{FBA}/\texttt{ACONTa} and respiratory \texttt{CYTBD} occupy the earliest ranks.
No false positives appear in the top-10 target-dependent genes.

\textbf{Contrast with succinate and extracellular $\alpha$-ketoglutarate.}
The shared core network therefore yields overlapping but reordered deletion priorities across the two cytosolic TCA targets.
Extracellular \texttt{akg\_e} and cytosolic \texttt{akg\_c} yield the same top-10 target-dependent gene set in the same order; the \texttt{akg\_c} list is shown in Table~\ref{tab:si-case-akg} and is not duplicated.
Under the Section~S3 protocol, neither thresholded predicted set is weakly growth-coupled, whereas both curated reference sets are, illustrating that ranking can transfer across related targets while binary FBA feasibility remains set-specific.

\subsection{Fumarate Production}
\label{si:case-fum}

\textbf{Biological background.}
Fumarate (\texttt{fum\_c}) is a TCA-cycle intermediate adjacent to succinate and provides a third ranking case on the same target-dependent-gene roster.

\textbf{Ranking.}
Metrics are summarized in Table~\ref{tab:si-case-overview}; Table~\ref{tab:si-case-fum} lists the ten lowest-retention calls.
The fumarate column of the main-text case-study heatmap (panel~C) uses the same retention scores.

\begin{table}[h]
  \centering
  \caption{Top-10 lowest-retention \proposed\ predicted deletions for \texttt{fum\_c} on \textit{e\_coli\_core} (target-dependent genes).}
  \label{tab:si-case-fum}
  \resizebox{0.85\linewidth}{!}{%
  \begin{tabular}{lllll}
    \toprule
    Gene & Retention score & Curated & Example reaction & GPR type \\
    \midrule
    \texttt{b0733} & 0.007 & Deleted & \texttt{CYTBD} & AND \\
    \texttt{b1276} & 0.008 & Deleted & \texttt{ACONTa} & OR \\
    \texttt{b3739} & 0.089 & Deleted & \texttt{ATPS4r} & AND \\
    \texttt{b3731} & 0.093 & Deleted & \texttt{ATPS4r} & AND \\
    \texttt{b0734} & 0.097 & Deleted & \texttt{CYTBD} & AND \\
    \texttt{b0727} & 0.104 & Deleted & \texttt{AKGDH} & AND \\
    \texttt{b3734} & 0.115 & Deleted & \texttt{ATPS4r} & AND \\
    \texttt{b3952} & 0.175 & Deleted & \texttt{PFL} & AND \\
    \texttt{b1602} & 0.210 & Deleted & \texttt{THD2} & AND \\
    \texttt{b0978} & 0.237 & Deleted & \texttt{CYTBD} & AND \\
    \bottomrule
  \end{tabular}
  }
\end{table}

\textbf{Pathway interpretation.}
The earliest ranks concentrate on respiratory and ATP-synthase AND complexes (\texttt{CYTBD}, multiple \texttt{ATPS4r} subunits) together with aconitase OR (\texttt{ACONTa}) and downstream TCA AND calls (\texttt{AKGDH}, \texttt{PFL}, \texttt{THD2}).
Relative to succinate (where \texttt{FBA} leads) and cytosolic $\alpha$-ketoglutarate (where \texttt{PFL}/\texttt{ALCD2x} lead), fumarate elevates oxidative-phosphorylation subunits, reinforcing that the shared core network yields target-dependent deletion orderings rather than a single fixed gene list.
No false positives appear in the top-10 target-dependent genes.

\section{Complexity Analysis}
\label{si:complexity}

\subsection{GPR Parser: Scope and Algorithmic Complexity}
\label{si:gpr-parser}

The \proposed\ GPRRuleParser accepts COBRA-style infix rules with \texttt{and}/\texttt{or}, parentheses, and whitespace-separated gene identifiers.
Negation (\texttt{not}) and nonstandard dialects are not supported.
Empty rules yield a zero reaction embedding; missing leaf genes are mapped to a zero vector.

Parsing a GPR rule of length $L$ (character count) and tree depth $D$ requires $O(L)$ time and $O(D)$ stack space.
For all three evaluated GEMs, maximum parse-tree depth is~4 (Table~\ref{tab:si-gpr-complexity}) and mean length is well below 200 characters; parsing overhead is negligible relative to forward-pass computation.

Forward-pass complexity of GPR-M for a single reaction with $K$ genes in its rule is $O(K \cdot d_g)$ for embedding aggregation plus $O(K \cdot d_g^2)$ for the MHA computation in OR nodes (gene embedding dimension $d_g$).
Because the OR query is a single pooled vector (Section~\ref{si:sba}), the attention score matrix is $1 \times K$ rather than $K \times K$: the query, key, value, and output projections contribute $O(K \cdot d_g^2)$, while forming and applying the attention weights contributes only $O(K \cdot d_g)$.
Across all reactions in a GEM, total GPR-M forward-pass complexity is $O(|\mathit{R}| \cdot \bar{K} \cdot d_g + |\mathit{R}| \cdot \bar{K} \cdot d_g^2)$, where $\bar{K}$ is the mean number of genes per rule.
For \textit{iML1515} ($|\mathit{R}|{=}2712$, $\bar{K} \approx 2.0$, $d_g{=}320$), this is fully dominated by the MHA projection term.

\subsection{ESM-2 Embedding Coverage}
\label{si:esm2-coverage}

All genes referenced in GPR rules across the three evaluated GEMs have corresponding ESM-2 embeddings (Table~\ref{tab:si-esm2-coverage}).
The zero-vector substitution mechanism (for genes without a sequence record) was therefore not invoked for any reaction in the experimental datasets; every GPR tree leaf receives a non-zero gene embedding during both training and evaluation.

\begin{table}[h]
  \centering
  \caption{ESM-2 embedding coverage for genes referenced in GPR rules. ``Reactions with complete GPR coverage'' counts reactions whose every gene has an ESM-2 embedding.}
  \label{tab:si-esm2-coverage}
  \resizebox{0.98\linewidth}{!}{%
  \begin{tabular}{lrrrr}
    \toprule
    GEM & GPR-referenced genes & Genes with ESM-2 & Gene coverage & Reactions (complete GPR) \\
    \midrule
    \textit{e\_coli\_core} & 137   & 137   & 100\% & 69\,/\,69\ (100\%)    \\
    \textit{iMM904}        & 905   & 905   & 100\% & 1{,}043\,/\,1{,}043\ (100\%) \\
    \textit{iML1515}       & 1{,}516 & 1{,}516 & 100\% & 2{,}266\,/\,2{,}266\ (100\%) \\
    \bottomrule
  \end{tabular}
  }
\end{table}

\subsection{Currency Metabolite Filtering}
\label{si:currency}

Following GraphGDel~\cite{yang2025graphgdel}, we exclude high-flux cofactors from compound--reaction hyperedges when building the incidence matrix $\mathbf{H}$, unless the metabolite is a supervised production target.
Matching is case-insensitive on the BiGG/KEGG base identifier (compartment suffixes such as \texttt{\_c} are stripped);
Table~\ref{tab:si-currency} lists the excluded cofactors and their per-GEM effect.

\begin{table}[H]
  \centering
  \caption{Currency metabolites excluded from hypergraph incidence construction (unless a supervised target). The 20 filtered base metabolites are \texttt{atp}, \texttt{adp}, \texttt{amp}, \texttt{nad}, \texttt{nadh}, \texttt{nadp}, \texttt{nadph}, \texttt{coa}, \texttt{h}, \texttt{h2o}, \texttt{pi}, \texttt{ppi}, \texttt{co2}, \texttt{nh4}, \texttt{o2}, \texttt{glutathione}, \texttt{na1}, \texttt{k}, \texttt{mg2}, and \texttt{ca2}. Effect on each GEM: incidence kept / total.}
  \label{tab:si-currency}
  \resizebox{0.98\linewidth}{!}{%
  \begin{tabular}{lrrrr}
    \toprule
    GEM & Unique metabolites & Currency matched & Non-currency & Incidence kept / total \\
    \midrule
    \textit{e\_coli\_core} & 72   & 20 & 52   & 175\,/\,360    \\
    \textit{iMM904}        & 1226 & 88 & 1138 & 3{,}749\,/\,6{,}517  \\
    \textit{iML1515}       & 1877 & 41 & 1836 & 6{,}333\,/\,10{,}575 \\
    \bottomrule
  \end{tabular}
  }
\end{table}

\section{Mathematical Details of Differentiable GPR}
\label{si:math}

This section provides the complete mathematical formulation of the SoftBooleanAggregator, including gradient expressions and numerical implementation.
The goal is to make the differentiable GPR construction self-contained without relying on the biological motivation in the main text.

\subsection{Soft Boolean Aggregation}
\label{si:sba}

Let $\{g_1, \ldots, g_K\}$ be the genes in a GPR rule, with projected embeddings $\mathbf{u}_i = W_{\text{proj}}\,\mathbf{e}_{g_i} \in \mathbb{R}^d$ (dimension $d = d_g = 320$).

\subsubsection*{AND Node (Enzyme Complexes)}

An AND node encodes a set of obligate subunits: all $K$ genes must be present for the reaction to be catalytically active.
The aggregation combines a \textit{mean branch} and a \textit{soft-min branch}.

\textbf{Mean branch.}
\begin{equation}
  \mathbf{m} = \frac{1}{K} \sum_{i=1}^{K} \mathbf{u}_i
\end{equation}
The mean captures the average embedding of all subunits, summarizing overall complex capacity.

\textbf{Soft-min branch.}
Applied independently to each embedding dimension $j \in \{1,\ldots,d\}$:
\begin{equation}
  s_j = -\log \sum_{i=1}^{K} \exp\!\left(-\frac{(u_i)_j}{\tau}\right)
\end{equation}
where $\tau = \mathrm{softplus}(\hat\tau) + \epsilon > 0$ is the effective soft-AND temperature derived from the learnable scalar $\hat\tau$ (Section~\ref{si:numerical-stability}).
This is a temperature-scaled log-sum-exp surrogate for $\min_i (u_i)_j / \tau$ over subunit $j$-coordinates; the aggregator consumes $\mathbf{s}$ directly, and $\tau s_j$ itself is the soft minimum.
Shifting by $\min_i (u_i)_j$ (Section~\ref{si:numerical-stability}) shows that every shifted exponent is non-positive, so the resulting sum lies in $[1, K]$ and $s_j - \min_i (u_i)_j / \tau \in [-\log K, 0]$.
Hence $\tau s_j \to \min_i (u_i)_j$ as $\tau \to 0$ (hard minimum, exact AND bottleneck).
This global $1/\tau$ factor is absorbed by the learned MLP $\phi_{\land}$ that consumes $\mathbf{s}$, leaving the AND bottleneck semantics unchanged.

The dimension-wise vector $\mathbf{s} = (s_1, \ldots, s_d)$ encodes the bottleneck constraint: the dimension in which the limiting subunit has the smallest embedding value dominates.

\textbf{AND output.}
\begin{equation}
  \mathbf{z}_{\text{and}} = \phi_{\land}\!\left([\mathbf{m} \,\|\, \mathbf{s}]\right)
\end{equation}
where $[\,\cdot\,\|\,\cdot\,]$ denotes concatenation and $\phi_{\land} : \mathbb{R}^{2d} \to \mathbb{R}^{d}$ is a two-layer MLP with ReLU activation (hidden width $256$).

\subsubsection*{OR Node (Isozyme Alternatives)}

An OR node encodes a set of isozymes: any one of the $K$ alternative genes is sufficient for catalytic activity.

\textbf{Multi-head attention (MHA) branch.}
The MHA mechanism computes a context-aware weighted combination of all isozyme embeddings:
\begin{equation}
  \mathbf{a} = \mathrm{MHA}(\mathbf{q}, \mathbf{U}, \mathbf{U}) \in \mathbb{R}^d,
  \qquad
  \mathbf{q} = \frac{1}{K}\sum_{i=1}^{K} \mathbf{u}_i,
  \qquad
  \mathbf{U} = [\mathbf{u}_1^\top; \cdots; \mathbf{u}_K^\top]
\end{equation}
with $H{=}4$ attention heads (main text).
The query is a single pooled vector rather than the full set of isozyme embeddings, so this is single-query cross-attention over $K$ keys and values.
Attention weights are data-dependent, allowing the aggregator to up-weight the isozyme most relevant to the deletion context.

\textbf{Log-sum-exp branch (soft-max).}
Applied dimension-wise:
\begin{equation}
  \ell_j = \log \sum_{i=1}^{K} \exp\!\left((u_i)_j\right)
\end{equation}
The OR branch carries no temperature parameter.
Shifting by $\max_i (u_i)_j$ gives $\ell_j - \max_i (u_i)_j \in [0, \log K]$, so $\ell_j$ is a smooth upper bound on the strongest isozyme coordinate, tight to within $\log K$.
The OR semantics, whereby \emph{any} sufficient alternative enables the reaction, are encoded by focusing representation on the strongest available isozyme.

\textbf{OR output.}
\begin{equation}
  \mathbf{z}_{\text{or}} = \phi_{\lor}\!\left([\mathbf{a} \,\|\, \boldsymbol{\ell}]\right)
\end{equation}
where $\phi_{\lor} : \mathbb{R}^{2d} \to \mathbb{R}^{d}$ is a two-layer MLP with ReLU activation (hidden width $256$).

Every node of the parse tree therefore maps $\mathbb{R}^{d}\to\mathbb{R}^{d}$ with $d=d_g=320$, so the reaction embedding produced by GPR-M has dimension $d_g$, not $D_{\text{rxn}}$.
The reduction to $D_{\text{rxn}}=128$ is performed downstream: Pred-M pools $\mathbf{Z}_{\text{rxn}}^{(r)}$ over $R_{g\cup m}$ and applies the association MLP $\mathbb{R}^{d_g}\to\mathbb{R}^{D_{\text{rxn}}}$, so that the classifier consumes $d_g + d_m + D_{\text{rxn}} = 704$ dimensions (Table~\ref{tab:si-hyperparams}).

\subsubsection*{Single-Gene and Empty Rules}

A single-gene rule ($K{=}1$) returns $\mathbf{u}_1$ directly (no aggregation needed).
An empty GPR string (no associated gene) returns $\mathbf{0} \in \mathbb{R}^d$.

\subsection{Gradient Propagation}
\label{si:gradients}

\subsubsection*{Soft-Min Gradient (AND Node)}

The gradient of $s_j$ with respect to $(u_k)_j$ is:
\begin{equation}
  \frac{\partial s_j}{\partial (u_k)_j}
  = \frac{1}{\tau}\cdot\frac{\exp\!\left(-\frac{(u_k)_j}{\tau}\right)}{\sum_{i=1}^K \exp\!\left(-\frac{(u_i)_j}{\tau}\right)}
  = \frac{1}{\tau}\,\mathrm{softmax}\!\left(-\frac{\mathbf{u}_{(\cdot)j}}{\tau}\right)_k
\end{equation}
This is the softmax of the \emph{negated} inputs: the gene with the \textbf{smallest} embedding value in dimension $j$ receives the \textbf{largest} gradient weight.
Consequently, during training, the loss signal flows predominantly through the subunit with the weakest representation, the bottleneck gene of the enzyme complex.
This gradient routing encourages the model to accurately represent the limiting subunit, which determines deletion feasibility for AND-typed reactions.

\subsubsection*{Log-Sum-Exp Gradient (OR Node)}

The gradient of $\ell_j$ with respect to $(u_k)_j$ is:
\begin{equation}
  \frac{\partial \ell_j}{\partial (u_k)_j}
  = \frac{\exp\!\left((u_k)_j\right)}{\sum_{i=1}^K \exp\!\left((u_i)_j\right)}
  = \mathrm{softmax}\!\left(\mathbf{u}_{(\cdot)j}\right)_k
\end{equation}
The gene with the \textbf{largest} embedding value receives the largest gradient weight.
Training therefore concentrates representation quality on the most capable isozyme, consistent with OR semantics: the reaction is blocked only if \emph{all} alternatives are deleted, making the strongest isozyme the critical blocking target.

\subsection{Numerical Stability}
\label{si:numerical-stability}

Direct computation of $\exp(-u/\tau)$ for small $\tau$ or large $|u|$ can overflow or underflow in single-precision arithmetic.
We apply the standard log-sum-exp identity:
\begin{equation}
  \log \sum_{i=1}^K \exp(x_i) = c + \log \sum_{i=1}^K \exp(x_i - c), \quad c = \max_i x_i
\end{equation}
For the soft-min branch, set $x_i = -(u_i)_j / \tau$, so $c = -\min_i(u_i)_j / \tau$:
\begin{equation}
  s_j = \frac{\min_i (u_i)_j}{\tau}
    - \log \sum_{i=1}^K \exp\!\left(\frac{\min_i(u_i)_j - (u_i)_j}{\tau}\right)
\end{equation}
All exponent arguments are non-positive after shifting by the maximum, preventing overflow.
The analogous shift applies to the log-sum-exp branch of OR nodes.

Temperature $\tau$ is parameterized as:
\begin{equation}
  \tau = \mathrm{softplus}(\hat\tau) + \epsilon, \qquad \epsilon = 10^{-6}
\end{equation}
where $\hat\tau$ is the unconstrained learnable scalar and $\mathrm{softplus}(x) = \log(1+e^x)$.
This guarantees $\tau > \epsilon > 0$ throughout training and prevents division by zero in the gradient computation.
The initial value of $\hat\tau$ is set so that $\tau \approx 1$ at the start of training.

\subsection{Empirical Fidelity of Soft Boolean Approximations}
\label{si:softbool-fidelity}

SoftBoolean fidelity is reported in the main-text mechanism figure (Panels~A--B); OR-branch attention entropy appears in Fig.~\ref{fig:si-umap-entropy}B.
All numbers below come from the per-fold model states of the single protocol, under the graph convention of Section~\ref{si:graph-convention}.
For completeness, the per-GEM Spearman correlations (soft vs.\ hard leaf $L_2$ scalars, i.e., the $L_2$ norm of each leaf embedding after soft vs.\ hard aggregation) are: AND soft-min vs.\ hard-min $r_s = 1.000$, $0.987$, $0.857$ on \textit{e\_coli\_core} ($n{=}10$), \textit{iMM904} ($n{=}86$), and \textit{iML1515} ($n{=}221$); OR log-sum-exp vs.\ hard-max $r_s = 0.992$, $0.998$, $0.924$ ($n{=}32$, $265$, $743$).
The AND correlation is lower on \textit{iML1515} than on the two smaller GEMs, which we report as observed rather than attribute to a mechanism: the AND stratum there is the largest of the three ($n{=}221$ versus $10$ and $86$) and the soft-min gate compresses leaf scalars whose hard-min values span the widest range.
OR-branch attention entropy $H_{\mathrm{att}}$ on \textit{iML1515} increases with isozyme group count $K$ but remains below $\ln K$ for most reactions, indicating non-uniform isozyme weighting rather than degenerate uniform attention.
Entropy below $\ln K$ establishes only that the weighting is non-uniform; it does not by itself demonstrate that the attention branch is functionally necessary.
All evaluations use frozen leaf embeddings at the final trained checkpoint of each GEM.

\section{Biological Interpretation}
\label{si:bio}

\subsection{Reaction Representation Analysis}
\label{si:embedding}

\begin{figure}[htbp]
  \centering
  \includegraphics[width=0.98\linewidth]{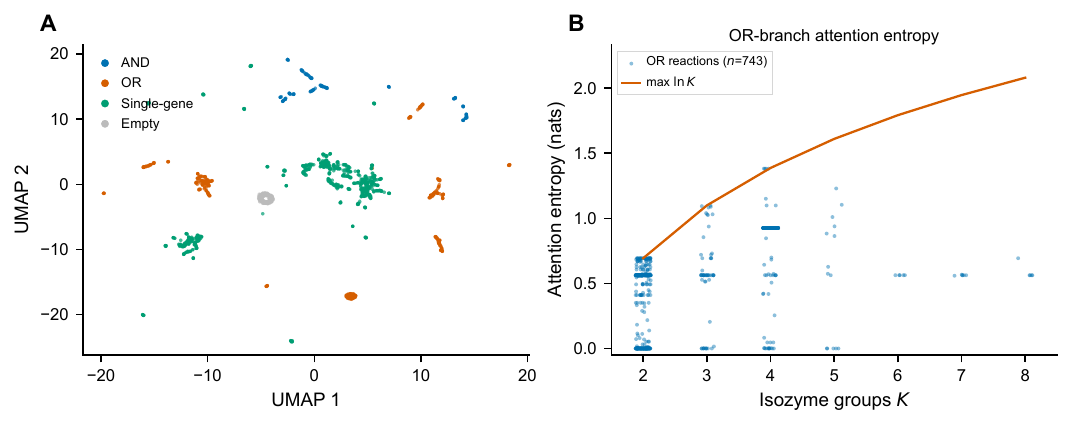}
  \caption{Supporting mechanism evidence.
  (\textbf{A}) UMAP of \textit{iML1515} reaction embeddings ($|\textit{R}|{=}2{,}712$) colored by top-level GPR type (AND, OR, Single-gene, Empty). Partial co-localization indicates that the learned representation encodes GPR type without explicit supervision; quantitative assessment is reported in main-text Fig.~4C,D.
  (\textbf{B}) OR-branch attention entropy $H_{\mathrm{att}}$ vs.\ isozyme group size $K$ (\textit{iML1515}); solid line $= \ln K$ (maximum entropy). Most OR reactions fall below $\ln K$, indicating non-uniform isozyme weighting rather than degenerate uniform attention.}
  \label{fig:si-umap-entropy}
\end{figure}

Reaction embeddings $\mathbf{Z}_{\text{rxn}}^{(r)}$ were extracted from trained \proposed\ models using the final reaction representations produced by GPR-M.
To avoid label leakage, embeddings were computed via fold-agnostic full-GEM forward passes with frozen ESM-2 and ChemBERTa encoders.
The main-text mechanism figure (Fig.~4) provides six panels of quantitative mechanistic evidence: SoftBoolean AND fidelity (Panel~A); OR fidelity (Panel~B); compact embedding diagnostics: linear probe AND/OR accuracy and nearest-neighbor subsystem enrichment (Panel~C); deletion F1 stratified by exclusive gene--GPR role (Panel~D); GPR-rule representation separation via per-reaction silhouette (Panel~E); and the association between the continuous deletion-priority score and reference-set membership, stratified by GPR role (Panel~F).
Panel~D deletion F1 (deletion as the positive class; MIXED-rule genes excluded) is $88.2$/$87.4$/$88.9\%$ (AND/OR/SINGLE) on \textit{e\_coli\_core}, $96.2$/$86.8$/$92.5\%$ on \textit{iMM904}, and $81.1$/$88.8$/$79.6\%$ on \textit{iML1515}; stratum sizes are $n_{\mathrm{genes}}{=}33/28/4$, $10/12/27$, and $17/5/11$, respectively.
Figure~\ref{fig:si-umap-entropy} provides additional supporting evidence: Panel~A shows a UMAP of \textit{iML1515} ($|\textit{R}|{=}2{,}712$) reaction embeddings colored by top-level GPR type, illustrating partial co-localization; Panel~B shows OR-branch attention entropy ($H_{\mathrm{att}}$) vs.\ isozyme group size $K$, confirming non-uniform attention weighting.
The linear probe underlying main-text Panel~C is restricted to AND- and OR-type reactions (i.e., multi-gene rules); single-gene and empty reactions are excluded because their GPR type is trivially recoverable from reaction degree, not from embedding geometry.
Embedding geometry was assessed using nearest-neighbor enrichment analysis ($k{=}10$) against $1{,}000$ randomly generated baselines (Table~\ref{tab:si-embedding-stats}).
Two-dimensional visualizations were generated using UMAP ($n_{\mathrm{neighbors}}{=}15$, $\mathrm{min\_dist}{=}0.1$, seed~42).

\begin{table}[h]
  \centering
  \caption{Reaction embedding organization tests. NN same-subsystem fraction compares observed neighbor purity to label-permutation baselines ($k{=}10$, $1{,}000$ shuffles of subsystem labels) in $\mathbf{Z}_{\text{rxn}}$ space; $z$-scores indicate standard deviations above the shuffled mean. NMI uses KMeans with one cluster per annotated subsystem (or per GPR type); NMI is reported descriptively, with no random baseline computed, and is not used as standalone evidence of embedding quality.}
  \label{tab:si-embedding-stats}
  \resizebox{0.98\linewidth}{!}{%
  \begin{tabular}{lrrrrrr}
    \toprule
    GEM & $|\mathit{R}|$ & NN same-subsystem & Random NN & $z$ & NMI (subsystem) & NMI (GPR type) \\
    \midrule
    \textit{e\_coli\_core} & 95   & 0.225 & $0.123 \pm 0.013$ & $7.7$   & 0.404 & 0.471 \\
    \textit{iMM904}        & 1577 & 0.196 & $0.042 \pm 0.004$ & $36.5$  & 0.357 & 0.411 \\
    \textit{iML1515}       & 2712 & 0.411 & $0.076 \pm 0.003$ & $104.8$ & 0.287 & 0.504 \\
    \bottomrule
  \end{tabular}
  }
\end{table}

Silhouette scores (cosine) are $-0.22$ (\textit{e\_coli\_core}), $-0.31$ (\textit{iMM904}), and $-0.25$ (\textit{iML1515}).
We report these scores for completeness, but do not use them as evidence of embedding quality: metabolic subsystems share intermediates and are not expected to form compact, disjoint clusters in embedding space, and the silhouette score penalizes such overlap regardless of whether it is biologically meaningful.
Nearest-neighbor enrichment and the linear probe are the primary quantitative evidence for embedding organization; they measure local neighborhood structure rather than global cluster compactness.
Nearest-neighbor subsystem enrichment is consistently above random across all three GEMs ($z{=}7.7$, $36.5$, and $104.8$ for \textit{e\_coli\_core}, \textit{iMM904}, and \textit{iML1515}, respectively), with the strongest enrichment on \textit{iML1515}.

\begin{figure}[t]
  \centering
  \includegraphics[width=0.5\linewidth]{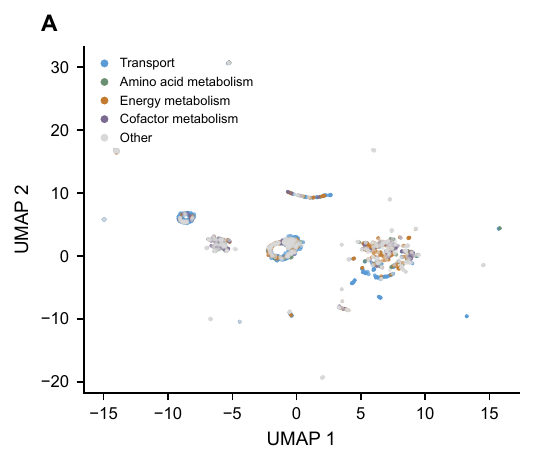}
  \caption{UMAP of GPR-M reaction embeddings on \textit{iML1515}, colored by four coarse metabolic categories (transport, amino-acid metabolism, energy metabolism, cofactor metabolism). Reactions outside these categories or without a coarse mapping are shown in grey ($n$ shown in figure). Global color mixing is expected because pathways share intermediates and is not used as quantitative evidence.}
  \label{fig:si-pathway-umap}
\end{figure}

Figure~\ref{fig:si-pathway-umap} shows a qualitative UMAP of \textit{iML1515} reaction embeddings colored by four coarse metabolic categories.
Global color mixing is expected because pathways share intermediates and is not used as quantitative evidence; organization is reported in the main-text mechanism figure (Panel~C diagnostics).

\end{document}